\documentclass[twocolumn,superscriptaddress,longbibliography]{revtex4-1}
\date{\today}
\usepackage{amsmath,amssymb}
\usepackage{epsfig}
\usepackage{epstopdf}
\usepackage{subfigure}
\usepackage{graphicx}
\usepackage{dcolumn}
\usepackage{bm}
\usepackage[colorlinks,citecolor=blue,linkcolor=blue,hyperindex,CJKbookmarks]{hyperref}
\usepackage{float}
\usepackage{comment}
\usepackage{xcolor}
\makeatletter

\newcommand{\pr}   {PrRhC$_2$}
\newcommand{\nd}   {NdRhC$_2$}
\newcommand{\la}   {LaRhC$_2$}
\newcommand{\yco}   {YCoC$_2$}
\newcommand{\luco}   {LuCoC$_2$}
\newcommand{\gdco}   {GdCoC$_2$}
\newcommand{\gdru}   {GdRuC$_2$}
\newcommand{\gdni}   {GdNiC$_2$}
\newcommand{\gen}   {RMC$_2$}

\begin{document}

\title{Tunable chirality of noncentrosymmetric magnetic Weyl semimetals}

\author{Rajyavardhan Ray}
 \email{r.ray@ifw-dresden.de}
\affiliation{Institute for Theoretical Solid State Physics, Leibniz IFW Dresden, Helmholtzstr. 20, 01069 Dresden, Germany}
 \affiliation{Dresden Center for Computational Materials Science (DCMS), TU Dresden, 01062 Dresden, Germany}
\author{Banasree Sadhukhan}
\affiliation{Institute for Theoretical Solid State Physics, Leibniz IFW Dresden, Helmholtzstr. 20, 01069 Dresden, Germany}
\author{Manuel Richter}
\affiliation{Institute for Theoretical Solid State Physics, Leibniz IFW Dresden, Helmholtzstr. 20, 01069 Dresden, Germany}
\affiliation{Dresden Center for Computational Materials Science (DCMS), TU Dresden, 01062 Dresden, Germany}
\author{Jorge I. Facio}
\affiliation{Institute for Theoretical Solid State Physics, Leibniz IFW Dresden, Helmholtzstr. 20, 01069 Dresden, Germany}
\author{Jeroen van den Brink}
\email{j.van.den.brink@ifw-dresden.de}
\affiliation{Institute for Theoretical Solid State Physics, Leibniz IFW Dresden, Helmholtzstr. 20, 01069 Dresden, Germany}
\affiliation{Dresden Center for Computational Materials Science (DCMS), TU Dresden, 01062 Dresden, Germany}
\affiliation{Institute of Theoretical Physics, Technische Universit{\"a}t Dresden, 01062 Dresden, Germany}

\date{\today}

\begin{abstract}
Even if Weyl semimetals are characterized by quasiparticles with well-defined chirality, exploiting
this experimentally is severely hampered by Weyl lattice-fermions coming in pairs with opposite
chirality, typically causing the net chirality picked up by experimental probes to vanish.
Here we show this issue can be circumvented in a controlled manner when both time-reversal-
and inversion-symmetry are broken.
To this end, we investigate chirality-disbalance in the carbide family {\gen} (R a rare-earth and M a 
transition metal), showing several members to be Weyl semimetals.
Using the noncentrosymmetric ferromagnet {\nd} as an illustrating example, we show that an
\textit{odd} number of Weyl nodes can be stabilized at its Fermi surface by properly tilting its
magnetization.
The tilt direction determines the sign of the resulting net chirality, opening up a simple route to
control it.
\end{abstract}

\pacs{ }
\maketitle

{\it Introduction ---}
Since their experimental discovery in the TaAs family, Weyl semimetals continue to gain interest. 
The non-trivial topology of this electronic phase follows from the
geometrical properties \cite{berry1984quantal} associated with 
electronic bands. Specifically, for band-crossing points, a topological invariant can be defined as the
flux of Berry curvature through a surface enclosing the point. The
low-energy effective theory around such a point corresponds to the Weyl
equation \cite{weyl1929}, identifying the topological invariant with the
corresponding Weyl fermion chirality. 
In lattice systems, however, Weyl nodes must come in pairs of opposite chirality
\cite{nielsen1981no}. Each pair having zero net chirality severely hampers experimental probes
sensitive to Brillouin-zone integrated quantities from picking up the Weyl-node chirality: for this
to work, one should create an overall chirality-imbalance.

The essential condition to enable the existence of Weyl nodes in the first place
is broken
spin degeneracy of Bloch states at a generic crystal momentum. This
requires time-reversal symmetry ($\Theta$) or inversion symmetry
($\mathcal{I}$) to be broken.
While the first experimental confirmations of topological semimetals
were achieved in $\Theta$-symmetric compounds
\cite{xu2015,yang2015,lv2015,di2015,souma2016,xu2015b,xu2016,haubold2017tairte4},
magnetic compounds are naturally appealing due to the broad prospects
that the interplay between the electronic structure and external
magnetic fields can offer
\cite{PhysRevB.84.235126,hirschberger2016chiral,PhysRevB.95.161306,borisenko2019time,Deng895,
PhysRevResearch.1.032044,PhysRevLett.122.206401,PhysRevX.9.041065,zou2019study,liao2020materials}.
Still, experimentally confirmed magnetic topological semimetals are rather scarce.
Two recently investigated cases are the nodal-line semimetal Co$_2$MnGa
\cite{PhysRevLett.119.156401,belopolski1278} and the Weyl semimetal
Co$_3$Sn$_2$S$_2$
\cite{liu2018giant,wang2018large,Morali1286,liu2019magnetic}, both
centrosymmetric ferromagnets with non-trivial topology. 

\begin{figure}[b!]
\centering
\includegraphics[width=0.48\textwidth,angle=0]{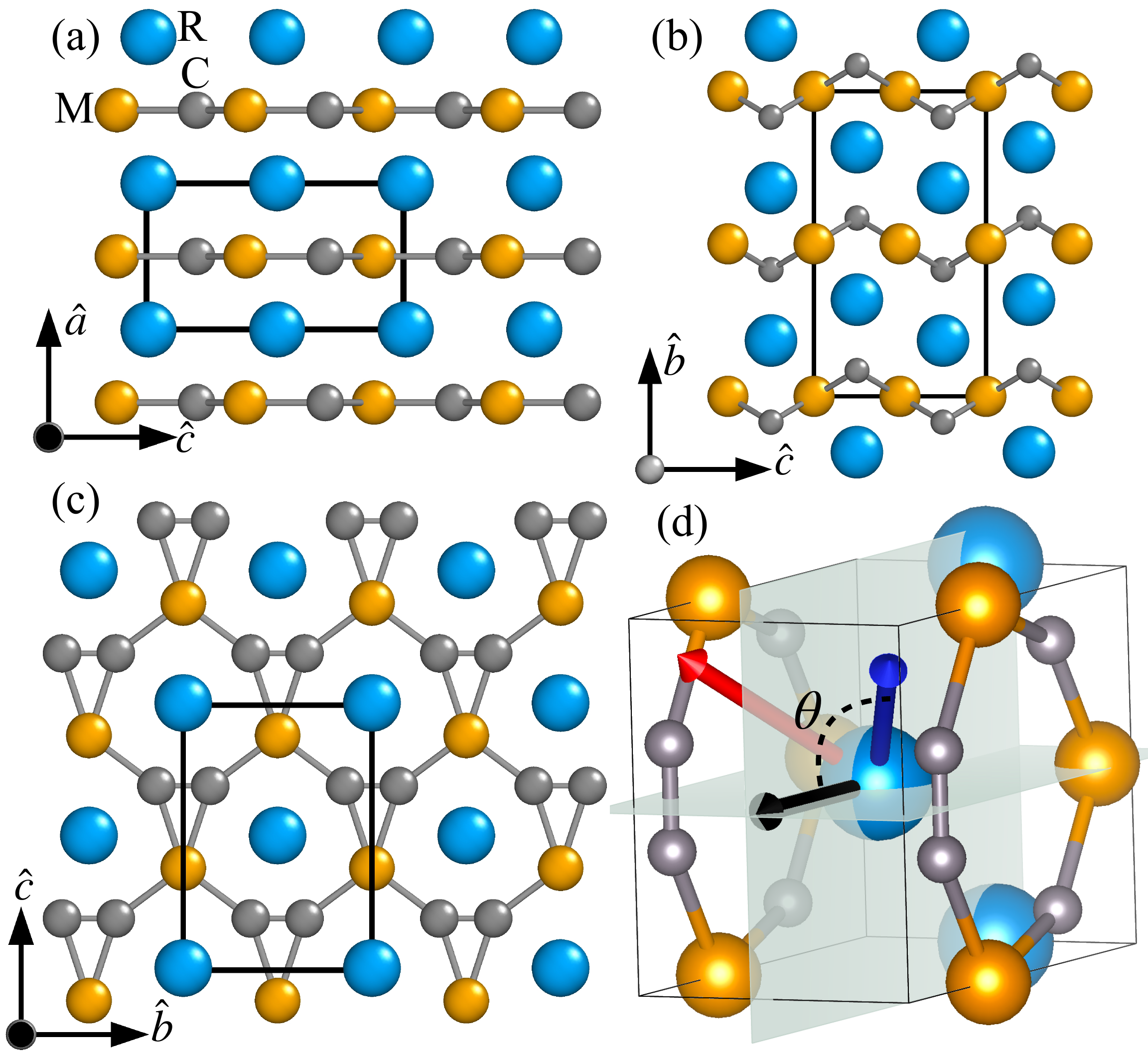}
	\caption{ Crystal structures of RMC$_2$. (a,c,d)  Compounds with noncentrosymmetric $Amm2$ space
        group, and (b) $Cmcm$ space group.
	(a) and (b) illustrate the layered structure, formed by a quasi-planar network spanned by the
    MC$_2$ complex while the rare-earth (R) ions occupy the interstitial space between the layers.
	In (d), reflection symmetry planes are depicted in green while the arrows indicate directions $[001]$ (black), $[111]$ (red) and $[-111]$ (blue).
	The conventional unit cells are indicated by black solid lines. 
	}
\label{fig:str} 
\end{figure}

Here we focus on the rare-earth carbides {\gen}, with R a rare-earth
metal and M a transition metal. 
This broad family of compounds exhibits a diversity of $\Theta$-breaking or $\mathcal{I}$-breaking phenomena
\cite{tsokol1988,hoffmann1989,hoffmann1995,matsuo1996,onodera1998magnetic,meng2016,meng2017,lee1996superconductivity,hirose2012fermi,PhysRevLett.102.117007,yanagisawa2012nonunitary,kolincio2016gdnic2,
steiner2018, kolincio2019ynic2,hanasaki2011successive}
both in long-range ordered magnetic compounds
\cite{tsokol1988,hoffmann1989,hoffmann1995,matsuo1996,onodera1998magnetic,meng2016,meng2017}
as well as in superconducting phases
\cite{lee1996superconductivity,hirose2012fermi,PhysRevLett.102.117007,yanagisawa2012nonunitary,yanagisawa2012nonunitary}
and a complex interplay between these phases and charge density waves
(CDW)
\cite{PhysRevLett.102.076404,PhysRevB.80.125111,hanasaki2012smnic2,prathiba2016tuning,hanasaki2011successive,kolincio2016gdnic2,kim2013chemical,hanasaki2017gdnic2,steiner2018,
kolincio2019ynic2}.
We show by consideration of available experimental information and own
density-functional calculations (DFT) that
{\gen} compounds
can be categorized in four classes:
(I) $\Theta$-symmetric and $\mathcal{I}$-broken semi-metals ({\yco} and {\luco}); 
(II) $\Theta$-broken and $\mathcal{I}$-symmetric metals ({\gdru});
(III) both $\Theta$- and $\mathcal{I}$-broken semi-metals ({\pr}, {\nd}, {\gdco} and {\gdni});
and (IV) insulators ({\la}), the latter being of secondary interest for this work.
We find that all the mentioned semi-metals possess Weyl nodes close to the 
Fermi energy, as summarized in Table \ref{table:WPs},
where they also tend to have relatively simple and uncluttered band structures. 
Figure \ref{fig:str} illustrates the two observed structure types of the 
(semi-)metallic compounds.

Systems belonging to class III are of particular interest as they allow for a specific demonstration of the 
unique interplay between topology and magnetism offered by $\mathcal{I}$-broken symmetry.
As a proof of principle,
we show for {\nd} how in ferromagnetic non-centrosymmetric phases
tilting of the magnetization ($\mathbf{m}$) along a low symmetry
direction produces a disbalance
in the number of opposite chirality Weyl fermions near the Fermi 
surface: of {\it all} Weyl nodes the degeneracy is lifted. 
Further, the direction in which the magnetization is canted
controls the sign of the chirality disbalance, allowing therefore to
\textit{switch} the dominant low-energy chirality of the electrons.

\begin{figure}[t!]
\centering
\includegraphics[width=0.5\textwidth,angle=0]{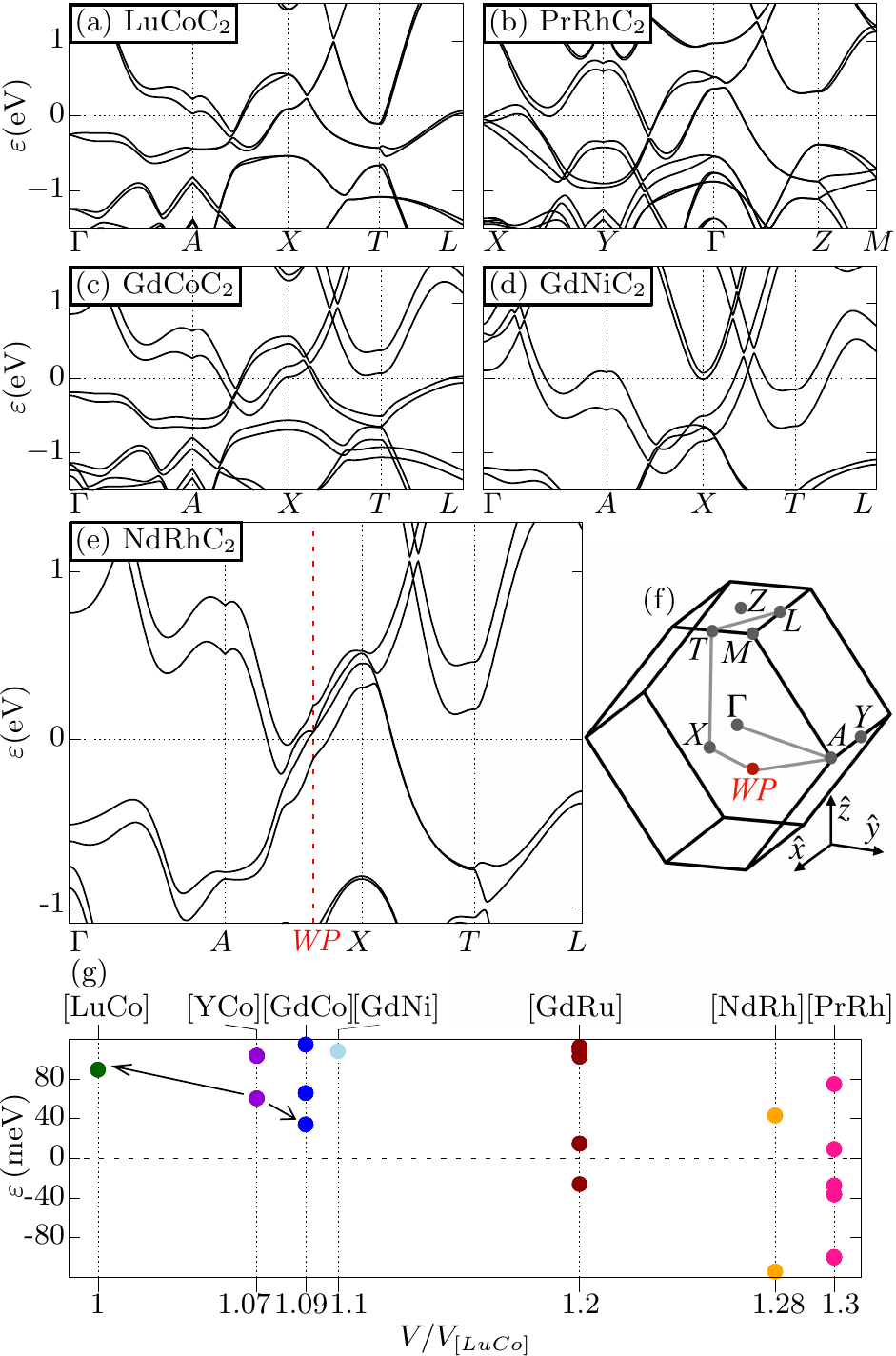}
	\caption{(a-e) Band structures of representative rare-earth dicarbide compounds.
(f) Brillouin zone. The red point corresponds to the Weyl node closest to the Fermi energy for NdRhC$_2$ and has coordinates $(0.665,0.309,-0.189)$\AA$^{-1}$.
For (b), the chosen high-symmetry points differ due to the AFM doubling of the unit-cell \cite{Note1}. 
	(g) Energy of the Weyl nodes vs. unit cell volume for all compounds considered in this work. 
	}
\label{fig:bands} 
\end{figure}

\begin{table*}[t!]
    \small
	\caption{Symmetry, magnetic ground state and, for the Weyl node
    closest to the Fermi energy, position, energy ($\varepsilon$),
degeneracy (Deg) and band index of the lower branch in the band
crossing ($N$ is the number of valence electrons).  Weyl nodes listed correspond to GGA results. For
class II and III, the open-core (OC)
approximation was applied with $\mathbf{m} \parallel [001]$. From top to bottom the compounds are presented in order of increasing unit cell volume. }
    \begin{tabular*}{0.98\textwidth}{ p{2.0cm} p{0.9cm} p{0.6cm} p{0.9cm}
	    p{2.4cm} p{1.0cm}  p{0.7cm}  p{0.3cm}  p{3.5cm} p{1.3 cm} p{0.6cm} p{0.75 cm}}
    \hline\hline
      {\bf Compound} & \multicolumn{3}{c}{\bf Symmetry} & \multicolumn{3}{c}{\bf Magnetic ground state}  & & \multicolumn{4}{c}{\bf Weyl nodes}\\
	    & Class & $\mathcal{I}$ & $\Theta$  & Expt.& \multicolumn{2}{c}{Theory}  & & \multicolumn{1}{c}{Position}  & $\varepsilon$ (meV) & Deg  &  Band \\
	    &  &  &  & & OC & $+U$ &  &  ($2\pi/a$, $2\pi/b$, $2\pi/c$) &  & & \\
    \hline
	    LuCoC$_2$ & I & $-$ & $+$ & NM$^a$ & NM  & NM                &  & ($0.362,0.163, 0$)       &
    $90$  & $4$ & $N-1$\\
	    YCoC$_2$ & I & $-$ & $+$ & NM$^b$ & NM  & NM                  & & ($0.366,0.173, 0$)       &
    $61$  & $4$ & $N-1$\\
	    GdCoC$_2$ & III  & $-$ & $-$ & $T_C$$=$$15$K$^c$  & FM & FM      &  & ($0.347, 0.206,0.171$)           & 
    $34$  & $4$ & $N-1$ \\
	    GdNiC$_2$ & III  & $-$ & $-$ & $T_N$$=$$20$K$^{d,e}$ & FM & FM  & & ($-1/2, 0, 0.487$)     &
    $108$  & $2$ & $N$ \\
	    GdRuC$_2$ &  II & $+$ & $-$ & $T_C$$=$45(3)K$^f$ & FM & FM  & & ($0.074, 0.178, 0$)  & 
    $15$  & $4$ & $N-2$ \\
	    NdRhC$_2$ & III  & $-$ & $-$ & $\theta_{CW}$$\sim$$0$K$^{g,h}$ & FM & FM   &  & ($0.393,0.233, -0.199$)  & 
    $43$  & $4$ & $N$  \\
	    PrRhC$_2$ & III  & $-$ & $-$ & $\theta_{CW}$$\sim$$-13$K$^{g,h}$ & AFM & FM          &  & ($0, 0.158,-0.221$) & 
    $9$  & $4$ & $N-1$ \\
    \hline
    \hline
    \multicolumn{10}{l}{$^a$ Ref. [\onlinecite{steiner2018}]; $^b$ Ref. [\onlinecite{xu2019}]; $^c$ Ref. [\onlinecite{meng2016}];$^d$
	    Ref. [\onlinecite{matsuo1996}]; $^e$ Ref. [\onlinecite{kolincio2016gdnic2}];  $^f$ Ref.
    [\onlinecite{hoffmann1995}]; $^g$ Ref. [\onlinecite{tsokol1988}]; $^h$ Ref. [\onlinecite{hoffmann1989}] } \\
    \end{tabular*}
    \label{table:WPs}
\end{table*}

\textit{From Y to Lu ---}
For compounds based on different R elements, we first present the main 
structural, electronic and magnetic properties of interest for this work.
Our DFT results were obtained using the Generalized Gradient
Approximation (GGA) \cite{pbe1996}, as implemented in FPLO-18
\cite{klaus1999,fplo_web} and considering for the treatment of the
$4f$-shell both the open-core approximation (OC)
with R-specific $4f$ spin moment but spherical orbital occupation \cite{Richter2001}
and the GGA+$U$ method with the full-localized limit for the double counting correction
\cite{Czyzyk1994} and parameters $U=7\,$eV and $J=1\,$eV.
We study compounds which have already been synthesized, including R = Y,
La, Nd, Pr, Gd or Lu, and M = Co, Rh, Ni or Ru \cite{tsokol1988,
hoffmann1989,matsuo1996,steiner2018}, and use the lattice parameters from {\it The
Materials Project} \cite{jain2013} (see the Supplementary Material (SM)
\cite{sm} for further details). Table \ref{table:WPs} includes 
available experimental data. For most of the compounds, we
find that the DFT calculations agree with the reported magnetic
properties. 
In order to analyze the electronic properties of the whole family on the same footing, we will
fix in the following the quantization axis along the in-plane direction $[001]$.

We find that the compounds involving rare earths with empty or
completely filled $f$-shells, R = Y, La or Lu, result in
$\Theta$-symmetric non-magnetic (NM) states, 
in agreement with experiment \cite{xu2019,hoffmann1989,steiner2018}. Among these, {\la} is the only 
system that crystallizes in the noncentrosymmetric tetragonal space group $P4_1$ \cite{hoffmann1989} 
and displays an insulating electronic structure (see SM \cite{sm}).
Opposed to this, {\yco} and {\luco} grow in the noncentrosymmetric orthorhombic space group 
$Amm2$ \cite{xu2019,steiner2018} and exhibit a semi-metallic band structure [see Fig.
    \ref{fig:bands}(a) and SM \cite{sm}].
Four bands dominate the energy spectrum of {\luco} near the Fermi energy.
These arise from hybridized Co-$d$, C-$p$ and R-$5d$ states (referred to as $pd$ 
states from now on).
The same characteristic four bands are present at low energy for other 
compounds with M = Co or Rh [Fig. \ref{fig:bands}(b), (c), (e)].
On the other hand, compounds based on transition metals not isoelectronic to Co,
namely, Ru or Ni, exhibit a shift of the Fermi energy of roughly $\pm 1$\,eV [Fig.
\ref{fig:bands}(d)]. This shift naturally changes the nature of the bands around the Fermi energy
and, as we will see, their electronic properties.

Members of the family based on magnetic R ions are characterized
as a lattice of localized moments on the R-$4f$ states coupled  with the
$pd$ states via an onsite (Kondo) exchange.  
Ruderman–Kittel–Kasuya–Yosida exchange interactions between the $4f$
moments can lead to long-range magnetic order including the induced moments on the $pd$ states.
Consequently, these compounds are an interesting platform to explore how
external magnetic fields, which couple primarily with the R-$4f$ states,
can tune Weyl-node properties of the low-energy electronic structure
associated with the $pd$ states.

Among the magnetic compounds, those obtained by replacing Lu by magnetic
rare earths while retaining the space group $Amm2$ [Fig. \ref{fig:str}(a),
(c)] belong to Class III, breaking both $\mathcal{I}$ and $\Theta$.
For these compounds, different magnetic states have been reported. 
{\gdco} was first described as antiferromagnetic (AFM) with in-plane
moments canted at $48^{\circ}$ from the $a$-axis and $T_{\rm N}= 15.6$ K
\cite{matsuo1996}.
However, a more recent study by Meng {\it et al.} finds a ferromagnetic
state (FM) below $\sim 15$\,K \cite{meng2016}.
GdNiC$_2$, on the other hand, as most of the Ni-based members in
the family \cite{onodera1998magnetic}, presents an AFM ground-state
\cite{matsuo1996}. 
For {\pr} and {\nd}, experimental data is limited to high-temperature
susceptibility measurements \cite{hoffmann1989}. These indicate small
Curie-Weiss temperatures ($\theta_{\rm CW}\sim-13\,$K and $\sim 0\,$K,
respectively) and magnetic moments in good agreement with the values
expected for the trivalent R ions.
{\gdru} exhibits an interesting contrast to the above, since it presents
a transition to a FM phase but crystallizes in the
$\mathcal{I}$-symmetric $Cmcm$ space group [Fig. \ref{fig:str}(b)], thus
belonging to Class II \cite{hoffmann1995}. 

Our DFT results correctly capture the NM states in {\yco} and {\luco} and the FM states in {\gdco} and {\gdru}.
For {\nd}, we find that both GGA+OC and GGA$+U$ predict an FM ground state.
Results for {\pr} are somewhat more complex, as GGA+OC and GGA$+U$ predict as ground state AFM and FM order, 
respectively. The only clear deviation from experiment is found in {\gdni}, where our calculations 
predict the ground state to be FM. One should, however, keep in mind
that in the Ni-based carbides, a strong interplay between the magnetic
state and a CDW (not explored in our calculations) has been established.
In fact, there are indications that the CDW tends to compete
against the FM phase
\cite{PhysRevLett.102.076404,hanasaki2012smnic2,kim2013chemical,kolincio2016gdnic2}.
Specifically in GdNiC$_2$, different metamagnetic transitions have been
observed under moderate external magnetic fields, yielding an
interesting and complex phase diagram
\cite{hanasaki2011successive,kolincio2016gdnic2, hanasaki2017gdnic2}.

\textit{Weyl nodes --- }
We now turn our attention to the topological properties of the electronic structure. 
For this, we will focus on the GGA+OC calculations. As a common
reference for the following, we define for each compound the number of
valence electrons as $N$ and search for Weyl points in a relevant
low-energy window $[-120,120]$\,meV, using the {\sc
Pyfplo} module of the FPLO package \cite{fplo_web}.

Figure \ref{fig:bands}(g) shows the energy of the identified Weyl nodes
for all the compounds considered in this work, ordered as a function of
their unit-cell volume ($V$).
Table \ref{table:WPs} includes the position and energy of the respective
lowest energy Weyl nodes (for a complete list, see SM \cite{sm}).
We also include YCoC$_2$ as a reference \footnote{For this compound, we have compared our results
with Ref. \cite{xu2019}. While we can reproduce the Weyl nodes in that work, additional search for
crossings between bands $N-1$ and $N$ discloses Weyl nodes at lower energy, as indicated in Fig.
\ref{fig:bands}(g)}.
Weyl nodes at low energy are found both for $\Theta$-symmetric and $\Theta$-breaking cases and,
among the latter, both for compounds with FM or AFM orderings.

Relations between the Weyl node structure of different compounds can also be established. 
For instance, an inspection of the node coordinates reveals that the
fourfold degenerate Weyl nodes of lowest energy in YCoC$_2$, which
lie in the $k_z=0$ plane, are also present in LuCoC$_2$ but are higher
in energy.
It further suggests that on replacing Y by Gd, the large exchange field
induced by the Gd-$4f$ spins moves these Weyl nodes away from the
$k_z=0$ plane and to a lower energy. To confirm this, we carried out 
explicit computation of Weyl points in {\yco} under external magnetic
field acting on the spin degrees of freedom. Indeed, we find that the field endows the nodes a finite $k_z$
component.

Regarding GdNiC$_2$ and GdRuC$_2$, the Fermi energy shift associated with their different number of
valence electrons with respect to GdCoC$_2$ is naturally in opposite directions and therefore has
different consequences. On GdRuC$_2$, it increases the density of states and the complexity of the
low-energy band structure. While it is interesting that it presents Weyl nodes close to the Fermi
surface, it should not be considered as a semi-metal \cite{Note1}.
On the other hand, the upward shift of Fermi energy makes the Ni-based compound of strong interest.
Indeed, upon this shift, a single pair of Weyl nodes -- the minimum possible in a periodic system --
lies $\sim 100\,$meV above the Fermi energy.
Our calculations performed on different compounds neatly explain the origin of these Weyl nodes: 
The isostructural but NM {\luco} features twofold
degenerate bands along the line $X$-$T$, forming a Dirac cone $\sim$1\,eV above the Fermi level
[Fig. \ref{fig:bands}(a)].
The band degeneracy is lifted in the isolectronic FM compounds 
[Fig. \ref{fig:bands}(c),(e)] and the Dirac cone is split into Weyl cones.
Substituting Co by Ni shifts these Weyl nodes closer to the Fermi level,
while keeping them along $X$-$T$ [Fig. \ref{fig:bands}(d)].
While the FM phase in GdNi$C_2$ has only been stabilized with an
external magnetic field
\cite{hanasaki2011successive,kolincio2016gdnic2}, it could be
interesting to study this phenomenology in SmNi$C_2$, where the
competing CDW is suppressed \cite{PhysRevLett.102.117007} leading to a FM
ground-state \cite{onodera1998magnetic}.

The rather large volume change caused by the substitutions Gd $\to$ Nd or
Pr and Co $\to$ Rh naturally induces sizable changes in the electronic
structure which remains, however, semi-metallic in NdRhC$_2$ and
PrRhC$_2$.
Also, low-energy Weyl nodes are present in these compounds [Fig. \ref{fig:bands}(g)].

{\it Pumping chirality to and through the Fermi surface --- }
Due to the Nielsen-Ninomiya ``no-go'' theorem for chiral lattice fermions
\cite{nielsen1981no,friedan1982proof}, Weyl nodes come in pairs of opposite chirality
\cite{witten2016three} (which can be broken up by very strong electron-electron interactions
\cite{PhysRevLett.122.046402,PhysRevResearch.2.012023}).
In a crystal, Weyl nodes occur in multiplets, their degeneracy being dictated by the Shubnikov group of the material which relates Weyl nodes of same or different chirality. Thus, the Weyl node degeneracy need not be even. 
The energy splitting between nodes of opposite chirality, a key quantity for the magnitude of different electromagnetic responses sensitive to the Weyl node chirality, actually depends on the material and external conditions. 
Low symmetry, in particular the absence of inversion and mirror symmetries is
of the essence. Here we build on the idea of using the magnetic degrees of freedom in a $\mathcal{I}$-broken material to reduce the symmetry such that the Shubnikov group contains only the identity.

\begin{figure}[t!]
\centering
\includegraphics[width=8. cm,angle=0]{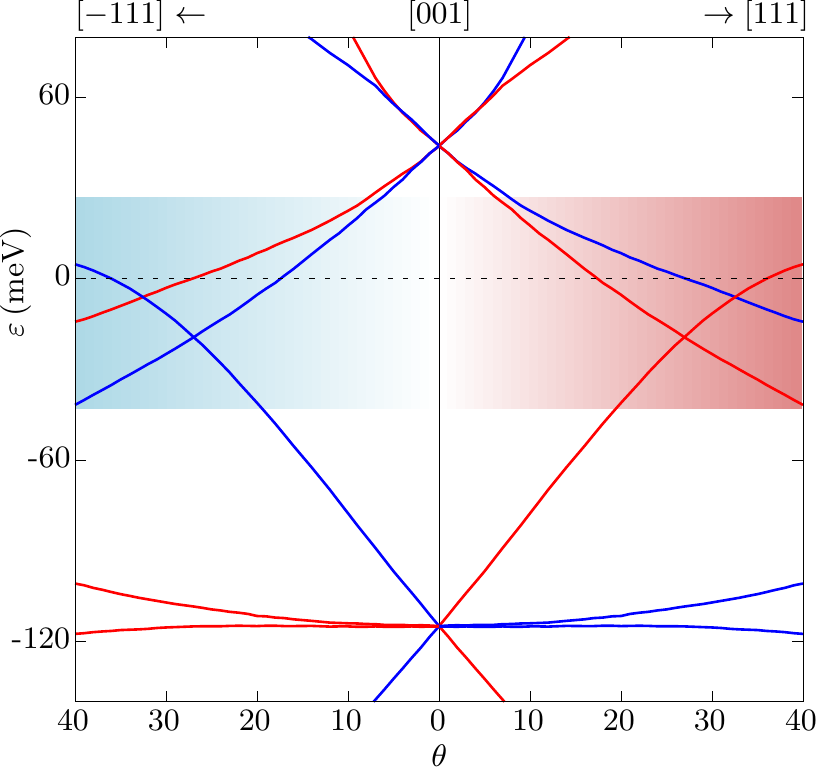}
	\caption{Energy of Weyl nodes in NdRhC$_2$ as a function of the canting angle $\theta$. 
	Starting from the magnetization along the direction $[001]$, on the
    left the canting is towards $[-111]$, while on the right is towards
    $[111]$.
	Blue and red lines correspond to Weyl nodes of positive and negative chirality, respectively.
The colored region highlights a wide angle range in which near the Fermi surface Weyl fermions of
one chirality are majority. }
\label{fig:wp_vs_angle} 
\end{figure}

As a proof of principle, we consider NdRhC$_2$, although the physics we discuss can be readily 
extended to other noncentrosymmetric magnetic compounds. 
In the FM ground-state, $\mathbf{m}$ points along the $[001]$ direction and the Shubnikov group contains
$\{E,\, m(x)\Theta,\, m(y)\Theta,\, C_2(z)\}$. 
Therefore, for each Weyl node away from high-symmetry lines, there are three degenerate symmetry-related partners.
Any component of $\mathbf{m}$ along a low-symmetry direction
does not only break the rotation $C_2(z)$ but also the symmetries involving mirrors. 
Thus, starting from $\mathbf{m}$ along $[001]$, a perturbation that cants $\mathbf{m}$ towards, \textit{e.g.}, $[111]$ leaves $\{E\}$ as the only symmetry element, removing all degeneracies among the Weyl nodes. The natural question is how large this effect is.

Figure \ref{fig:wp_vs_angle} shows the energy of Weyl
nodes in NdRhC$_2$ as $\mathbf{m}$ is canted at an angle $\theta$ towards $[-111]$
(on the left) or $[111]$ (right).
At $\theta=0$, there are two sets of fourfold degenerate Weyl nodes in the energy range $[-120,80]\,$meV.
Remarkably, even a moderate canting of $\mathbf{m}$ is enough to produce experimentally meaningful energy splittings between Weyl nodes of opposite chirality, of the order of tens of meV.
Similar results for GdCoC$_2$ (see SM \cite{sm}) indicate that such splitting need not scale proportionally to the
magnetic anisotropy field, expected to be smaller in the Gd compound than in {\nd} due to the
$S=J=7/2$ state of Gd$^{3+}$.

The tuning of Weyl nodes to the Fermi surface at specific angles resembles the prediction for Co$_3$Sn$_2$S$_2$ \cite{PhysRevResearch.1.032044}, 
with the difference that in the latter case the crystal structure is centrosymmetric and each Weyl node is degenerate.
This difference is crucial as $\mathcal{I}$ always connects Weyl nodes of opposite chirality and, therefore, enforces a vanishing total chirality of the Weyl nodes at the Fermi surface.
While mirror symmetries can be dynamically broken by external magnetic fields, 
breaking $\mathcal{I}$ is the structural prerequisite for tuning an odd number of Weyl nodes to the Fermi surface.

The two chosen canting directions are related by a
crystal mirror symmetry [Fig. \ref{fig:str}(d)], and therefore, the chirality of the Weyl
nodes reaching the Fermi surface at a certain $\theta$ is opposite for
$[-111]$ and $[111]$.
Thus, the canting direction provides an experimentally viable way of
discarding effects not originated in the chirality disbalance.
It is worth noting that this \textit{chirality valve} is effective not only
at the specific angle at which a Weyl node crosses the Fermi surface. 
Indeed, due to the particularly large energy trajectory of one of the nodes 
below the Fermi energy, in a wide range of angles for $\theta>20^\circ$, 
NdRhC$_2$ exhibits two Weyl fermions of negative (positive) and only one
of positive (negative)
chirality in a transport-relevant energy window of $\pm 40\,$meV around
the Fermi energy for $\theta$ towards $[111]$ ($[-111]$). 
This may generate ideal conditions to experimentally explore the role of Weyl nodes in gyrotropic and chiral magnetic effects \cite{PhysRevLett.116.077201,PhysRevB.92.235205}, in the nonlinear electric response of magnetic materials \cite{de2017quantized,holder2019consequences}, and in effects recently associated with the chiral anomaly like the planar Hall effect \cite{PhysRevLett.119.176804,nag2018transport}.

\textit{Conclusions ---} 
Combination of broken inversion and broken time-reversal symmetry with
a general orientation of the magnetization can lift all degeneracies
among Weyl points. 
We have demonstrated that this allows to tune an odd number of Weyl nodes to the Fermi surface
for a specific member of the {\gen} family, which we have also shown to host various Weyl semi-metals.
Our ideas can be naturally tested in all noncentrosymmetric compounds exhibiting
magnetic order and provide straightforward experimental access to chiral transport.

\vspace{2cm}

{\it Acknowledgments.}
We thank Ulrike Nitzsche for technical assistance and Klaus Koepernik for discussion.
We acknowledge financial support from the German Research Foundation (Deutsche Forschungsgemeinschaft, DFG) via SFB1143 Project No. A5 and under Germany's Excellence Strategy through Würzburg‐Dresden Cluster of Excellence on Complexity and Topology in Quantum Matter ‐ \textit{ct.qmat} (EXC 2147, Project No. 390858490).
RR and MR acknowledge partial financial support from the European Union
(ERDF) and the Free State of Saxony via the ESF Projects No. 100231947 and 
No. 100339533 (Young Investigators Group Computer Simulations for Materials
Design $-$ {\it CoSiMa}) during the early stages of the project. JIF acknowledges the support from the Alexander von Humboldt foundation.

\bibliography{rmc2}

\begin{thebibliography}{77}%
\makeatletter
\providecommand \@ifxundefined [1]{%
 \@ifx{#1\undefined}
}%
\providecommand \@ifnum [1]{%
 \ifnum #1\expandafter \@firstoftwo
 \else \expandafter \@secondoftwo
 \fi
}%
\providecommand \@ifx [1]{%
 \ifx #1\expandafter \@firstoftwo
 \else \expandafter \@secondoftwo
 \fi
}%
\providecommand \natexlab [1]{#1}%
\providecommand \enquote  [1]{``#1''}%
\providecommand \bibnamefont  [1]{#1}%
\providecommand \bibfnamefont [1]{#1}%
\providecommand \citenamefont [1]{#1}%
\providecommand \href@noop [0]{\@secondoftwo}%
\providecommand \href [0]{\begingroup \@sanitize@url \@href}%
\providecommand \@href[1]{\@@startlink{#1}\@@href}%
\providecommand \@@href[1]{\endgroup#1\@@endlink}%
\providecommand \@sanitize@url [0]{\catcode `\\12\catcode `\$12\catcode
  `\&12\catcode `\#12\catcode `\^12\catcode `\_12\catcode `\%12\relax}%
\providecommand \@@startlink[1]{}%
\providecommand \@@endlink[0]{}%
\providecommand \url  [0]{\begingroup\@sanitize@url \@url }%
\providecommand \@url [1]{\endgroup\@href {#1}{\urlprefix }}%
\providecommand \urlprefix  [0]{URL }%
\providecommand \Eprint [0]{\href }%
\providecommand \doibase [0]{https://doi.org/}%
\providecommand \selectlanguage [0]{\@gobble}%
\providecommand \bibinfo  [0]{\@secondoftwo}%
\providecommand \bibfield  [0]{\@secondoftwo}%
\providecommand \translation [1]{[#1]}%
\providecommand \BibitemOpen [0]{}%
\providecommand \bibitemStop [0]{}%
\providecommand \bibitemNoStop [0]{.\EOS\space}%
\providecommand \EOS [0]{\spacefactor3000\relax}%
\providecommand \BibitemShut  [1]{\csname bibitem#1\endcsname}%
\let\auto@bib@innerbib\@empty
\bibitem [{\citenamefont {Berry}(1984)}]{berry1984quantal}%
  \BibitemOpen
  \bibfield  {author} {\bibinfo {author} {\bibfnamefont {M.~V.}\ \bibnamefont
  {Berry}},\ }\bibfield  {title} {\bibinfo {title} {Quantal phase factors
  accompanying adiabatic changes},\ }\href
  {https://royalsocietypublishing.org/doi/abs/10.1098/rspa.1984.0023}
  {\bibfield  {journal} {\bibinfo  {journal} {Proceedings of the Royal Society
  of London. A. Mathematical and Physical Sciences}\ }\textbf {\bibinfo
  {volume} {392}},\ \bibinfo {pages} {45} (\bibinfo {year} {1984})}\BibitemShut
  {NoStop}%
\bibitem [{\citenamefont {Weyl}(1929)}]{weyl1929}%
  \BibitemOpen
  \bibfield  {author} {\bibinfo {author} {\bibfnamefont {H.}~\bibnamefont
  {Weyl}},\ }\bibfield  {title} {\bibinfo {title} {Elektron und gravitation.
  {I}},\ }\href {https://doi.org/https://doi.org/10.1007/BF01339504} {\bibfield
   {journal} {\bibinfo  {journal} {Zeitschrift f{\"u}r Physik A Hadrons and
  Nuclei}\ }\textbf {\bibinfo {volume} {56}},\ \bibinfo {pages} {330} (\bibinfo
  {year} {1929})}\BibitemShut {NoStop}%
\bibitem [{\citenamefont {Nielsen}\ and\ \citenamefont
  {Ninomiya}(1981)}]{nielsen1981no}%
  \BibitemOpen
  \bibfield  {author} {\bibinfo {author} {\bibfnamefont {H.~B.}\ \bibnamefont
  {Nielsen}}\ and\ \bibinfo {author} {\bibfnamefont {M.}~\bibnamefont
  {Ninomiya}},\ }\href
  {https://inis.iaea.org/search/searchsinglerecord.aspx?recordsFor=SingleRecord&RN=13643721}
  {\emph {\bibinfo {title} {No-go theorum for regularizing chiral fermions}}},\
  \bibinfo {type} {Tech. Rep.}\ (\bibinfo  {institution} {Science Research
  Council},\ \bibinfo {year} {1981})\BibitemShut {NoStop}%
\bibitem [{\citenamefont {Xu}\ \emph {et~al.}(2015{\natexlab{a}})\citenamefont
  {Xu}, \citenamefont {Belopolski}, \citenamefont {Alidoust}, \citenamefont
  {Neupane}, \citenamefont {Bian}, \citenamefont {Zhang}, \citenamefont
  {Sankar}, \citenamefont {Chang}, \citenamefont {Yuan}, \citenamefont {Lee}
  \emph {et~al.}}]{xu2015}%
  \BibitemOpen
  \bibfield  {author} {\bibinfo {author} {\bibfnamefont {S.-Y.}\ \bibnamefont
  {Xu}}, \bibinfo {author} {\bibfnamefont {I.}~\bibnamefont {Belopolski}},
  \bibinfo {author} {\bibfnamefont {N.}~\bibnamefont {Alidoust}}, \bibinfo
  {author} {\bibfnamefont {M.}~\bibnamefont {Neupane}}, \bibinfo {author}
  {\bibfnamefont {G.}~\bibnamefont {Bian}}, \bibinfo {author} {\bibfnamefont
  {C.}~\bibnamefont {Zhang}}, \bibinfo {author} {\bibfnamefont
  {R.}~\bibnamefont {Sankar}}, \bibinfo {author} {\bibfnamefont
  {G.}~\bibnamefont {Chang}}, \bibinfo {author} {\bibfnamefont
  {Z.}~\bibnamefont {Yuan}}, \bibinfo {author} {\bibfnamefont {C.-C.}\
  \bibnamefont {Lee}}, \emph {et~al.},\ }\bibfield  {title} {\bibinfo {title}
  {{Discovery of a Weyl fermion semimetal and topological Fermi arcs}},\ }\href
  {https://science.sciencemag.org/content/349/6248/613} {\bibfield  {journal}
  {\bibinfo  {journal} {Science}\ }\textbf {\bibinfo {volume} {349}},\ \bibinfo
  {pages} {613} (\bibinfo {year} {2015}{\natexlab{a}})}\BibitemShut {NoStop}%
\bibitem [{\citenamefont {Yang}\ \emph {et~al.}(2015)\citenamefont {Yang},
  \citenamefont {Liu}, \citenamefont {Sun}, \citenamefont {Peng}, \citenamefont
  {Yang}, \citenamefont {Zhang}, \citenamefont {Zhou}, \citenamefont {Zhang},
  \citenamefont {Guo}, \citenamefont {Rahn} \emph {et~al.}}]{yang2015}%
  \BibitemOpen
  \bibfield  {author} {\bibinfo {author} {\bibfnamefont {L.}~\bibnamefont
  {Yang}}, \bibinfo {author} {\bibfnamefont {Z.}~\bibnamefont {Liu}}, \bibinfo
  {author} {\bibfnamefont {Y.}~\bibnamefont {Sun}}, \bibinfo {author}
  {\bibfnamefont {H.}~\bibnamefont {Peng}}, \bibinfo {author} {\bibfnamefont
  {H.}~\bibnamefont {Yang}}, \bibinfo {author} {\bibfnamefont {T.}~\bibnamefont
  {Zhang}}, \bibinfo {author} {\bibfnamefont {B.}~\bibnamefont {Zhou}},
  \bibinfo {author} {\bibfnamefont {Y.}~\bibnamefont {Zhang}}, \bibinfo
  {author} {\bibfnamefont {Y.}~\bibnamefont {Guo}}, \bibinfo {author}
  {\bibfnamefont {M.}~\bibnamefont {Rahn}}, \emph {et~al.},\ }\bibfield
  {title} {\bibinfo {title} {{Weyl semimetal phase in the non-centrosymmetric
  compound TaAs}},\ }\href {https://www.nature.com/articles/nphys3425}
  {\bibfield  {journal} {\bibinfo  {journal} {Nature physics}\ }\textbf
  {\bibinfo {volume} {11}},\ \bibinfo {pages} {728} (\bibinfo {year}
  {2015})}\BibitemShut {NoStop}%
\bibitem [{\citenamefont {Lv}\ \emph {et~al.}(2015)\citenamefont {Lv},
  \citenamefont {Weng}, \citenamefont {Fu}, \citenamefont {Wang}, \citenamefont
  {Miao}, \citenamefont {Ma}, \citenamefont {Richard}, \citenamefont {Huang},
  \citenamefont {Zhao}, \citenamefont {Chen}, \citenamefont {Fang},
  \citenamefont {Dai}, \citenamefont {Qian},\ and\ \citenamefont
  {Ding}}]{lv2015}%
  \BibitemOpen
  \bibfield  {author} {\bibinfo {author} {\bibfnamefont {B.~Q.}\ \bibnamefont
  {Lv}}, \bibinfo {author} {\bibfnamefont {H.~M.}\ \bibnamefont {Weng}},
  \bibinfo {author} {\bibfnamefont {B.~B.}\ \bibnamefont {Fu}}, \bibinfo
  {author} {\bibfnamefont {X.~P.}\ \bibnamefont {Wang}}, \bibinfo {author}
  {\bibfnamefont {H.}~\bibnamefont {Miao}}, \bibinfo {author} {\bibfnamefont
  {J.}~\bibnamefont {Ma}}, \bibinfo {author} {\bibfnamefont {P.}~\bibnamefont
  {Richard}}, \bibinfo {author} {\bibfnamefont {X.~C.}\ \bibnamefont {Huang}},
  \bibinfo {author} {\bibfnamefont {L.~X.}\ \bibnamefont {Zhao}}, \bibinfo
  {author} {\bibfnamefont {G.~F.}\ \bibnamefont {Chen}}, \bibinfo {author}
  {\bibfnamefont {Z.}~\bibnamefont {Fang}}, \bibinfo {author} {\bibfnamefont
  {X.}~\bibnamefont {Dai}}, \bibinfo {author} {\bibfnamefont {T.}~\bibnamefont
  {Qian}},\ and\ \bibinfo {author} {\bibfnamefont {H.}~\bibnamefont {Ding}},\
  }\bibfield  {title} {\bibinfo {title} {{Experimental Discovery of Weyl
  Semimetal TaAs}},\ }\href {https://doi.org/10.1103/PhysRevX.5.031013}
  {\bibfield  {journal} {\bibinfo  {journal} {Phys. Rev. X}\ }\textbf {\bibinfo
  {volume} {5}},\ \bibinfo {pages} {031013} (\bibinfo {year}
  {2015})}\BibitemShut {NoStop}%
\bibitem [{\citenamefont {Di-Fei}\ \emph {et~al.}(2015)\citenamefont {Di-Fei},
  \citenamefont {Yong-Ping}, \citenamefont {Zhen}, \citenamefont {Yu-Peng},
  \citenamefont {Xiao-Hai}, \citenamefont {Qi}, \citenamefont {Pavel},
  \citenamefont {Zhu-An}, \citenamefont {Xian-Gang},\ and\ \citenamefont
  {Dong-Lai}}]{di2015}%
  \BibitemOpen
  \bibfield  {author} {\bibinfo {author} {\bibfnamefont {X.}~\bibnamefont
  {Di-Fei}}, \bibinfo {author} {\bibfnamefont {D.}~\bibnamefont {Yong-Ping}},
  \bibinfo {author} {\bibfnamefont {W.}~\bibnamefont {Zhen}}, \bibinfo {author}
  {\bibfnamefont {L.}~\bibnamefont {Yu-Peng}}, \bibinfo {author} {\bibfnamefont
  {N.}~\bibnamefont {Xiao-Hai}}, \bibinfo {author} {\bibfnamefont
  {Y.}~\bibnamefont {Qi}}, \bibinfo {author} {\bibfnamefont {D.}~\bibnamefont
  {Pavel}}, \bibinfo {author} {\bibfnamefont {X.}~\bibnamefont {Zhu-An}},
  \bibinfo {author} {\bibfnamefont {W.}~\bibnamefont {Xian-Gang}},\ and\
  \bibinfo {author} {\bibfnamefont {F.}~\bibnamefont {Dong-Lai}},\ }\bibfield
  {title} {\bibinfo {title} {{Observation of Fermi arcs in non-centrosymmetric
  Weyl semi-metal candidate NbP}},\ }\href
  {https://iopscience.iop.org/article/10.1088/0256-307X/32/10/107101}
  {\bibfield  {journal} {\bibinfo  {journal} {Chinese Physics Letters}\
  }\textbf {\bibinfo {volume} {32}},\ \bibinfo {pages} {107101} (\bibinfo
  {year} {2015})}\BibitemShut {NoStop}%
\bibitem [{\citenamefont {Souma}\ \emph {et~al.}(2016)\citenamefont {Souma},
  \citenamefont {Wang}, \citenamefont {Kotaka}, \citenamefont {Sato},
  \citenamefont {Nakayama}, \citenamefont {Tanaka}, \citenamefont {Kimizuka},
  \citenamefont {Takahashi}, \citenamefont {Yamauchi}, \citenamefont {Oguchi}
  \emph {et~al.}}]{souma2016}%
  \BibitemOpen
  \bibfield  {author} {\bibinfo {author} {\bibfnamefont {S.}~\bibnamefont
  {Souma}}, \bibinfo {author} {\bibfnamefont {Z.}~\bibnamefont {Wang}},
  \bibinfo {author} {\bibfnamefont {H.}~\bibnamefont {Kotaka}}, \bibinfo
  {author} {\bibfnamefont {T.}~\bibnamefont {Sato}}, \bibinfo {author}
  {\bibfnamefont {K.}~\bibnamefont {Nakayama}}, \bibinfo {author}
  {\bibfnamefont {Y.}~\bibnamefont {Tanaka}}, \bibinfo {author} {\bibfnamefont
  {H.}~\bibnamefont {Kimizuka}}, \bibinfo {author} {\bibfnamefont
  {T.}~\bibnamefont {Takahashi}}, \bibinfo {author} {\bibfnamefont
  {K.}~\bibnamefont {Yamauchi}}, \bibinfo {author} {\bibfnamefont
  {T.}~\bibnamefont {Oguchi}}, \emph {et~al.},\ }\bibfield  {title} {\bibinfo
  {title} {{Direct observation of nonequivalent Fermi-arc states of opposite
  surfaces in the noncentrosymmetric Weyl semimetal NbP}},\ }\href
  {https://journals.aps.org/prb/abstract/10.1103/PhysRevB.93.161112} {\bibfield
   {journal} {\bibinfo  {journal} {Physical Review B}\ }\textbf {\bibinfo
  {volume} {93}},\ \bibinfo {pages} {161112(R)} (\bibinfo {year}
  {2016})}\BibitemShut {NoStop}%
\bibitem [{\citenamefont {Xu}\ \emph {et~al.}(2015{\natexlab{b}})\citenamefont
  {Xu}, \citenamefont {Belopolski}, \citenamefont {Sanchez}, \citenamefont
  {Zhang}, \citenamefont {Chang}, \citenamefont {Guo}, \citenamefont {Bian},
  \citenamefont {Yuan}, \citenamefont {Lu}, \citenamefont {Chang} \emph
  {et~al.}}]{xu2015b}%
  \BibitemOpen
  \bibfield  {author} {\bibinfo {author} {\bibfnamefont {S.-Y.}\ \bibnamefont
  {Xu}}, \bibinfo {author} {\bibfnamefont {I.}~\bibnamefont {Belopolski}},
  \bibinfo {author} {\bibfnamefont {D.~S.}\ \bibnamefont {Sanchez}}, \bibinfo
  {author} {\bibfnamefont {C.}~\bibnamefont {Zhang}}, \bibinfo {author}
  {\bibfnamefont {G.}~\bibnamefont {Chang}}, \bibinfo {author} {\bibfnamefont
  {C.}~\bibnamefont {Guo}}, \bibinfo {author} {\bibfnamefont {G.}~\bibnamefont
  {Bian}}, \bibinfo {author} {\bibfnamefont {Z.}~\bibnamefont {Yuan}}, \bibinfo
  {author} {\bibfnamefont {H.}~\bibnamefont {Lu}}, \bibinfo {author}
  {\bibfnamefont {T.-R.}\ \bibnamefont {Chang}}, \emph {et~al.},\ }\bibfield
  {title} {\bibinfo {title} {{Experimental discovery of a topological Weyl
  semimetal state in TaP}},\ }\href
  {https://advances.sciencemag.org/content/1/10/e1501092} {\bibfield  {journal}
  {\bibinfo  {journal} {Science advances}\ }\textbf {\bibinfo {volume} {1}},\
  \bibinfo {pages} {e1501092} (\bibinfo {year}
  {2015}{\natexlab{b}})}\BibitemShut {NoStop}%
\bibitem [{\citenamefont {Xu}\ \emph {et~al.}(2016)\citenamefont {Xu},
  \citenamefont {Weng}, \citenamefont {Lv}, \citenamefont {Matt}, \citenamefont
  {Park}, \citenamefont {Bisti}, \citenamefont {Strocov}, \citenamefont
  {Gawryluk}, \citenamefont {Pomjakushina}, \citenamefont {Conder} \emph
  {et~al.}}]{xu2016}%
  \BibitemOpen
  \bibfield  {author} {\bibinfo {author} {\bibfnamefont {N.}~\bibnamefont
  {Xu}}, \bibinfo {author} {\bibfnamefont {H.}~\bibnamefont {Weng}}, \bibinfo
  {author} {\bibfnamefont {B.}~\bibnamefont {Lv}}, \bibinfo {author}
  {\bibfnamefont {C.~E.}\ \bibnamefont {Matt}}, \bibinfo {author}
  {\bibfnamefont {J.}~\bibnamefont {Park}}, \bibinfo {author} {\bibfnamefont
  {F.}~\bibnamefont {Bisti}}, \bibinfo {author} {\bibfnamefont {V.~N.}\
  \bibnamefont {Strocov}}, \bibinfo {author} {\bibfnamefont {D.}~\bibnamefont
  {Gawryluk}}, \bibinfo {author} {\bibfnamefont {E.}~\bibnamefont
  {Pomjakushina}}, \bibinfo {author} {\bibfnamefont {K.}~\bibnamefont
  {Conder}}, \emph {et~al.},\ }\bibfield  {title} {\bibinfo {title}
  {Observation of { Weyl nodes and Fermi} arcs in tantalum phosphide},\ }\href
  {https://www.nature.com/articles/ncomms11006} {\bibfield  {journal} {\bibinfo
   {journal} {Nature communications}\ }\textbf {\bibinfo {volume} {7}},\
  \bibinfo {pages} {11006} (\bibinfo {year} {2016})}\BibitemShut {NoStop}%
\bibitem [{\citenamefont {Haubold}\ \emph {et~al.}(2017)\citenamefont
  {Haubold}, \citenamefont {Koepernik}, \citenamefont {Efremov}, \citenamefont
  {Khim}, \citenamefont {Fedorov}, \citenamefont {Kushnirenko}, \citenamefont
  {van~den Brink}, \citenamefont {Wurmehl}, \citenamefont {B\"uchner},
  \citenamefont {Kim}, \citenamefont {Hoesch}, \citenamefont {Sumida},
  \citenamefont {Taguchi}, \citenamefont {Yoshikawa}, \citenamefont {Kimura},
  \citenamefont {Okuda},\ and\ \citenamefont {Borisenko}}]{haubold2017tairte4}%
  \BibitemOpen
  \bibfield  {author} {\bibinfo {author} {\bibfnamefont {E.}~\bibnamefont
  {Haubold}}, \bibinfo {author} {\bibfnamefont {K.}~\bibnamefont {Koepernik}},
  \bibinfo {author} {\bibfnamefont {D.}~\bibnamefont {Efremov}}, \bibinfo
  {author} {\bibfnamefont {S.}~\bibnamefont {Khim}}, \bibinfo {author}
  {\bibfnamefont {A.}~\bibnamefont {Fedorov}}, \bibinfo {author} {\bibfnamefont
  {Y.}~\bibnamefont {Kushnirenko}}, \bibinfo {author} {\bibfnamefont
  {J.}~\bibnamefont {van~den Brink}}, \bibinfo {author} {\bibfnamefont
  {S.}~\bibnamefont {Wurmehl}}, \bibinfo {author} {\bibfnamefont
  {B.}~\bibnamefont {B\"uchner}}, \bibinfo {author} {\bibfnamefont {T.~K.}\
  \bibnamefont {Kim}}, \bibinfo {author} {\bibfnamefont {M.}~\bibnamefont
  {Hoesch}}, \bibinfo {author} {\bibfnamefont {K.}~\bibnamefont {Sumida}},
  \bibinfo {author} {\bibfnamefont {K.}~\bibnamefont {Taguchi}}, \bibinfo
  {author} {\bibfnamefont {T.}~\bibnamefont {Yoshikawa}}, \bibinfo {author}
  {\bibfnamefont {A.}~\bibnamefont {Kimura}}, \bibinfo {author} {\bibfnamefont
  {T.}~\bibnamefont {Okuda}},\ and\ \bibinfo {author} {\bibfnamefont {S.~V.}\
  \bibnamefont {Borisenko}},\ }\bibfield  {title} {\bibinfo {title}
  {{Experimental realization of type-II Weyl state in noncentrosymmetric
  TaIrTe$_4$}},\ }\href {https://doi.org/10.1103/PhysRevB.95.241108} {\bibfield
   {journal} {\bibinfo  {journal} {Phys. Rev. B}\ }\textbf {\bibinfo {volume}
  {95}},\ \bibinfo {pages} {241108(R)} (\bibinfo {year} {2017})}\BibitemShut
  {NoStop}%
\bibitem [{\citenamefont {Burkov}\ \emph {et~al.}(2011)\citenamefont {Burkov},
  \citenamefont {Hook},\ and\ \citenamefont {Balents}}]{PhysRevB.84.235126}%
  \BibitemOpen
  \bibfield  {author} {\bibinfo {author} {\bibfnamefont {A.~A.}\ \bibnamefont
  {Burkov}}, \bibinfo {author} {\bibfnamefont {M.~D.}\ \bibnamefont {Hook}},\
  and\ \bibinfo {author} {\bibfnamefont {L.}~\bibnamefont {Balents}},\
  }\bibfield  {title} {\bibinfo {title} {Topological nodal semimetals},\ }\href
  {https://doi.org/10.1103/PhysRevB.84.235126} {\bibfield  {journal} {\bibinfo
  {journal} {Phys. Rev. B}\ }\textbf {\bibinfo {volume} {84}},\ \bibinfo
  {pages} {235126} (\bibinfo {year} {2011})}\BibitemShut {NoStop}%
\bibitem [{\citenamefont {Hirschberger}\ \emph {et~al.}(2016)\citenamefont
  {Hirschberger}, \citenamefont {Kushwaha}, \citenamefont {Wang}, \citenamefont
  {Gibson}, \citenamefont {Liang}, \citenamefont {Belvin}, \citenamefont
  {Bernevig}, \citenamefont {Cava},\ and\ \citenamefont
  {Ong}}]{hirschberger2016chiral}%
  \BibitemOpen
  \bibfield  {author} {\bibinfo {author} {\bibfnamefont {M.}~\bibnamefont
  {Hirschberger}}, \bibinfo {author} {\bibfnamefont {S.}~\bibnamefont
  {Kushwaha}}, \bibinfo {author} {\bibfnamefont {Z.}~\bibnamefont {Wang}},
  \bibinfo {author} {\bibfnamefont {Q.}~\bibnamefont {Gibson}}, \bibinfo
  {author} {\bibfnamefont {S.}~\bibnamefont {Liang}}, \bibinfo {author}
  {\bibfnamefont {C.~A.}\ \bibnamefont {Belvin}}, \bibinfo {author}
  {\bibfnamefont {B.~A.}\ \bibnamefont {Bernevig}}, \bibinfo {author}
  {\bibfnamefont {R.~J.}\ \bibnamefont {Cava}},\ and\ \bibinfo {author}
  {\bibfnamefont {N.~P.}\ \bibnamefont {Ong}},\ }\bibfield  {title} {\bibinfo
  {title} {{The chiral anomaly and thermopower of Weyl fermions in the
  half-Heusler GdPtBi}},\ }\href {https://www.nature.com/articles/nmat4684}
  {\bibfield  {journal} {\bibinfo  {journal} {Nature materials}\ }\textbf
  {\bibinfo {volume} {15}},\ \bibinfo {pages} {1161} (\bibinfo {year}
  {2016})}\BibitemShut {NoStop}%
\bibitem [{\citenamefont {Cano}\ \emph {et~al.}(2017)\citenamefont {Cano},
  \citenamefont {Bradlyn}, \citenamefont {Wang}, \citenamefont {Hirschberger},
  \citenamefont {Ong},\ and\ \citenamefont {Bernevig}}]{PhysRevB.95.161306}%
  \BibitemOpen
  \bibfield  {author} {\bibinfo {author} {\bibfnamefont {J.}~\bibnamefont
  {Cano}}, \bibinfo {author} {\bibfnamefont {B.}~\bibnamefont {Bradlyn}},
  \bibinfo {author} {\bibfnamefont {Z.}~\bibnamefont {Wang}}, \bibinfo {author}
  {\bibfnamefont {M.}~\bibnamefont {Hirschberger}}, \bibinfo {author}
  {\bibfnamefont {N.~P.}\ \bibnamefont {Ong}},\ and\ \bibinfo {author}
  {\bibfnamefont {B.~A.}\ \bibnamefont {Bernevig}},\ }\bibfield  {title}
  {\bibinfo {title} {Chiral anomaly factory: Creating {Weyl} fermions with a
  magnetic field},\ }\href {https://doi.org/10.1103/PhysRevB.95.161306}
  {\bibfield  {journal} {\bibinfo  {journal} {Phys. Rev. B}\ }\textbf {\bibinfo
  {volume} {95}},\ \bibinfo {pages} {161306(R)} (\bibinfo {year}
  {2017})}\BibitemShut {NoStop}%
\bibitem [{\citenamefont {Borisenko}\ \emph {et~al.}(2019)\citenamefont
  {Borisenko}, \citenamefont {Evtushinsky}, \citenamefont {Gibson},
  \citenamefont {Yaresko}, \citenamefont {Koepernik}, \citenamefont {Kim},
  \citenamefont {Ali}, \citenamefont {van~den Brink}, \citenamefont {Hoesch},
  \citenamefont {Fedorov} \emph {et~al.}}]{borisenko2019time}%
  \BibitemOpen
  \bibfield  {author} {\bibinfo {author} {\bibfnamefont {S.}~\bibnamefont
  {Borisenko}}, \bibinfo {author} {\bibfnamefont {D.}~\bibnamefont
  {Evtushinsky}}, \bibinfo {author} {\bibfnamefont {Q.}~\bibnamefont {Gibson}},
  \bibinfo {author} {\bibfnamefont {A.}~\bibnamefont {Yaresko}}, \bibinfo
  {author} {\bibfnamefont {K.}~\bibnamefont {Koepernik}}, \bibinfo {author}
  {\bibfnamefont {T.}~\bibnamefont {Kim}}, \bibinfo {author} {\bibfnamefont
  {M.}~\bibnamefont {Ali}}, \bibinfo {author} {\bibfnamefont {J.}~\bibnamefont
  {van~den Brink}}, \bibinfo {author} {\bibfnamefont {M.}~\bibnamefont
  {Hoesch}}, \bibinfo {author} {\bibfnamefont {A.}~\bibnamefont {Fedorov}},
  \emph {et~al.},\ }\bibfield  {title} {\bibinfo {title} {{Time-reversal
  symmetry breaking type-II Weyl state in YbMnBi$_2$}},\ }\href
  {https://www.nature.com/articles/s41467-019-11393-5} {\bibfield  {journal}
  {\bibinfo  {journal} {Nature communications}\ }\textbf {\bibinfo {volume}
  {10}},\ \bibinfo {pages} {3424} (\bibinfo {year} {2019})}\BibitemShut
  {NoStop}%
\bibitem [{\citenamefont {Deng}\ \emph {et~al.}(2020)\citenamefont {Deng},
  \citenamefont {Yu}, \citenamefont {Shi}, \citenamefont {Guo}, \citenamefont
  {Xu}, \citenamefont {Wang}, \citenamefont {Chen},\ and\ \citenamefont
  {Zhang}}]{Deng895}%
  \BibitemOpen
  \bibfield  {author} {\bibinfo {author} {\bibfnamefont {Y.}~\bibnamefont
  {Deng}}, \bibinfo {author} {\bibfnamefont {Y.}~\bibnamefont {Yu}}, \bibinfo
  {author} {\bibfnamefont {M.~Z.}\ \bibnamefont {Shi}}, \bibinfo {author}
  {\bibfnamefont {Z.}~\bibnamefont {Guo}}, \bibinfo {author} {\bibfnamefont
  {Z.}~\bibnamefont {Xu}}, \bibinfo {author} {\bibfnamefont {J.}~\bibnamefont
  {Wang}}, \bibinfo {author} {\bibfnamefont {X.~H.}\ \bibnamefont {Chen}},\
  and\ \bibinfo {author} {\bibfnamefont {Y.}~\bibnamefont {Zhang}},\ }\bibfield
   {title} {\bibinfo {title} {{Quantum anomalous Hall effect in intrinsic
  magnetic topological insulator MnBi$_2$Te$_4$}},\ }\href
  {https://doi.org/10.1126/science.aax8156} {\bibfield  {journal} {\bibinfo
  {journal} {Science}\ }\textbf {\bibinfo {volume} {367}},\ \bibinfo {pages}
  {895} (\bibinfo {year} {2020})}\BibitemShut {NoStop}%
\bibitem [{\citenamefont {Ghimire}\ \emph {et~al.}(2019)\citenamefont
  {Ghimire}, \citenamefont {Facio}, \citenamefont {You}, \citenamefont {Ye},
  \citenamefont {Checkelsky}, \citenamefont {Fang}, \citenamefont {Kaxiras},
  \citenamefont {Richter},\ and\ \citenamefont {van~den
  Brink}}]{PhysRevResearch.1.032044}%
  \BibitemOpen
  \bibfield  {author} {\bibinfo {author} {\bibfnamefont {M.~P.}\ \bibnamefont
  {Ghimire}}, \bibinfo {author} {\bibfnamefont {J.~I.}\ \bibnamefont {Facio}},
  \bibinfo {author} {\bibfnamefont {J.-S.}\ \bibnamefont {You}}, \bibinfo
  {author} {\bibfnamefont {L.}~\bibnamefont {Ye}}, \bibinfo {author}
  {\bibfnamefont {J.~G.}\ \bibnamefont {Checkelsky}}, \bibinfo {author}
  {\bibfnamefont {S.}~\bibnamefont {Fang}}, \bibinfo {author} {\bibfnamefont
  {E.}~\bibnamefont {Kaxiras}}, \bibinfo {author} {\bibfnamefont
  {M.}~\bibnamefont {Richter}},\ and\ \bibinfo {author} {\bibfnamefont
  {J.}~\bibnamefont {van~den Brink}},\ }\bibfield  {title} {\bibinfo {title}
  {Creating {Weyl} nodes and controlling their energy by magnetization
  rotation},\ }\href {https://doi.org/10.1103/PhysRevResearch.1.032044}
  {\bibfield  {journal} {\bibinfo  {journal} {Phys. Rev. Research}\ }\textbf
  {\bibinfo {volume} {1}},\ \bibinfo {pages} {032044} (\bibinfo {year}
  {2019})}\BibitemShut {NoStop}%
\bibitem [{\citenamefont {Zhang}\ \emph {et~al.}(2019)\citenamefont {Zhang},
  \citenamefont {Shi}, \citenamefont {Zhu}, \citenamefont {Xing}, \citenamefont
  {Zhang},\ and\ \citenamefont {Wang}}]{PhysRevLett.122.206401}%
  \BibitemOpen
  \bibfield  {author} {\bibinfo {author} {\bibfnamefont {D.}~\bibnamefont
  {Zhang}}, \bibinfo {author} {\bibfnamefont {M.}~\bibnamefont {Shi}}, \bibinfo
  {author} {\bibfnamefont {T.}~\bibnamefont {Zhu}}, \bibinfo {author}
  {\bibfnamefont {D.}~\bibnamefont {Xing}}, \bibinfo {author} {\bibfnamefont
  {H.}~\bibnamefont {Zhang}},\ and\ \bibinfo {author} {\bibfnamefont
  {J.}~\bibnamefont {Wang}},\ }\bibfield  {title} {\bibinfo {title}
  {{Topological Axion States in the Magnetic Insulator MnBi$_2$Te$_4$ with the
  Quantized Magnetoelectric Effect}},\ }\href
  {https://doi.org/10.1103/PhysRevLett.122.206401} {\bibfield  {journal}
  {\bibinfo  {journal} {Phys. Rev. Lett.}\ }\textbf {\bibinfo {volume} {122}},\
  \bibinfo {pages} {206401} (\bibinfo {year} {2019})}\BibitemShut {NoStop}%
\bibitem [{\citenamefont {Vidal}\ \emph {et~al.}(2019)\citenamefont {Vidal},
  \citenamefont {Zeugner}, \citenamefont {Facio}, \citenamefont {Ray},
  \citenamefont {Haghighi}, \citenamefont {Wolter}, \citenamefont
  {Corredor~Bohorquez}, \citenamefont {Caglieris}, \citenamefont {Moser},
  \citenamefont {Figgemeier}, \citenamefont {Peixoto}, \citenamefont {Vasili},
  \citenamefont {Valvidares}, \citenamefont {Jung}, \citenamefont {Cacho},
  \citenamefont {Alfonsov}, \citenamefont {Mehlawat}, \citenamefont {Kataev},
  \citenamefont {Hess}, \citenamefont {Richter}, \citenamefont {B\"uchner},
  \citenamefont {van~den Brink}, \citenamefont {Ruck}, \citenamefont {Reinert},
  \citenamefont {Bentmann},\ and\ \citenamefont {Isaeva}}]{PhysRevX.9.041065}%
  \BibitemOpen
  \bibfield  {author} {\bibinfo {author} {\bibfnamefont {R.~C.}\ \bibnamefont
  {Vidal}}, \bibinfo {author} {\bibfnamefont {A.}~\bibnamefont {Zeugner}},
  \bibinfo {author} {\bibfnamefont {J.~I.}\ \bibnamefont {Facio}}, \bibinfo
  {author} {\bibfnamefont {R.}~\bibnamefont {Ray}}, \bibinfo {author}
  {\bibfnamefont {M.~H.}\ \bibnamefont {Haghighi}}, \bibinfo {author}
  {\bibfnamefont {A.~U.~B.}\ \bibnamefont {Wolter}}, \bibinfo {author}
  {\bibfnamefont {L.~T.}\ \bibnamefont {Corredor~Bohorquez}}, \bibinfo {author}
  {\bibfnamefont {F.}~\bibnamefont {Caglieris}}, \bibinfo {author}
  {\bibfnamefont {S.}~\bibnamefont {Moser}}, \bibinfo {author} {\bibfnamefont
  {T.}~\bibnamefont {Figgemeier}}, \bibinfo {author} {\bibfnamefont {T.~R.~F.}\
  \bibnamefont {Peixoto}}, \bibinfo {author} {\bibfnamefont {H.~B.}\
  \bibnamefont {Vasili}}, \bibinfo {author} {\bibfnamefont {M.}~\bibnamefont
  {Valvidares}}, \bibinfo {author} {\bibfnamefont {S.}~\bibnamefont {Jung}},
  \bibinfo {author} {\bibfnamefont {C.}~\bibnamefont {Cacho}}, \bibinfo
  {author} {\bibfnamefont {A.}~\bibnamefont {Alfonsov}}, \bibinfo {author}
  {\bibfnamefont {K.}~\bibnamefont {Mehlawat}}, \bibinfo {author}
  {\bibfnamefont {V.}~\bibnamefont {Kataev}}, \bibinfo {author} {\bibfnamefont
  {C.}~\bibnamefont {Hess}}, \bibinfo {author} {\bibfnamefont {M.}~\bibnamefont
  {Richter}}, \bibinfo {author} {\bibfnamefont {B.}~\bibnamefont {B\"uchner}},
  \bibinfo {author} {\bibfnamefont {J.}~\bibnamefont {van~den Brink}}, \bibinfo
  {author} {\bibfnamefont {M.}~\bibnamefont {Ruck}}, \bibinfo {author}
  {\bibfnamefont {F.}~\bibnamefont {Reinert}}, \bibinfo {author} {\bibfnamefont
  {H.}~\bibnamefont {Bentmann}},\ and\ \bibinfo {author} {\bibfnamefont
  {A.}~\bibnamefont {Isaeva}},\ }\bibfield  {title} {\bibinfo {title}
  {{Topological Electronic Structure and Intrinsic Magnetization in
  MnBi$_4$Te$_7$: A Bi$_2$Te$_3$ Derivative with a Periodic Mn Sublattice}},\
  }\href {https://doi.org/10.1103/PhysRevX.9.041065} {\bibfield  {journal}
  {\bibinfo  {journal} {Phys. Rev. X}\ }\textbf {\bibinfo {volume} {9}},\
  \bibinfo {pages} {041065} (\bibinfo {year} {2019})}\BibitemShut {NoStop}%
\bibitem [{\citenamefont {Zou}\ \emph {et~al.}(2019)\citenamefont {Zou},
  \citenamefont {He},\ and\ \citenamefont {Xu}}]{zou2019study}%
  \BibitemOpen
  \bibfield  {author} {\bibinfo {author} {\bibfnamefont {J.}~\bibnamefont
  {Zou}}, \bibinfo {author} {\bibfnamefont {Z.}~\bibnamefont {He}},\ and\
  \bibinfo {author} {\bibfnamefont {G.}~\bibnamefont {Xu}},\ }\bibfield
  {title} {\bibinfo {title} {The study of magnetic topological semimetals by
  first principles calculations},\ }\href
  {https://www.nature.com/articles/s41524-019-0237-5} {\bibfield  {journal}
  {\bibinfo  {journal} {npj Computational Materials}\ }\textbf {\bibinfo
  {volume} {5}},\ \bibinfo {pages} {1} (\bibinfo {year} {2019})}\BibitemShut
  {NoStop}%
\bibitem [{\citenamefont {Liao}\ \emph {et~al.}(2020)\citenamefont {Liao},
  \citenamefont {Jiang}, \citenamefont {Zhong},\ and\ \citenamefont
  {Li}}]{liao2020materials}%
  \BibitemOpen
  \bibfield  {author} {\bibinfo {author} {\bibfnamefont {Z.}~\bibnamefont
  {Liao}}, \bibinfo {author} {\bibfnamefont {P.}~\bibnamefont {Jiang}},
  \bibinfo {author} {\bibfnamefont {Z.}~\bibnamefont {Zhong}},\ and\ \bibinfo
  {author} {\bibfnamefont {R.-W.}\ \bibnamefont {Li}},\ }\bibfield  {title}
  {\bibinfo {title} {Materials with strong spin-textured bands},\ }\href
  {https://www.nature.com/articles/s41535-020-0233-5} {\bibfield  {journal}
  {\bibinfo  {journal} {npj Quantum Materials}\ }\textbf {\bibinfo {volume}
  {5}},\ \bibinfo {pages} {1} (\bibinfo {year} {2020})}\BibitemShut {NoStop}%
\bibitem [{\citenamefont {Chang}\ \emph {et~al.}(2017)\citenamefont {Chang},
  \citenamefont {Xu}, \citenamefont {Zhou}, \citenamefont {Huang},
  \citenamefont {Singh}, \citenamefont {Wang}, \citenamefont {Belopolski},
  \citenamefont {Yin}, \citenamefont {Zhang}, \citenamefont {Bansil},
  \citenamefont {Lin},\ and\ \citenamefont {Hasan}}]{PhysRevLett.119.156401}%
  \BibitemOpen
  \bibfield  {author} {\bibinfo {author} {\bibfnamefont {G.}~\bibnamefont
  {Chang}}, \bibinfo {author} {\bibfnamefont {S.-Y.}\ \bibnamefont {Xu}},
  \bibinfo {author} {\bibfnamefont {X.}~\bibnamefont {Zhou}}, \bibinfo {author}
  {\bibfnamefont {S.-M.}\ \bibnamefont {Huang}}, \bibinfo {author}
  {\bibfnamefont {B.}~\bibnamefont {Singh}}, \bibinfo {author} {\bibfnamefont
  {B.}~\bibnamefont {Wang}}, \bibinfo {author} {\bibfnamefont {I.}~\bibnamefont
  {Belopolski}}, \bibinfo {author} {\bibfnamefont {J.}~\bibnamefont {Yin}},
  \bibinfo {author} {\bibfnamefont {S.}~\bibnamefont {Zhang}}, \bibinfo
  {author} {\bibfnamefont {A.}~\bibnamefont {Bansil}}, \bibinfo {author}
  {\bibfnamefont {H.}~\bibnamefont {Lin}},\ and\ \bibinfo {author}
  {\bibfnamefont {M.~Z.}\ \bibnamefont {Hasan}},\ }\bibfield  {title} {\bibinfo
  {title} {Topological hopf and chain link semimetal states and their
  application to {Co$_2$MnGa}},\ }\href
  {https://doi.org/10.1103/PhysRevLett.119.156401} {\bibfield  {journal}
  {\bibinfo  {journal} {Phys. Rev. Lett.}\ }\textbf {\bibinfo {volume} {119}},\
  \bibinfo {pages} {156401} (\bibinfo {year} {2017})}\BibitemShut {NoStop}%
\bibitem [{\citenamefont {Belopolski}\ \emph {et~al.}(2019)\citenamefont
  {Belopolski}, \citenamefont {Manna}, \citenamefont {Sanchez}, \citenamefont
  {Chang}, \citenamefont {Ernst}, \citenamefont {Yin}, \citenamefont {Zhang},
  \citenamefont {Cochran}, \citenamefont {Shumiya}, \citenamefont {Zheng},
  \citenamefont {Singh}, \citenamefont {Bian}, \citenamefont {Multer},
  \citenamefont {Litskevich}, \citenamefont {Zhou}, \citenamefont {Huang},
  \citenamefont {Wang}, \citenamefont {Chang}, \citenamefont {Xu},
  \citenamefont {Bansil}, \citenamefont {Felser}, \citenamefont {Lin},\ and\
  \citenamefont {Hasan}}]{belopolski1278}%
  \BibitemOpen
  \bibfield  {author} {\bibinfo {author} {\bibfnamefont {I.}~\bibnamefont
  {Belopolski}}, \bibinfo {author} {\bibfnamefont {K.}~\bibnamefont {Manna}},
  \bibinfo {author} {\bibfnamefont {D.~S.}\ \bibnamefont {Sanchez}}, \bibinfo
  {author} {\bibfnamefont {G.}~\bibnamefont {Chang}}, \bibinfo {author}
  {\bibfnamefont {B.}~\bibnamefont {Ernst}}, \bibinfo {author} {\bibfnamefont
  {J.}~\bibnamefont {Yin}}, \bibinfo {author} {\bibfnamefont {S.~S.}\
  \bibnamefont {Zhang}}, \bibinfo {author} {\bibfnamefont {T.}~\bibnamefont
  {Cochran}}, \bibinfo {author} {\bibfnamefont {N.}~\bibnamefont {Shumiya}},
  \bibinfo {author} {\bibfnamefont {H.}~\bibnamefont {Zheng}}, \bibinfo
  {author} {\bibfnamefont {B.}~\bibnamefont {Singh}}, \bibinfo {author}
  {\bibfnamefont {G.}~\bibnamefont {Bian}}, \bibinfo {author} {\bibfnamefont
  {D.}~\bibnamefont {Multer}}, \bibinfo {author} {\bibfnamefont
  {M.}~\bibnamefont {Litskevich}}, \bibinfo {author} {\bibfnamefont
  {X.}~\bibnamefont {Zhou}}, \bibinfo {author} {\bibfnamefont {S.-M.}\
  \bibnamefont {Huang}}, \bibinfo {author} {\bibfnamefont {B.}~\bibnamefont
  {Wang}}, \bibinfo {author} {\bibfnamefont {T.-R.}\ \bibnamefont {Chang}},
  \bibinfo {author} {\bibfnamefont {S.-Y.}\ \bibnamefont {Xu}}, \bibinfo
  {author} {\bibfnamefont {A.}~\bibnamefont {Bansil}}, \bibinfo {author}
  {\bibfnamefont {C.}~\bibnamefont {Felser}}, \bibinfo {author} {\bibfnamefont
  {H.}~\bibnamefont {Lin}},\ and\ \bibinfo {author} {\bibfnamefont {M.~Z.}\
  \bibnamefont {Hasan}},\ }\bibfield  {title} {\bibinfo {title} {Discovery of
  topological {Weyl} fermion lines and drumhead surface states in a room
  temperature magnet},\ }\href {https://doi.org/10.1126/science.aav2327}
  {\bibfield  {journal} {\bibinfo  {journal} {Science}\ }\textbf {\bibinfo
  {volume} {365}},\ \bibinfo {pages} {1278} (\bibinfo {year}
  {2019})}\BibitemShut {NoStop}%
\bibitem [{\citenamefont {Liu}\ \emph {et~al.}(2018)\citenamefont {Liu},
  \citenamefont {Sun}, \citenamefont {Kumar}, \citenamefont {Muechler},
  \citenamefont {Sun}, \citenamefont {Jiao}, \citenamefont {Yang},
  \citenamefont {Liu}, \citenamefont {Liang}, \citenamefont {Xu}, \citenamefont
  {Kroder}, \citenamefont {Süß}, \citenamefont {Borrmann}, \citenamefont
  {Shekhar}, \citenamefont {Wang}, \citenamefont {Xi}, \citenamefont {Wang},
  \citenamefont {Schnelle}, \citenamefont {Wirth}, \citenamefont {Chen},
  \citenamefont {Goennenwein},\ and\ \citenamefont {Felser}}]{liu2018giant}%
  \BibitemOpen
  \bibfield  {author} {\bibinfo {author} {\bibfnamefont {E.}~\bibnamefont
  {Liu}}, \bibinfo {author} {\bibfnamefont {Y.}~\bibnamefont {Sun}}, \bibinfo
  {author} {\bibfnamefont {N.}~\bibnamefont {Kumar}}, \bibinfo {author}
  {\bibfnamefont {L.}~\bibnamefont {Muechler}}, \bibinfo {author}
  {\bibfnamefont {A.}~\bibnamefont {Sun}}, \bibinfo {author} {\bibfnamefont
  {L.}~\bibnamefont {Jiao}}, \bibinfo {author} {\bibfnamefont {S.-Y.}\
  \bibnamefont {Yang}}, \bibinfo {author} {\bibfnamefont {D.}~\bibnamefont
  {Liu}}, \bibinfo {author} {\bibfnamefont {A.}~\bibnamefont {Liang}}, \bibinfo
  {author} {\bibfnamefont {Q.}~\bibnamefont {Xu}}, \bibinfo {author}
  {\bibfnamefont {J.}~\bibnamefont {Kroder}}, \bibinfo {author} {\bibfnamefont
  {V.}~\bibnamefont {Süß}}, \bibinfo {author} {\bibfnamefont
  {H.}~\bibnamefont {Borrmann}}, \bibinfo {author} {\bibfnamefont
  {C.}~\bibnamefont {Shekhar}}, \bibinfo {author} {\bibfnamefont
  {Z.}~\bibnamefont {Wang}}, \bibinfo {author} {\bibfnamefont {C.}~\bibnamefont
  {Xi}}, \bibinfo {author} {\bibfnamefont {W.}~\bibnamefont {Wang}}, \bibinfo
  {author} {\bibfnamefont {W.}~\bibnamefont {Schnelle}}, \bibinfo {author}
  {\bibfnamefont {S.}~\bibnamefont {Wirth}}, \bibinfo {author} {\bibfnamefont
  {Y.}~\bibnamefont {Chen}}, \bibinfo {author} {\bibfnamefont {S.~T.~B.}\
  \bibnamefont {Goennenwein}},\ and\ \bibinfo {author} {\bibfnamefont
  {C.}~\bibnamefont {Felser}},\ }\bibfield  {title} {\bibinfo {title} {Giant
  anomalous {Hall} effect in a ferromagnetic kagome-lattice semimetal},\ }\href
  {https://doi.org/10.1038/s41567-018-0234-5} {\bibfield  {journal} {\bibinfo
  {journal} {Nature Physics}\ ,\ \bibinfo {pages} {1}} (\bibinfo {year}
  {2018})}\BibitemShut {NoStop}%
\bibitem [{\citenamefont {Wang}\ \emph {et~al.}(2018)\citenamefont {Wang},
  \citenamefont {Xu}, \citenamefont {Lou}, \citenamefont {Liu}, \citenamefont
  {Li}, \citenamefont {Huang}, \citenamefont {Shen}, \citenamefont {Weng},
  \citenamefont {Wang},\ and\ \citenamefont {Lei}}]{wang2018large}%
  \BibitemOpen
  \bibfield  {author} {\bibinfo {author} {\bibfnamefont {Q.}~\bibnamefont
  {Wang}}, \bibinfo {author} {\bibfnamefont {Y.}~\bibnamefont {Xu}}, \bibinfo
  {author} {\bibfnamefont {R.}~\bibnamefont {Lou}}, \bibinfo {author}
  {\bibfnamefont {Z.}~\bibnamefont {Liu}}, \bibinfo {author} {\bibfnamefont
  {M.}~\bibnamefont {Li}}, \bibinfo {author} {\bibfnamefont {Y.}~\bibnamefont
  {Huang}}, \bibinfo {author} {\bibfnamefont {D.}~\bibnamefont {Shen}},
  \bibinfo {author} {\bibfnamefont {H.}~\bibnamefont {Weng}}, \bibinfo {author}
  {\bibfnamefont {S.}~\bibnamefont {Wang}},\ and\ \bibinfo {author}
  {\bibfnamefont {H.}~\bibnamefont {Lei}},\ }\bibfield  {title} {\bibinfo
  {title} {Large intrinsic anomalous {Hall} effect in half-metallic ferromagnet
  {Co$_3$Sn$_2$S$_2$} with magnetic {Weyl} fermions},\ }\href
  {https://www.nature.com/articles/s41467-018-06088-2} {\bibfield  {journal}
  {\bibinfo  {journal} {Nature Communications}\ }\textbf {\bibinfo {volume}
  {9}},\ \bibinfo {pages} {3681} (\bibinfo {year} {2018})}\BibitemShut
  {NoStop}%
\bibitem [{\citenamefont {Morali}\ \emph {et~al.}(2019)\citenamefont {Morali},
  \citenamefont {Batabyal}, \citenamefont {Nag}, \citenamefont {Liu},
  \citenamefont {Xu}, \citenamefont {Sun}, \citenamefont {Yan}, \citenamefont
  {Felser}, \citenamefont {Avraham},\ and\ \citenamefont
  {Beidenkopf}}]{Morali1286}%
  \BibitemOpen
  \bibfield  {author} {\bibinfo {author} {\bibfnamefont {N.}~\bibnamefont
  {Morali}}, \bibinfo {author} {\bibfnamefont {R.}~\bibnamefont {Batabyal}},
  \bibinfo {author} {\bibfnamefont {P.~K.}\ \bibnamefont {Nag}}, \bibinfo
  {author} {\bibfnamefont {E.}~\bibnamefont {Liu}}, \bibinfo {author}
  {\bibfnamefont {Q.}~\bibnamefont {Xu}}, \bibinfo {author} {\bibfnamefont
  {Y.}~\bibnamefont {Sun}}, \bibinfo {author} {\bibfnamefont {B.}~\bibnamefont
  {Yan}}, \bibinfo {author} {\bibfnamefont {C.}~\bibnamefont {Felser}},
  \bibinfo {author} {\bibfnamefont {N.}~\bibnamefont {Avraham}},\ and\ \bibinfo
  {author} {\bibfnamefont {H.}~\bibnamefont {Beidenkopf}},\ }\bibfield  {title}
  {\bibinfo {title} {{Fermi-arc diversity on surface terminations of the
  magnetic Weyl semimetal Co$_3$Sn$_2$S$_2$}},\ }\href
  {https://doi.org/10.1126/science.aav2334} {\bibfield  {journal} {\bibinfo
  {journal} {Science}\ }\textbf {\bibinfo {volume} {365}},\ \bibinfo {pages}
  {1286} (\bibinfo {year} {2019})}\BibitemShut {NoStop}%
\bibitem [{\citenamefont {Liu}\ \emph {et~al.}(2019)\citenamefont {Liu},
  \citenamefont {Liang}, \citenamefont {Liu}, \citenamefont {Xu}, \citenamefont
  {Li}, \citenamefont {Chen}, \citenamefont {Pei}, \citenamefont {Shi},
  \citenamefont {Mo}, \citenamefont {Dudin} \emph {et~al.}}]{liu2019magnetic}%
  \BibitemOpen
  \bibfield  {author} {\bibinfo {author} {\bibfnamefont {D.}~\bibnamefont
  {Liu}}, \bibinfo {author} {\bibfnamefont {A.}~\bibnamefont {Liang}}, \bibinfo
  {author} {\bibfnamefont {E.}~\bibnamefont {Liu}}, \bibinfo {author}
  {\bibfnamefont {Q.}~\bibnamefont {Xu}}, \bibinfo {author} {\bibfnamefont
  {Y.}~\bibnamefont {Li}}, \bibinfo {author} {\bibfnamefont {C.}~\bibnamefont
  {Chen}}, \bibinfo {author} {\bibfnamefont {D.}~\bibnamefont {Pei}}, \bibinfo
  {author} {\bibfnamefont {W.}~\bibnamefont {Shi}}, \bibinfo {author}
  {\bibfnamefont {S.}~\bibnamefont {Mo}}, \bibinfo {author} {\bibfnamefont
  {P.}~\bibnamefont {Dudin}}, \emph {et~al.},\ }\bibfield  {title} {\bibinfo
  {title} {{Magnetic Weyl semimetal phase in a Kagom{\'e} crystal}},\ }\href
  {https://science.sciencemag.org/content/365/6459/1282} {\bibfield  {journal}
  {\bibinfo  {journal} {Science}\ }\textbf {\bibinfo {volume} {365}},\ \bibinfo
  {pages} {1282} (\bibinfo {year} {2019})}\BibitemShut {NoStop}%
\bibitem [{\citenamefont {Tsokol'}\ \emph {et~al.}(1988)\citenamefont
  {Tsokol'}, \citenamefont {Bodak}, \citenamefont {Marusin},\ and\
  \citenamefont {Zavdonik}}]{tsokol1988}%
  \BibitemOpen
  \bibfield  {author} {\bibinfo {author} {\bibfnamefont {A.~O.}\ \bibnamefont
  {Tsokol'}}, \bibinfo {author} {\bibfnamefont {O.~I.}\ \bibnamefont {Bodak}},
  \bibinfo {author} {\bibfnamefont {E.~P.}\ \bibnamefont {Marusin}},\ and\
  \bibinfo {author} {\bibfnamefont {V.~E.}\ \bibnamefont {Zavdonik}},\
  }\bibfield  {title} {\bibinfo {title} {{X-ray diffraction studies of ternary
  RRhC$_2$ (R=La, Ce, Pr, Nd, Sm) compounds)}},\ }\href@noop {} {\bibfield
  {journal} {\bibinfo  {journal} {Kristallografiya}\ }\textbf {\bibinfo
  {volume} {33}},\ \bibinfo {pages} {345} (\bibinfo {year} {1988})}\BibitemShut
  {NoStop}%
\bibitem [{\citenamefont {Hoffmann}\ \emph {et~al.}(1989)\citenamefont
  {Hoffmann}, \citenamefont {Jeitschko},\ and\ \citenamefont
  {Boonk}}]{hoffmann1989}%
  \BibitemOpen
  \bibfield  {author} {\bibinfo {author} {\bibfnamefont {R.~D.}\ \bibnamefont
  {Hoffmann}}, \bibinfo {author} {\bibfnamefont {W.}~\bibnamefont
  {Jeitschko}},\ and\ \bibinfo {author} {\bibfnamefont {L.}~\bibnamefont
  {Boonk}},\ }\bibfield  {title} {\bibinfo {title} {{Structural, chemical, and
  physical properties of rare-earth metal rhodium carbides LnRhC$_2$ (Ln= La,
  Ce, Pr, Nd, Sm)}},\ }\href {http://pubs.acs.org/doi/full/10.1021/cm00006a007}
  {\bibfield  {journal} {\bibinfo  {journal} {Chemistry of Materials}\ }\textbf
  {\bibinfo {volume} {1}},\ \bibinfo {pages} {580} (\bibinfo {year}
  {1989})}\BibitemShut {NoStop}%
\bibitem [{\citenamefont {Hoffmann}\ \emph {et~al.}(1995)\citenamefont
  {Hoffmann}, \citenamefont {Wachtmann}, \citenamefont {Ebel},\ and\
  \citenamefont {Jeitschko}}]{hoffmann1995}%
  \BibitemOpen
  \bibfield  {author} {\bibinfo {author} {\bibfnamefont {R.-D.}\ \bibnamefont
  {Hoffmann}}, \bibinfo {author} {\bibfnamefont {K.~H.}\ \bibnamefont
  {Wachtmann}}, \bibinfo {author} {\bibfnamefont {T.}~\bibnamefont {Ebel}},\
  and\ \bibinfo {author} {\bibfnamefont {W.}~\bibnamefont {Jeitschko}},\
  }\bibfield  {title} {\bibinfo {title} {{GdRuC$_2$, a ternary carbide with
  filled NiAs structure}},\ }\href
  {https://www.sciencedirect.com/science/article/pii/S0022459685713254}
  {\bibfield  {journal} {\bibinfo  {journal} {Journal of Solid State
  Chemistry}\ }\textbf {\bibinfo {volume} {118}},\ \bibinfo {pages} {158}
  (\bibinfo {year} {1995})}\BibitemShut {NoStop}%
\bibitem [{\citenamefont {Matsuo}\ \emph {et~al.}(1996)\citenamefont {Matsuo},
  \citenamefont {Onodera}, \citenamefont {Kosaka}, \citenamefont {Kobayashi},
  \citenamefont {Ohashi}, \citenamefont {Yamauchi},\ and\ \citenamefont
  {Yamaguchi}}]{matsuo1996}%
  \BibitemOpen
  \bibfield  {author} {\bibinfo {author} {\bibfnamefont {S.}~\bibnamefont
  {Matsuo}}, \bibinfo {author} {\bibfnamefont {H.}~\bibnamefont {Onodera}},
  \bibinfo {author} {\bibfnamefont {M.}~\bibnamefont {Kosaka}}, \bibinfo
  {author} {\bibfnamefont {H.}~\bibnamefont {Kobayashi}}, \bibinfo {author}
  {\bibfnamefont {M.}~\bibnamefont {Ohashi}}, \bibinfo {author} {\bibfnamefont
  {H.}~\bibnamefont {Yamauchi}},\ and\ \bibinfo {author} {\bibfnamefont
  {Y.}~\bibnamefont {Yamaguchi}},\ }\bibfield  {title} {\bibinfo {title}
  {{Antiferromagnetism of GdCoC$_2$ and GdNiC$_2$ intermetallics studied by
  magnetization measurement and $^{155}$Gd M{\"o}ssbauer spectroscopy}},\
  }\href
  {https://www.sciencedirect.com/science/article/abs/pii/S0304885396012826}
  {\bibfield  {journal} {\bibinfo  {journal} {Journal of Magnetism and Magnetic
  Materials}\ }\textbf {\bibinfo {volume} {161}},\ \bibinfo {pages} {255}
  (\bibinfo {year} {1996})}\BibitemShut {NoStop}%
\bibitem [{\citenamefont {Onodera}\ \emph {et~al.}(1998)\citenamefont
  {Onodera}, \citenamefont {Koshikawa}, \citenamefont {Kosaka}, \citenamefont
  {Ohashi}, \citenamefont {Yamauchi},\ and\ \citenamefont
  {Yamaguchi}}]{onodera1998magnetic}%
  \BibitemOpen
  \bibfield  {author} {\bibinfo {author} {\bibfnamefont {H.}~\bibnamefont
  {Onodera}}, \bibinfo {author} {\bibfnamefont {Y.}~\bibnamefont {Koshikawa}},
  \bibinfo {author} {\bibfnamefont {M.}~\bibnamefont {Kosaka}}, \bibinfo
  {author} {\bibfnamefont {M.}~\bibnamefont {Ohashi}}, \bibinfo {author}
  {\bibfnamefont {H.}~\bibnamefont {Yamauchi}},\ and\ \bibinfo {author}
  {\bibfnamefont {Y.}~\bibnamefont {Yamaguchi}},\ }\bibfield  {title} {\bibinfo
  {title} {{Magnetic properties of single-crystalline RNiC$_2$ compounds (R=
  Ce, Pr, Nd and Sm)}},\ }\href
  {https://www.sciencedirect.com/science/article/abs/pii/S0304885397010111?via%3Dihub}
  {\bibfield  {journal} {\bibinfo  {journal} {Journal of Magnetism and Magnetic
  Materials}\ }\textbf {\bibinfo {volume} {182}},\ \bibinfo {pages} {161}
  (\bibinfo {year} {1998})}\BibitemShut {NoStop}%
\bibitem [{\citenamefont {Meng}\ \emph {et~al.}(2016)\citenamefont {Meng},
  \citenamefont {Xu}, \citenamefont {Yuan}, \citenamefont {Qi}, \citenamefont
  {Zhou},\ and\ \citenamefont {Li}}]{meng2016}%
  \BibitemOpen
  \bibfield  {author} {\bibinfo {author} {\bibfnamefont {L.}~\bibnamefont
  {Meng}}, \bibinfo {author} {\bibfnamefont {C.}~\bibnamefont {Xu}}, \bibinfo
  {author} {\bibfnamefont {Y.}~\bibnamefont {Yuan}}, \bibinfo {author}
  {\bibfnamefont {Y.}~\bibnamefont {Qi}}, \bibinfo {author} {\bibfnamefont
  {S.}~\bibnamefont {Zhou}},\ and\ \bibinfo {author} {\bibfnamefont
  {L.}~\bibnamefont {Li}},\ }\bibfield  {title} {\bibinfo {title} {{Magnetic
  properties and giant reversible magnetocaloric effect in GdCoC$_2$}},\ }\href
  {https://pubs.rsc.org/en/content/articlelanding/2016/ra/c6ra16486b#!divAbstract}
  {\bibfield  {journal} {\bibinfo  {journal} {RSC Advances}\ }\textbf {\bibinfo
  {volume} {6}},\ \bibinfo {pages} {74765} (\bibinfo {year}
  {2016})}\BibitemShut {NoStop}%
\bibitem [{\citenamefont {Meng}\ \emph {et~al.}(2017)\citenamefont {Meng},
  \citenamefont {Jia},\ and\ \citenamefont {Li}}]{meng2017}%
  \BibitemOpen
  \bibfield  {author} {\bibinfo {author} {\bibfnamefont {L.}~\bibnamefont
  {Meng}}, \bibinfo {author} {\bibfnamefont {Y.}~\bibnamefont {Jia}},\ and\
  \bibinfo {author} {\bibfnamefont {L.}~\bibnamefont {Li}},\ }\bibfield
  {title} {\bibinfo {title} {{Large reversible magnetocaloric effect in the
  RECoC$_2$ (RE= Ho and Er) compounds}},\ }\href
  {https://www.sciencedirect.com/science/article/pii/S0966979517300018?via%3Dihub}
  {\bibfield  {journal} {\bibinfo  {journal} {Intermetallics}\ }\textbf
  {\bibinfo {volume} {85}},\ \bibinfo {pages} {69} (\bibinfo {year}
  {2017})}\BibitemShut {NoStop}%
\bibitem [{\citenamefont {Lee}\ \emph {et~al.}(1996)\citenamefont {Lee},
  \citenamefont {Zeng}, \citenamefont {Yao},\ and\ \citenamefont
  {Chen}}]{lee1996superconductivity}%
  \BibitemOpen
  \bibfield  {author} {\bibinfo {author} {\bibfnamefont {W.}~\bibnamefont
  {Lee}}, \bibinfo {author} {\bibfnamefont {H.}~\bibnamefont {Zeng}}, \bibinfo
  {author} {\bibfnamefont {Y.}~\bibnamefont {Yao}},\ and\ \bibinfo {author}
  {\bibfnamefont {Y.}~\bibnamefont {Chen}},\ }\bibfield  {title} {\bibinfo
  {title} {{Superconductivity in the Ni based ternary carbide LaNiC$_2$}},\
  }\href
  {https://www.sciencedirect.com/science/article/abs/pii/0921453496003097}
  {\bibfield  {journal} {\bibinfo  {journal} {Physica C: Superconductivity}\
  }\textbf {\bibinfo {volume} {266}},\ \bibinfo {pages} {138} (\bibinfo {year}
  {1996})}\BibitemShut {NoStop}%
\bibitem [{\citenamefont {Hirose}\ \emph {et~al.}(2012)\citenamefont {Hirose},
  \citenamefont {Kishino}, \citenamefont {Sakaguchi}, \citenamefont {Miura},
  \citenamefont {Honda}, \citenamefont {Takeuchi}, \citenamefont {Yamamoto},
  \citenamefont {Haga}, \citenamefont {Harima}, \citenamefont {Settai},\ and\
  \citenamefont {Ōnuki}}]{hirose2012fermi}%
  \BibitemOpen
  \bibfield  {author} {\bibinfo {author} {\bibfnamefont {Y.}~\bibnamefont
  {Hirose}}, \bibinfo {author} {\bibfnamefont {T.}~\bibnamefont {Kishino}},
  \bibinfo {author} {\bibfnamefont {J.}~\bibnamefont {Sakaguchi}}, \bibinfo
  {author} {\bibfnamefont {Y.}~\bibnamefont {Miura}}, \bibinfo {author}
  {\bibfnamefont {F.}~\bibnamefont {Honda}}, \bibinfo {author} {\bibfnamefont
  {T.}~\bibnamefont {Takeuchi}}, \bibinfo {author} {\bibfnamefont
  {E.}~\bibnamefont {Yamamoto}}, \bibinfo {author} {\bibfnamefont
  {Y.}~\bibnamefont {Haga}}, \bibinfo {author} {\bibfnamefont {H.}~\bibnamefont
  {Harima}}, \bibinfo {author} {\bibfnamefont {R.}~\bibnamefont {Settai}},\
  and\ \bibinfo {author} {\bibfnamefont {Y.}~\bibnamefont {Ōnuki}},\
  }\bibfield  {title} {\bibinfo {title} {Fermi surface and superconducting
  properties of non-centrosymmetric {LaNiC}$_2$},\ }\href
  {https://doi.org/10.1143/JPSJ.81.113703} {\bibfield  {journal} {\bibinfo
  {journal} {Journal of the Physical Society of Japan}\ }\textbf {\bibinfo
  {volume} {81}},\ \bibinfo {pages} {113703} (\bibinfo {year}
  {2012})}\BibitemShut {NoStop}%
\bibitem [{\citenamefont {Hillier}\ \emph {et~al.}(2009)\citenamefont
  {Hillier}, \citenamefont {Quintanilla},\ and\ \citenamefont
  {Cywinski}}]{PhysRevLett.102.117007}%
  \BibitemOpen
  \bibfield  {author} {\bibinfo {author} {\bibfnamefont {A.~D.}\ \bibnamefont
  {Hillier}}, \bibinfo {author} {\bibfnamefont {J.}~\bibnamefont
  {Quintanilla}},\ and\ \bibinfo {author} {\bibfnamefont {R.}~\bibnamefont
  {Cywinski}},\ }\bibfield  {title} {\bibinfo {title} {{Evidence for
  Time-Reversal Symmetry Breaking in the Noncentrosymmetric Superconductor
  LaNiC$_2$}},\ }\href {https://doi.org/10.1103/PhysRevLett.102.117007}
  {\bibfield  {journal} {\bibinfo  {journal} {Phys. Rev. Lett.}\ }\textbf
  {\bibinfo {volume} {102}},\ \bibinfo {pages} {117007} (\bibinfo {year}
  {2009})}\BibitemShut {NoStop}%
\bibitem [{\citenamefont {Yanagisawa}\ and\ \citenamefont
  {Hase}(2012)}]{yanagisawa2012nonunitary}%
  \BibitemOpen
  \bibfield  {author} {\bibinfo {author} {\bibfnamefont {T.}~\bibnamefont
  {Yanagisawa}}\ and\ \bibinfo {author} {\bibfnamefont {I.}~\bibnamefont
  {Hase}},\ }\bibfield  {title} {\bibinfo {title} {Nonunitary triplet
  superconductivity in the noncentrosymmetric rare-earth compound
  {LaNiC$_2$}},\ }\href {https://journals.jps.jp/doi/10.1143/JPSJS.81SB.SB039}
  {\bibfield  {journal} {\bibinfo  {journal} {Journal of the Physical Society
  of Japan}\ }\textbf {\bibinfo {volume} {81}},\ \bibinfo {pages} {SB039}
  (\bibinfo {year} {2012})}\BibitemShut {NoStop}%
\bibitem [{\citenamefont {Kolincio}\ \emph {et~al.}(2016)\citenamefont
  {Kolincio}, \citenamefont {G\'ornicka}, \citenamefont {Winiarski},
  \citenamefont {Strychalska-Nowak},\ and\ \citenamefont
  {Klimczuk}}]{kolincio2016gdnic2}%
  \BibitemOpen
  \bibfield  {author} {\bibinfo {author} {\bibfnamefont {K.~K.}\ \bibnamefont
  {Kolincio}}, \bibinfo {author} {\bibfnamefont {K.}~\bibnamefont
  {G\'ornicka}}, \bibinfo {author} {\bibfnamefont {M.~J.}\ \bibnamefont
  {Winiarski}}, \bibinfo {author} {\bibfnamefont {J.}~\bibnamefont
  {Strychalska-Nowak}},\ and\ \bibinfo {author} {\bibfnamefont
  {T.}~\bibnamefont {Klimczuk}},\ }\bibfield  {title} {\bibinfo {title}
  {Field-induced suppression of charge density wave in {GdNiC$_2$}},\ }\href
  {https://doi.org/10.1103/PhysRevB.94.195149} {\bibfield  {journal} {\bibinfo
  {journal} {Phys. Rev. B}\ }\textbf {\bibinfo {volume} {94}},\ \bibinfo
  {pages} {195149} (\bibinfo {year} {2016})}\BibitemShut {NoStop}%
\bibitem [{\citenamefont {Steiner}\ \emph {et~al.}(2018)\citenamefont
  {Steiner}, \citenamefont {Michor}, \citenamefont {Sologub}, \citenamefont
  {Hinterleitner}, \citenamefont {H\"ofenstock}, \citenamefont {Waas},
  \citenamefont {Bauer}, \citenamefont {St\"oger}, \citenamefont
  {Babizhetskyy}, \citenamefont {Levytskyy},\ and\ \citenamefont
  {Kotur}}]{steiner2018}%
  \BibitemOpen
  \bibfield  {author} {\bibinfo {author} {\bibfnamefont {S.}~\bibnamefont
  {Steiner}}, \bibinfo {author} {\bibfnamefont {H.}~\bibnamefont {Michor}},
  \bibinfo {author} {\bibfnamefont {O.}~\bibnamefont {Sologub}}, \bibinfo
  {author} {\bibfnamefont {B.}~\bibnamefont {Hinterleitner}}, \bibinfo {author}
  {\bibfnamefont {F.}~\bibnamefont {H\"ofenstock}}, \bibinfo {author}
  {\bibfnamefont {M.}~\bibnamefont {Waas}}, \bibinfo {author} {\bibfnamefont
  {E.}~\bibnamefont {Bauer}}, \bibinfo {author} {\bibfnamefont
  {B.}~\bibnamefont {St\"oger}}, \bibinfo {author} {\bibfnamefont
  {V.}~\bibnamefont {Babizhetskyy}}, \bibinfo {author} {\bibfnamefont
  {V.}~\bibnamefont {Levytskyy}},\ and\ \bibinfo {author} {\bibfnamefont
  {B.}~\bibnamefont {Kotur}},\ }\bibfield  {title} {\bibinfo {title}
  {Single-crystal study of the charge density wave metal {LuNiC$_2$}},\ }\href
  {https://link.aps.org/doi/10.1103/PhysRevB.97.205115} {\bibfield  {journal}
  {\bibinfo  {journal} {Phys. Rev. B}\ }\textbf {\bibinfo {volume} {97}},\
  \bibinfo {pages} {205115} (\bibinfo {year} {2018})}\BibitemShut {NoStop}%
\bibitem [{\citenamefont {Kolincio}\ \emph {et~al.}(2019)\citenamefont
  {Kolincio}, \citenamefont {Roman},\ and\ \citenamefont
  {Klimczuk}}]{kolincio2019ynic2}%
  \BibitemOpen
  \bibfield  {author} {\bibinfo {author} {\bibfnamefont {K.~K.}\ \bibnamefont
  {Kolincio}}, \bibinfo {author} {\bibfnamefont {M.}~\bibnamefont {Roman}},\
  and\ \bibinfo {author} {\bibfnamefont {T.}~\bibnamefont {Klimczuk}},\
  }\bibfield  {title} {\bibinfo {title} {{Charge density wave and large
  nonsaturating magnetoresistance in {YNiC}$_2$ and {LuNiC}$_2$}},\ }\href
  {https://doi.org/10.1103/PhysRevB.99.205127} {\bibfield  {journal} {\bibinfo
  {journal} {Phys. Rev. B}\ }\textbf {\bibinfo {volume} {99}},\ \bibinfo
  {pages} {205127} (\bibinfo {year} {2019})}\BibitemShut {NoStop}%
\bibitem [{\citenamefont {Hanasaki}\ \emph {et~al.}(2011)\citenamefont
  {Hanasaki}, \citenamefont {Mikami}, \citenamefont {Torigoe}, \citenamefont
  {Nogami}, \citenamefont {Shimomura}, \citenamefont {Kosaka},\ and\
  \citenamefont {Onodera}}]{hanasaki2011successive}%
  \BibitemOpen
  \bibfield  {author} {\bibinfo {author} {\bibfnamefont {N.}~\bibnamefont
  {Hanasaki}}, \bibinfo {author} {\bibfnamefont {K.}~\bibnamefont {Mikami}},
  \bibinfo {author} {\bibfnamefont {S.}~\bibnamefont {Torigoe}}, \bibinfo
  {author} {\bibfnamefont {Y.}~\bibnamefont {Nogami}}, \bibinfo {author}
  {\bibfnamefont {S.}~\bibnamefont {Shimomura}}, \bibinfo {author}
  {\bibfnamefont {M.}~\bibnamefont {Kosaka}},\ and\ \bibinfo {author}
  {\bibfnamefont {H.}~\bibnamefont {Onodera}},\ }\bibfield  {title} {\bibinfo
  {title} {{Successive Transition in Rare-earth Intermetallic Compound
  GdNiC$_2$}},\ }\href
  {https://iopscience.iop.org/article/10.1088/1742-6596/320/1/012072}
  {\bibfield  {journal} {\bibinfo  {journal} {Journal of Physics: Conference
  Series}\ }\textbf {\bibinfo {volume} {320}},\ \bibinfo {pages} {012072}
  (\bibinfo {year} {2011})}\BibitemShut {NoStop}%
\bibitem [{\citenamefont {Shimomura}\ \emph {et~al.}(2009)\citenamefont
  {Shimomura}, \citenamefont {Hayashi}, \citenamefont {Asaka}, \citenamefont
  {Wakabayashi}, \citenamefont {Mizumaki},\ and\ \citenamefont
  {Onodera}}]{PhysRevLett.102.076404}%
  \BibitemOpen
  \bibfield  {author} {\bibinfo {author} {\bibfnamefont {S.}~\bibnamefont
  {Shimomura}}, \bibinfo {author} {\bibfnamefont {C.}~\bibnamefont {Hayashi}},
  \bibinfo {author} {\bibfnamefont {G.}~\bibnamefont {Asaka}}, \bibinfo
  {author} {\bibfnamefont {N.}~\bibnamefont {Wakabayashi}}, \bibinfo {author}
  {\bibfnamefont {M.}~\bibnamefont {Mizumaki}},\ and\ \bibinfo {author}
  {\bibfnamefont {H.}~\bibnamefont {Onodera}},\ }\bibfield  {title} {\bibinfo
  {title} {{Charge-Density-Wave Destruction and Ferromagnetic Order in
  SmNiC$_2$}},\ }\href {https://doi.org/10.1103/PhysRevLett.102.076404}
  {\bibfield  {journal} {\bibinfo  {journal} {Phys. Rev. Lett.}\ }\textbf
  {\bibinfo {volume} {102}},\ \bibinfo {pages} {076404} (\bibinfo {year}
  {2009})}\BibitemShut {NoStop}%
\bibitem [{\citenamefont {Laverock}\ \emph {et~al.}(2009)\citenamefont
  {Laverock}, \citenamefont {Haynes}, \citenamefont {Utfeld},\ and\
  \citenamefont {Dugdale}}]{PhysRevB.80.125111}%
  \BibitemOpen
  \bibfield  {author} {\bibinfo {author} {\bibfnamefont {J.}~\bibnamefont
  {Laverock}}, \bibinfo {author} {\bibfnamefont {T.~D.}\ \bibnamefont
  {Haynes}}, \bibinfo {author} {\bibfnamefont {C.}~\bibnamefont {Utfeld}},\
  and\ \bibinfo {author} {\bibfnamefont {S.~B.}\ \bibnamefont {Dugdale}},\
  }\bibfield  {title} {\bibinfo {title} {Electronic structure of {RNiC$_2$
  (R=Sm, Gd, and Nd)} intermetallic compounds},\ }\href
  {https://doi.org/10.1103/PhysRevB.80.125111} {\bibfield  {journal} {\bibinfo
  {journal} {Phys. Rev. B}\ }\textbf {\bibinfo {volume} {80}},\ \bibinfo
  {pages} {125111} (\bibinfo {year} {2009})}\BibitemShut {NoStop}%
\bibitem [{\citenamefont {Hanasaki}\ \emph {et~al.}(2012)\citenamefont
  {Hanasaki}, \citenamefont {Nogami}, \citenamefont {Kakinuma}, \citenamefont
  {Shimomura}, \citenamefont {Kosaka},\ and\ \citenamefont
  {Onodera}}]{hanasaki2012smnic2}%
  \BibitemOpen
  \bibfield  {author} {\bibinfo {author} {\bibfnamefont {N.}~\bibnamefont
  {Hanasaki}}, \bibinfo {author} {\bibfnamefont {Y.}~\bibnamefont {Nogami}},
  \bibinfo {author} {\bibfnamefont {M.}~\bibnamefont {Kakinuma}}, \bibinfo
  {author} {\bibfnamefont {S.}~\bibnamefont {Shimomura}}, \bibinfo {author}
  {\bibfnamefont {M.}~\bibnamefont {Kosaka}},\ and\ \bibinfo {author}
  {\bibfnamefont {H.}~\bibnamefont {Onodera}},\ }\bibfield  {title} {\bibinfo
  {title} {Magnetic field switching of the charge-density-wave state in the
  lanthanide intermetallic {SmNiC$_{2}$}},\ }\href
  {https://doi.org/10.1103/PhysRevB.85.092402} {\bibfield  {journal} {\bibinfo
  {journal} {Phys. Rev. B}\ }\textbf {\bibinfo {volume} {85}},\ \bibinfo
  {pages} {092402} (\bibinfo {year} {2012})}\BibitemShut {NoStop}%
\bibitem [{\citenamefont {Prathiba}\ \emph {et~al.}(2016)\citenamefont
  {Prathiba}, \citenamefont {Kim}, \citenamefont {Shin}, \citenamefont
  {Strychalska}, \citenamefont {Klimczuk},\ and\ \citenamefont
  {Park}}]{prathiba2016tuning}%
  \BibitemOpen
  \bibfield  {author} {\bibinfo {author} {\bibfnamefont {G.}~\bibnamefont
  {Prathiba}}, \bibinfo {author} {\bibfnamefont {I.}~\bibnamefont {Kim}},
  \bibinfo {author} {\bibfnamefont {S.}~\bibnamefont {Shin}}, \bibinfo {author}
  {\bibfnamefont {J.}~\bibnamefont {Strychalska}}, \bibinfo {author}
  {\bibfnamefont {T.}~\bibnamefont {Klimczuk}},\ and\ \bibinfo {author}
  {\bibfnamefont {T.}~\bibnamefont {Park}},\ }\bibfield  {title} {\bibinfo
  {title} {{Tuning the ferromagnetic phase in the CDW compound SmNiC$_2$ via
  chemical alloying}},\ }\href {https://www.nature.com/articles/srep26530}
  {\bibfield  {journal} {\bibinfo  {journal} {Scientific reports}\ }\textbf
  {\bibinfo {volume} {6}},\ \bibinfo {pages} {26530} (\bibinfo {year}
  {2016})}\BibitemShut {NoStop}%
\bibitem [{\citenamefont {Kim}\ \emph {et~al.}(2013)\citenamefont {Kim},
  \citenamefont {Lee},\ and\ \citenamefont {Shim}}]{kim2013chemical}%
  \BibitemOpen
  \bibfield  {author} {\bibinfo {author} {\bibfnamefont {J.~N.}\ \bibnamefont
  {Kim}}, \bibinfo {author} {\bibfnamefont {C.}~\bibnamefont {Lee}},\ and\
  \bibinfo {author} {\bibfnamefont {J.-H.}\ \bibnamefont {Shim}},\ }\bibfield
  {title} {\bibinfo {title} {Chemical and hydrostatic pressure effect on charge
  density waves of {SmNiC$_2$}},\ }\href
  {https://iopscience.iop.org/article/10.1088/1367-2630/15/12/123018}
  {\bibfield  {journal} {\bibinfo  {journal} {New Journal of Physics}\ }\textbf
  {\bibinfo {volume} {15}},\ \bibinfo {pages} {123018} (\bibinfo {year}
  {2013})}\BibitemShut {NoStop}%
\bibitem [{\citenamefont {Hanasaki}\ \emph {et~al.}(2017)\citenamefont
  {Hanasaki}, \citenamefont {Shimomura}, \citenamefont {Mikami}, \citenamefont
  {Nogami}, \citenamefont {Nakao},\ and\ \citenamefont
  {Onodera}}]{hanasaki2017gdnic2}%
  \BibitemOpen
  \bibfield  {author} {\bibinfo {author} {\bibfnamefont {N.}~\bibnamefont
  {Hanasaki}}, \bibinfo {author} {\bibfnamefont {S.}~\bibnamefont {Shimomura}},
  \bibinfo {author} {\bibfnamefont {K.}~\bibnamefont {Mikami}}, \bibinfo
  {author} {\bibfnamefont {Y.}~\bibnamefont {Nogami}}, \bibinfo {author}
  {\bibfnamefont {H.}~\bibnamefont {Nakao}},\ and\ \bibinfo {author}
  {\bibfnamefont {H.}~\bibnamefont {Onodera}},\ }\bibfield  {title} {\bibinfo
  {title} {Interplay between charge density wave and antiferromagnetic order in
  {GdNiC$_2$}},\ }\href {https://doi.org/10.1103/PhysRevB.95.085103} {\bibfield
   {journal} {\bibinfo  {journal} {Phys. Rev. B}\ }\textbf {\bibinfo {volume}
  {95}},\ \bibinfo {pages} {085103} (\bibinfo {year} {2017})}\BibitemShut
  {NoStop}%
\bibitem [{Note1()}]{Note1}%
  \BibitemOpen
  \bibinfo {note} {For this compound, we have compared our results with Ref.
  \cite {xu2019}. While we can reproduce the Weyl nodes in that work,
  additional search for crossings between bands $N-1$ and $N$ discloses Weyl
  nodes at lower energy, as indicated in Fig. \ref {fig:bands}(g)}\BibitemShut
  {NoStop}%
\bibitem [{\citenamefont {Xu}\ \emph {et~al.}(2019)\citenamefont {Xu},
  \citenamefont {Gu}, \citenamefont {Zhang}, \citenamefont {Fang},
  \citenamefont {Fang}, \citenamefont {Sheng},\ and\ \citenamefont
  {Weng}}]{xu2019}%
  \BibitemOpen
  \bibfield  {author} {\bibinfo {author} {\bibfnamefont {Y.}~\bibnamefont
  {Xu}}, \bibinfo {author} {\bibfnamefont {Y.}~\bibnamefont {Gu}}, \bibinfo
  {author} {\bibfnamefont {T.}~\bibnamefont {Zhang}}, \bibinfo {author}
  {\bibfnamefont {C.}~\bibnamefont {Fang}}, \bibinfo {author} {\bibfnamefont
  {Z.}~\bibnamefont {Fang}}, \bibinfo {author} {\bibfnamefont {X.-L.}\
  \bibnamefont {Sheng}},\ and\ \bibinfo {author} {\bibfnamefont
  {H.}~\bibnamefont {Weng}},\ }\bibfield  {title} {\bibinfo {title}
  {{Topological nodal lines and hybrid Weyl nodes in YCoC$_2$}},\ }\href
  {https://doi.org/10.1063/1.5123222} {\bibfield  {journal} {\bibinfo
  {journal} {APL Materials}\ }\textbf {\bibinfo {volume} {7}},\ \bibinfo
  {pages} {101109} (\bibinfo {year} {2019})}\BibitemShut {NoStop}%
\bibitem [{\citenamefont {Perdew}\ \emph {et~al.}(1996)\citenamefont {Perdew},
  \citenamefont {Burke},\ and\ \citenamefont {Ernzerhof}}]{pbe1996}%
  \BibitemOpen
  \bibfield  {author} {\bibinfo {author} {\bibfnamefont {J.~P.}\ \bibnamefont
  {Perdew}}, \bibinfo {author} {\bibfnamefont {K.}~\bibnamefont {Burke}},\ and\
  \bibinfo {author} {\bibfnamefont {M.}~\bibnamefont {Ernzerhof}},\ }\bibfield
  {title} {\bibinfo {title} {Generalized gradient approximation made simple},\
  }\href {https://doi.org/10.1103/PhysRevLett.77.3865} {\bibfield  {journal}
  {\bibinfo  {journal} {Phys. Rev. Lett.}\ }\textbf {\bibinfo {volume} {77}},\
  \bibinfo {pages} {3865} (\bibinfo {year} {1996})}\BibitemShut {NoStop}%
\bibitem [{\citenamefont {Koepernik}\ and\ \citenamefont
  {Eschrig}(1999)}]{klaus1999}%
  \BibitemOpen
  \bibfield  {author} {\bibinfo {author} {\bibfnamefont {K.}~\bibnamefont
  {Koepernik}}\ and\ \bibinfo {author} {\bibfnamefont {H.}~\bibnamefont
  {Eschrig}},\ }\bibfield  {title} {\bibinfo {title} {Full-potential
  nonorthogonal local-orbital minimum-basis band-structure scheme},\ }\href
  {https://doi.org/10.1103/PhysRevB.59.1743} {\bibfield  {journal} {\bibinfo
  {journal} {Phys. Rev. B}\ }\textbf {\bibinfo {volume} {59}},\ \bibinfo
  {pages} {1743} (\bibinfo {year} {1999})}\BibitemShut {NoStop}%
\bibitem [{fpl()}]{fplo_web}%
  \BibitemOpen
  \href {https://www.fplo.de} {}\bibinfo {note}
  {\href{https://www.fplo.de}{https://www.fplo.de}}\BibitemShut {NoStop}%
\bibitem [{\citenamefont {Richter}(2001)}]{Richter2001}%
  \BibitemOpen
  \bibfield  {author} {\bibinfo {author} {\bibfnamefont {M.}~\bibnamefont
  {Richter}},\ }\bibinfo {title} {{Density Functional Theory Applied to $4f$
  and $5f$ Elements and Metallic Compounds}}\ (\bibinfo  {publisher}
  {Elsevier},\ \bibinfo {address} {Amsterdam},\ \bibinfo {year} {2001})\
  Chap.~\bibinfo {chapter} {2}, pp.\ \bibinfo {pages} {87--228}\BibitemShut
  {NoStop}%
\bibitem [{\citenamefont {Czyzyk}\ and\ \citenamefont
  {Sawatzky}(1994)}]{Czyzyk1994}%
  \BibitemOpen
  \bibfield  {author} {\bibinfo {author} {\bibfnamefont {M.~T.}\ \bibnamefont
  {Czyzyk}}\ and\ \bibinfo {author} {\bibfnamefont {G.~A.}\ \bibnamefont
  {Sawatzky}},\ }\bibfield  {title} {\bibinfo {title} {Local-density functional
  and on-site correlations: The electronic structure of {La}$_{2}${CuO}$_{4}$
  and {LaCuO}$_{3}$},\ }\href {https://doi.org/10.1103/PhysRevB.49.14211}
  {\bibfield  {journal} {\bibinfo  {journal} {Phys. Rev. B}\ }\textbf {\bibinfo
  {volume} {49}},\ \bibinfo {pages} {14211} (\bibinfo {year}
  {1994})}\BibitemShut {NoStop}%
\bibitem [{\citenamefont {Jain}\ \emph {et~al.}(2013)\citenamefont {Jain},
  \citenamefont {Ong}, \citenamefont {Hautier}, \citenamefont {Chen},
  \citenamefont {Richards}, \citenamefont {Dacek}, \citenamefont {Cholia},
  \citenamefont {Gunter}, \citenamefont {Skinner}, \citenamefont {Ceder} \emph
  {et~al.}}]{jain2013}%
  \BibitemOpen
  \bibfield  {author} {\bibinfo {author} {\bibfnamefont {A.}~\bibnamefont
  {Jain}}, \bibinfo {author} {\bibfnamefont {S.~P.}\ \bibnamefont {Ong}},
  \bibinfo {author} {\bibfnamefont {G.}~\bibnamefont {Hautier}}, \bibinfo
  {author} {\bibfnamefont {W.}~\bibnamefont {Chen}}, \bibinfo {author}
  {\bibfnamefont {W.~D.}\ \bibnamefont {Richards}}, \bibinfo {author}
  {\bibfnamefont {S.}~\bibnamefont {Dacek}}, \bibinfo {author} {\bibfnamefont
  {S.}~\bibnamefont {Cholia}}, \bibinfo {author} {\bibfnamefont
  {D.}~\bibnamefont {Gunter}}, \bibinfo {author} {\bibfnamefont
  {D.}~\bibnamefont {Skinner}}, \bibinfo {author} {\bibfnamefont
  {G.}~\bibnamefont {Ceder}}, \emph {et~al.},\ }\bibfield  {title} {\bibinfo
  {title} {Commentary: The materials project: A materials genome approach to
  accelerating materials innovation},\ }\href
  {https://doi.org/10.1063/1.4812323} {\bibfield  {journal} {\bibinfo
  {journal} {APL Materials}\ }\textbf {\bibinfo {volume} {1}},\ \bibinfo
  {pages} {011002} (\bibinfo {year} {2013})}\BibitemShut {NoStop}%
\bibitem [{sm()}]{sm}%
  \BibitemOpen
  \href@noop {} {}\bibinfo {note} {Supplemental material contains further
  information on (i) crystal structures used, (ii) methods used, (iii) effects
  of canting the magnetization in GdCoC$_2$, (iv) band structures and density
  of states, and (v) Weyl nodes found.}\BibitemShut {Stop}%
\bibitem [{\citenamefont {Friedan}(1982)}]{friedan1982proof}%
  \BibitemOpen
  \bibfield  {author} {\bibinfo {author} {\bibfnamefont {D.}~\bibnamefont
  {Friedan}},\ }\bibfield  {title} {\bibinfo {title} {{A proof of the
  Nielsen-Ninomiya theorem}},\ }\href
  {https://link.springer.com/article/10.1007/BF01403500} {\bibfield  {journal}
  {\bibinfo  {journal} {Communications in Mathematical Physics}\ }\textbf
  {\bibinfo {volume} {85}},\ \bibinfo {pages} {481} (\bibinfo {year}
  {1982})}\BibitemShut {NoStop}%
\bibitem [{\citenamefont {Witten}(2016)}]{witten2016three}%
  \BibitemOpen
  \bibfield  {author} {\bibinfo {author} {\bibfnamefont {E.}~\bibnamefont
  {Witten}},\ }\bibfield  {title} {\bibinfo {title} {Three lectures on
  topological phases of matter},\ }\href
  {https://www.sif.it/riviste/sif/ncr/econtents/2016/039/07/article/0}
  {\bibfield  {journal} {\bibinfo  {journal} {Rivista del Nuovo Cimento della
  Societa Italiana di Fisica}\ }\textbf {\bibinfo {volume} {39}},\ \bibinfo
  {pages} {313} (\bibinfo {year} {2016})}\BibitemShut {NoStop}%
\bibitem [{\citenamefont {Meng}\ and\ \citenamefont
  {Budich}(2019)}]{PhysRevLett.122.046402}%
  \BibitemOpen
  \bibfield  {author} {\bibinfo {author} {\bibfnamefont {T.}~\bibnamefont
  {Meng}}\ and\ \bibinfo {author} {\bibfnamefont {J.~C.}\ \bibnamefont
  {Budich}},\ }\bibfield  {title} {\bibinfo {title} {{Unpaired Weyl Nodes from
  Long-Ranged Interactions: Fate of Quantum Anomalies}},\ }\href
  {https://doi.org/10.1103/PhysRevLett.122.046402} {\bibfield  {journal}
  {\bibinfo  {journal} {Phys. Rev. Lett.}\ }\textbf {\bibinfo {volume} {122}},\
  \bibinfo {pages} {046402} (\bibinfo {year} {2019})}\BibitemShut {NoStop}%
\bibitem [{\citenamefont {Crippa}\ \emph {et~al.}(2020)\citenamefont {Crippa},
  \citenamefont {Amaricci}, \citenamefont {Wagner}, \citenamefont
  {Sangiovanni}, \citenamefont {Budich},\ and\ \citenamefont
  {Capone}}]{PhysRevResearch.2.012023}%
  \BibitemOpen
  \bibfield  {author} {\bibinfo {author} {\bibfnamefont {L.}~\bibnamefont
  {Crippa}}, \bibinfo {author} {\bibfnamefont {A.}~\bibnamefont {Amaricci}},
  \bibinfo {author} {\bibfnamefont {N.}~\bibnamefont {Wagner}}, \bibinfo
  {author} {\bibfnamefont {G.}~\bibnamefont {Sangiovanni}}, \bibinfo {author}
  {\bibfnamefont {J.~C.}\ \bibnamefont {Budich}},\ and\ \bibinfo {author}
  {\bibfnamefont {M.}~\bibnamefont {Capone}},\ }\bibfield  {title} {\bibinfo
  {title} {{Nonlocal annihilation of Weyl fermions in correlated systems}},\
  }\href {https://doi.org/10.1103/PhysRevResearch.2.012023} {\bibfield
  {journal} {\bibinfo  {journal} {Phys. Rev. Research}\ }\textbf {\bibinfo
  {volume} {2}},\ \bibinfo {pages} {012023} (\bibinfo {year}
  {2020})}\BibitemShut {NoStop}%
\bibitem [{\citenamefont {Zhong}\ \emph {et~al.}(2016)\citenamefont {Zhong},
  \citenamefont {Moore},\ and\ \citenamefont {Souza}}]{PhysRevLett.116.077201}%
  \BibitemOpen
  \bibfield  {author} {\bibinfo {author} {\bibfnamefont {S.}~\bibnamefont
  {Zhong}}, \bibinfo {author} {\bibfnamefont {J.~E.}\ \bibnamefont {Moore}},\
  and\ \bibinfo {author} {\bibfnamefont {I.}~\bibnamefont {Souza}},\ }\bibfield
   {title} {\bibinfo {title} {{Gyrotropic Magnetic Effect and the Magnetic
  Moment on the Fermi Surface}},\ }\href
  {https://doi.org/10.1103/PhysRevLett.116.077201} {\bibfield  {journal}
  {\bibinfo  {journal} {Phys. Rev. Lett.}\ }\textbf {\bibinfo {volume} {116}},\
  \bibinfo {pages} {077201} (\bibinfo {year} {2016})}\BibitemShut {NoStop}%
\bibitem [{\citenamefont {Ma}\ and\ \citenamefont
  {Pesin}(2015)}]{PhysRevB.92.235205}%
  \BibitemOpen
  \bibfield  {author} {\bibinfo {author} {\bibfnamefont {J.}~\bibnamefont
  {Ma}}\ and\ \bibinfo {author} {\bibfnamefont {D.~A.}\ \bibnamefont {Pesin}},\
  }\bibfield  {title} {\bibinfo {title} {Chiral magnetic effect and natural
  optical activity in metals with or without {Weyl} points},\ }\href
  {https://doi.org/10.1103/PhysRevB.92.235205} {\bibfield  {journal} {\bibinfo
  {journal} {Phys. Rev. B}\ }\textbf {\bibinfo {volume} {92}},\ \bibinfo
  {pages} {235205} (\bibinfo {year} {2015})}\BibitemShut {NoStop}%
\bibitem [{\citenamefont {de~Juan}\ \emph {et~al.}(2017)\citenamefont
  {de~Juan}, \citenamefont {Grushin}, \citenamefont {Morimoto},\ and\
  \citenamefont {Moore}}]{de2017quantized}%
  \BibitemOpen
  \bibfield  {author} {\bibinfo {author} {\bibfnamefont {F.}~\bibnamefont
  {de~Juan}}, \bibinfo {author} {\bibfnamefont {A.~G.}\ \bibnamefont
  {Grushin}}, \bibinfo {author} {\bibfnamefont {T.}~\bibnamefont {Morimoto}},\
  and\ \bibinfo {author} {\bibfnamefont {J.~E.}\ \bibnamefont {Moore}},\
  }\bibfield  {title} {\bibinfo {title} {Quantized circular photogalvanic
  effect in {Weyl} semimetals},\ }\href
  {https://www.nature.com/articles/ncomms15995} {\bibfield  {journal} {\bibinfo
   {journal} {Nature communications}\ }\textbf {\bibinfo {volume} {8}},\
  \bibinfo {pages} {15995} (\bibinfo {year} {2017})}\BibitemShut {NoStop}%
\bibitem [{\citenamefont {Holder}\ \emph {et~al.}(2019)\citenamefont {Holder},
  \citenamefont {Kaplan},\ and\ \citenamefont {Yan}}]{holder2019consequences}%
  \BibitemOpen
  \bibfield  {author} {\bibinfo {author} {\bibfnamefont {T.}~\bibnamefont
  {Holder}}, \bibinfo {author} {\bibfnamefont {D.}~\bibnamefont {Kaplan}},\
  and\ \bibinfo {author} {\bibfnamefont {B.}~\bibnamefont {Yan}},\ }\bibfield
  {title} {\bibinfo {title} {{Consequences of Time-reversal-symmetry Breaking
  in the Light-Matter Interaction: Berry Curvature, Quantum Metric and Diabatic
  Motion}},\ }\href {https://arxiv.org/pdf/1911.05667} {\bibfield  {journal}
  {\bibinfo  {journal} {arXiv preprint arXiv:1911.05667}\ } (\bibinfo {year}
  {2019})}\BibitemShut {NoStop}%
\bibitem [{\citenamefont {Nandy}\ \emph {et~al.}(2017)\citenamefont {Nandy},
  \citenamefont {Sharma}, \citenamefont {Taraphder},\ and\ \citenamefont
  {Tewari}}]{PhysRevLett.119.176804}%
  \BibitemOpen
  \bibfield  {author} {\bibinfo {author} {\bibfnamefont {S.}~\bibnamefont
  {Nandy}}, \bibinfo {author} {\bibfnamefont {G.}~\bibnamefont {Sharma}},
  \bibinfo {author} {\bibfnamefont {A.}~\bibnamefont {Taraphder}},\ and\
  \bibinfo {author} {\bibfnamefont {S.}~\bibnamefont {Tewari}},\ }\bibfield
  {title} {\bibinfo {title} {{Chiral Anomaly as the Origin of the Planar Hall
  Effect in Weyl Semimetals}},\ }\href
  {https://doi.org/10.1103/PhysRevLett.119.176804} {\bibfield  {journal}
  {\bibinfo  {journal} {Phys. Rev. Lett.}\ }\textbf {\bibinfo {volume} {119}},\
  \bibinfo {pages} {176804} (\bibinfo {year} {2017})}\BibitemShut {NoStop}%
\bibitem [{\citenamefont {Nag}\ and\ \citenamefont
  {Nandy}(2018)}]{nag2018transport}%
  \BibitemOpen
  \bibfield  {author} {\bibinfo {author} {\bibfnamefont {T.}~\bibnamefont
  {Nag}}\ and\ \bibinfo {author} {\bibfnamefont {S.}~\bibnamefont {Nandy}},\
  }\bibfield  {title} {\bibinfo {title} {{Transport phenomena of multi-Weyl
  semimetals in co-planar setups}},\ }\href {https://arxiv.org/abs/1812.08322}
  {\bibfield  {journal} {\bibinfo  {journal} {arXiv preprint arXiv:1812.08322}\
  } (\bibinfo {year} {2018})}\BibitemShut {NoStop}%
\bibitem [{\citenamefont {Persson}(2016{\natexlab{a}})}]{ycoc2_mp}%
  \BibitemOpen
  \bibfield  {author} {\bibinfo {author} {\bibfnamefont {K.}~\bibnamefont
  {Persson}},\ }\href {https://doi.org/10.17188/1207998} {\bibinfo {title}
  {Materials data on {YCoC}$_2$ ({SG}:38) by materials project}} (\bibinfo
  {year} {2016}{\natexlab{a}})\BibitemShut {NoStop}%
\bibitem [{\citenamefont {Persson}(2016{\natexlab{b}})}]{lucoc2_mp}%
  \BibitemOpen
  \bibfield  {author} {\bibinfo {author} {\bibfnamefont {K.}~\bibnamefont
  {Persson}},\ }\href {https://doi.org/10.17188/1273031} {\bibinfo {title}
  {Materials data on {LuCoC}$_2$ ({SG}:38) by materials project}} (\bibinfo
  {year} {2016}{\natexlab{b}})\BibitemShut {NoStop}%
\bibitem [{\citenamefont {Persson}(2017)}]{gdcoc2_mp}%
  \BibitemOpen
  \bibfield  {author} {\bibinfo {author} {\bibfnamefont {K.}~\bibnamefont
  {Persson}},\ }\href {https://doi.org/10.17188/1350255} {\bibinfo {title}
  {Materials data on {GdCoC}$_2$ ({SG}:38) by materials project}} (\bibinfo
  {year} {2017})\BibitemShut {NoStop}%
\bibitem [{\citenamefont {Persson}()}]{gdnic2_mp}%
  \BibitemOpen
  \bibfield  {author} {\bibinfo {author} {\bibfnamefont {K.}~\bibnamefont
  {Persson}},\ }\href {https://materialsproject.org/materials/mp-1065668/}
  {\bibinfo {title} {Materials data on {GdNiC}$_2$ ({SG}:38) by materials
  project}}\BibitemShut {NoStop}%
\bibitem [{\citenamefont {Persson}(2016{\natexlab{c}})}]{ndrhc2_mp}%
  \BibitemOpen
  \bibfield  {author} {\bibinfo {author} {\bibfnamefont {K.}~\bibnamefont
  {Persson}},\ }\href {https://doi.org/10.17188/1309147} {\bibinfo {title}
  {Materials data on {NdRhC}$_2$ ({SG}:38) by materials project}} (\bibinfo
  {year} {2016}{\natexlab{c}})\BibitemShut {NoStop}%
\bibitem [{\citenamefont {Persson}(2016{\natexlab{d}})}]{prrhc2_mp}%
  \BibitemOpen
  \bibfield  {author} {\bibinfo {author} {\bibfnamefont {K.}~\bibnamefont
  {Persson}},\ }\href {https://doi.org/10.17188/1317526} {\bibinfo {title}
  {Materials data on {PrRhC}$_2$ ({SG}:38) by materials project}} (\bibinfo
  {year} {2016}{\natexlab{d}})\BibitemShut {NoStop}%
\bibitem [{\citenamefont {Persson}(2016{\natexlab{e}})}]{gdruc2_mp}%
  \BibitemOpen
  \bibfield  {author} {\bibinfo {author} {\bibfnamefont {K.}~\bibnamefont
  {Persson}},\ }\href {https://doi.org/10.17188/1277011} {\bibinfo {title}
  {Materials data on {GdRuC}$_2$ ({SG}:63) by materials project}} (\bibinfo
  {year} {2016}{\natexlab{e}})\BibitemShut {NoStop}%
\bibitem [{\citenamefont {Persson}(2016{\natexlab{f}})}]{larhc2_mp}%
  \BibitemOpen
  \bibfield  {author} {\bibinfo {author} {\bibfnamefont {K.}~\bibnamefont
  {Persson}},\ }\href {https://doi.org/10.17188/1206662} {\bibinfo {title}
  {Materials data on {LaRhC}$_2$ ({SG}:76) by materials project}} (\bibinfo
  {year} {2016}{\natexlab{f}})\BibitemShut {NoStop}%
\bibitem [{\citenamefont {Dorado}\ \emph {et~al.}(2013)\citenamefont {Dorado},
  \citenamefont {Freyss}, \citenamefont {Amadon}, \citenamefont {Bertolus},
  \citenamefont {Jomard},\ and\ \citenamefont {Garcia}}]{dorado2013}%
  \BibitemOpen
  \bibfield  {author} {\bibinfo {author} {\bibfnamefont {B.}~\bibnamefont
  {Dorado}}, \bibinfo {author} {\bibfnamefont {M.}~\bibnamefont {Freyss}},
  \bibinfo {author} {\bibfnamefont {B.}~\bibnamefont {Amadon}}, \bibinfo
  {author} {\bibfnamefont {M.}~\bibnamefont {Bertolus}}, \bibinfo {author}
  {\bibfnamefont {G.}~\bibnamefont {Jomard}},\ and\ \bibinfo {author}
  {\bibfnamefont {P.}~\bibnamefont {Garcia}},\ }\bibfield  {title} {\bibinfo
  {title} {Advances in first-principles modelling of point defects in {UO}$_2$:
  f electron correlations and the issue of local energy minima},\ }\href
  {https://doi.org/10.1088/0953-8984/25/33/333201} {\bibfield  {journal}
  {\bibinfo  {journal} {Journal of Physics: Condensed Matter}\ }\textbf
  {\bibinfo {volume} {25}},\ \bibinfo {pages} {333201} (\bibinfo {year}
  {2013})}\BibitemShut {NoStop}%
\bibitem [{\citenamefont {Allen}\ and\ \citenamefont
  {Watson}(2014)}]{allen2014}%
  \BibitemOpen
  \bibfield  {author} {\bibinfo {author} {\bibfnamefont {J.~P.}\ \bibnamefont
  {Allen}}\ and\ \bibinfo {author} {\bibfnamefont {G.~W.}\ \bibnamefont
  {Watson}},\ }\bibfield  {title} {\bibinfo {title} {Occupation matrix control
  of d- and f-electron localisations using {DFT +$U$}},\ }\href
  {https://doi.org/10.1039/C4CP01083C} {\bibfield  {journal} {\bibinfo
  {journal} {Phys. Chem. Chem. Phys.}\ }\textbf {\bibinfo {volume} {16}},\
  \bibinfo {pages} {21016} (\bibinfo {year} {2014})}\BibitemShut {NoStop}%
\end{thebibliography}%

\clearpage
\newpage
\renewcommand{\thesection}{S\arabic{section}}
\renewcommand{\thesubsection}{S\arabic{subsection}}
\renewcommand{\thetable}{S\Roman{table}}
\renewcommand{\thefigure}{S\arabic{figure}}
\setcounter{section}{0}
\setcounter{subsection}{0}
\setcounter{figure}{0}
\setcounter{table}{0}
\setcounter{equation}{0}
\widetext
\section*{Supplemental Information}

Section \ref{sec_CS} contains details about the crystal structures used in the calculations.
Section \ref{sec_M} explains the methodological aspects of the calculations.
Section \ref{sec_wn_gdcoc2} presents the results of canting the magnetization in GdCoC$_2$.
Section \ref{sec_EP} exhibits, in Figs. \ref{fig:elprop_ycoc2} $-$ \ref{fig:elprop_larhc2}, the
bandstructures of the considered materials on a larger Brillouin zone path and larger energy window
than in the main text. Tables \ref{table:WPs_ycoc2} $-$ \ref{table:WPs_prrhc2} contain further
information about the Weyl nodes found in the considered semimetals.

\section{Crystal structures}
\label{sec_CS}

The structural parameters for all
the compounds considered in this study were obtained from {\it The Materials Project} \cite{jain2013}.
Table \ref{table:str} lists the space groups, lattice constants ($a$, $b$, $c$), primitive unit cell volumes
(Vol), corresponding formula units (f.u.) per unit cell ($Z$), and
selected bond lengths ($d$) for these compounds.

\begin{table*}[h!]
    \small
    \caption{Structural parameters for the considered compounds with the
        general formula {\gen}. Note that only the distance between intra-layer R-ions is listed, since the
        layers are separated by the lattice vector $a$. For systems belonging to classes I-III, the layers are
        defined according to Fig. 1 of the main text. }

    \begin{tabular*}{0.98\textwidth}{ p{2.5cm} p{1.72cm} p{1.72cm} p{1.72cm} p{1.72cm} p{1.72cm} p{1.72cm} p{1.72cm} p{1.72cm} }
    
        \hline\hline
        {\bf Parameter} & {\yco}\cite{ycoc2_mp} & {\luco}\cite{lucoc2_mp}
                        & {\gdco}\cite{gdcoc2_mp}  &
                        {\gdni}\cite{gdnic2_mp} & {\nd}\cite{ndrhc2_mp}  & {\pr}\cite{prrhc2_mp}
  & {\gdru}\cite{gdruc2_mp} 
                        & {\la}\cite{larhc2_mp} \\
        \hline
        Class & I & I & III & III & III & III & II & IV\\
        \\
        Space Gr. & $Amm2$ & $Amm2$ & $Amm2$ & $Amm2$ & $Amm2$ & $Amm2$ & $Cmcm$ & $P4_1$ \\
        $a$ ({\AA}) & 3.5880 & 3.4336 & 3.6336 & 3.6597  & 3.7184 & 3.7760 & 4.4766 & 3.9913   \\
        $b$ ({\AA}) & 4.5141 & 3.4759 & 4.5185 & 4.5335 & 4.7308 & 4.7367 & 9.3054 & 3.9913  \\
        $c$ ({\AA}) & 5.9955 & 5.9272 & 6.0277 & 6.03826 & 6.6371 & 6.6528 & 5.2503  & 15.4194 \\
        $Z$  & 1 & 1 & 1 & 1 & 1 & 1  & 2& 4 \\
        Vol ({\AA}$^3$) & 48.56 & 45.55 & 49.48 & 50.10 & 58.38 & 59.50  & 109.36 & 245.64 \\
        \\
        $d_{\rm R-R}^{\rm intra-layer}$ ({\AA}) & 3.7524 & 3.7137 & 3.7666 & 3.7754 & 4.0753 & 4.0834 & 3.6681 & 4.0335 \\
                              & 4.5141 & 4.4759 & 4.5185 & 4.5335 & 4.7308 & 4.7367 & 4.4766 & 4.7669 \\
                              
        $d_{\rm C-C}$ ({\AA}) & 1.3794 & 1.3796 & 1.3776 & 1.3733 & 1.3498 & 1.3496 & 1.4078 & 1.3552 \\
        $d_{\rm M-C}$ ({\AA}) & 1.9345 & 1.9092 & 1.9462 & 1.9476 & 2.1075 & 2.1135 & 2.16734 & 2.0746 \\
                              & 1.9873 & 1.9710 & 1.9875 & 2.0020 & 2.1679 & 2.1696 & $-$ & 2.0867  \\
                              &        &        &        &        &        &        &       & 2.1292 \\
                              &        &        &        &        &        &        &       & 2.1662 \\
        \hline
    \hline
    \end{tabular*}
    \label{table:str}
\end{table*}

\section{Methods}
\label{sec_M}

\begin{figure*}[hb!]
    \centering
        \includegraphics[angle=0,width=0.855\textwidth]{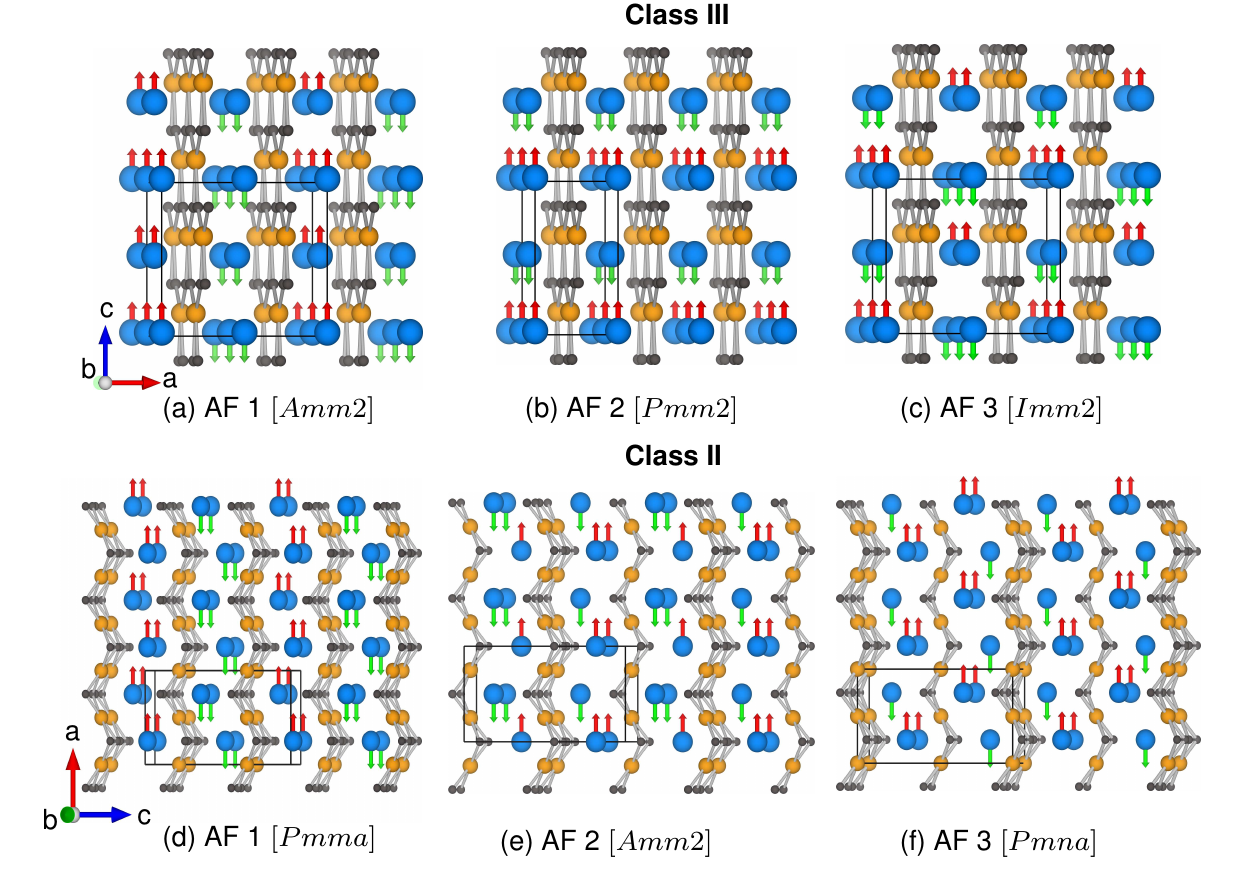}
        \caption{AFM configurations considered for compounds belonging
            to the symmetry classes II and III. 
            The R atoms, transition metal atoms and carbon atoms are, respectively, shown
            with blue, yellow and grey spheres. The relative spin directions on the R atoms
            are represented by arrows. The choice of quantization axis is only schematic. 
            For each case, the resulting lower symmetry space groups are also indicated.}
    \label{fig:str_af}
\end{figure*}

Density functional theory (DFT) calculations were carried out using the
Perdew-Burke-Ernzerhof (PBE) implementation \cite{pbe1996} of the generalized gradient approximation (GGA)
using the full-potential local-orbital (FPLO) code\cite{klaus1999}, version 18.00-57 \cite{fplo_web}. 
A $k$-mesh with $12 \times 12 \times 12$ subdivisions was used for
numerical integration in the Brillouin zone (BZ) along with a linear
tetrahedron method. 
Spin-orbit effects were included in the self-consistent calculations via the 4-spinor formalism implemented
in the FPLO code. 
To study the relative stability of different collinear long-range ordered states (applicable to materials belonging to classes II \& III),
different lower-symmetry antiferromagnetic (AFM) configurations were
generated:
\begin{itemize}
    \item[-] {\bf AF1}, with AFM interlayer and ferromagnetic (FM) intralayer
        couplings,
      \item[-] {\bf AF2}, with FM interlayer and AFM intralayer interaction,
       \item[-] {\bf AF3}, with AFM interlayer as well as AFM intralayer interaction.
\end{itemize}
Fig. \ref{fig:str_af} shows these AFM configurations for compounds
belonging to classes II and III.

The description of the
$4f$ elements within the framework of DFT is still a subtle problem \cite{Richter2001}.
In this work, we considered both of the widely-used approaches to treat the $4f$-shell: the open-core (OC) 
approximation, and the GGA+$U$ scheme. In the first approach, we set the spin for R$^{3+}$
ions according to
the Hund's first rule while maintaining a spherically averaged distribution of electrons in the $4f$
shell \cite{Richter2001}. In the second approach, to circumvent the usual problem of multiple
metastable solutions, we used the occupation matrix control \cite{dorado2013,allen2014}, whereby
several different initial $4f$ density matrices, corresponding to ${+3}$ valence of R ions, were
considered to explore the energy landscape. We used the full-localized limit of the double counting
scheme with $F^0=7.0\,$eV, $F^2=11.92\,$eV, $F^4=7.96\,$eV,
and $F^6=5.89\,$eV, leading to $U=7\,$eV and $J=1\,$eV.  For most of the compounds, both GGA+OC and GGA+$U$
favors the magnetic moments to be in-plane. Therefore, we consider $\mathbf{m} \parallel [001]$ for all
the compounds for brevity and comparison. 
While we can anticipate that the existence of Weyl nodes at low energies is a finding robust to
these explored choices, we will discuss to what extent details such as position in energy and
momentum can be affected in a forthcoming publication.

To study the topological properties, we constructed a tight-binding model based on maximally
projected Wannier functions. In the basis set, states lying in the energy window $-9.5\,$eV
to $10\,$eV were considered. The Wannier basis set typically included the $5d$ and $6s$
states for
R (respectively, $3d$ and $4s$ for Y), valence $d$ and $s$ states for M, and $2s$ and $2p$ states for C. The accuracy of the resulting tight-binding
models was typically $\lesssim 15$ meV  compared to
the self-consistent bandstructures.

\clearpage
\newpage

\section{Weyl nodes in ${\boldmath{\texorpdfstring{{\rm GdCoC}_2}{gdco}}}$}
\label{sec_wn_gdcoc2}

\begin{figure}[h!]
    \centering
        \includegraphics[angle=0,width=0.4\textwidth]{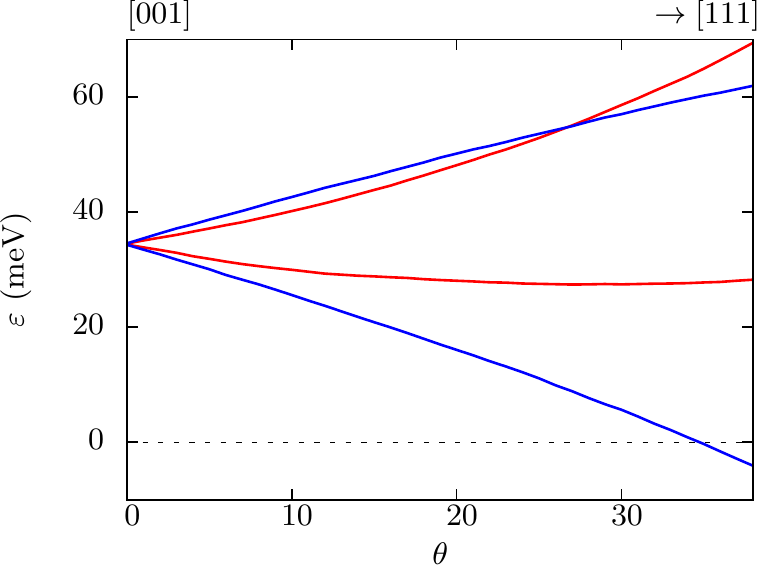}
        \caption{Energy of the Weyl nodes as a function of the canting angle $\theta$ for
        GdCoC$_2$. Blue and red lines, respectively, correspond to Weyl nodes of positive and negative chirality.}
    \label{fig:wn_gdcoc2}
\end{figure}

\begin{figure*}[h!]
    \centering
        \includegraphics[angle=0,width=0.895\textwidth]{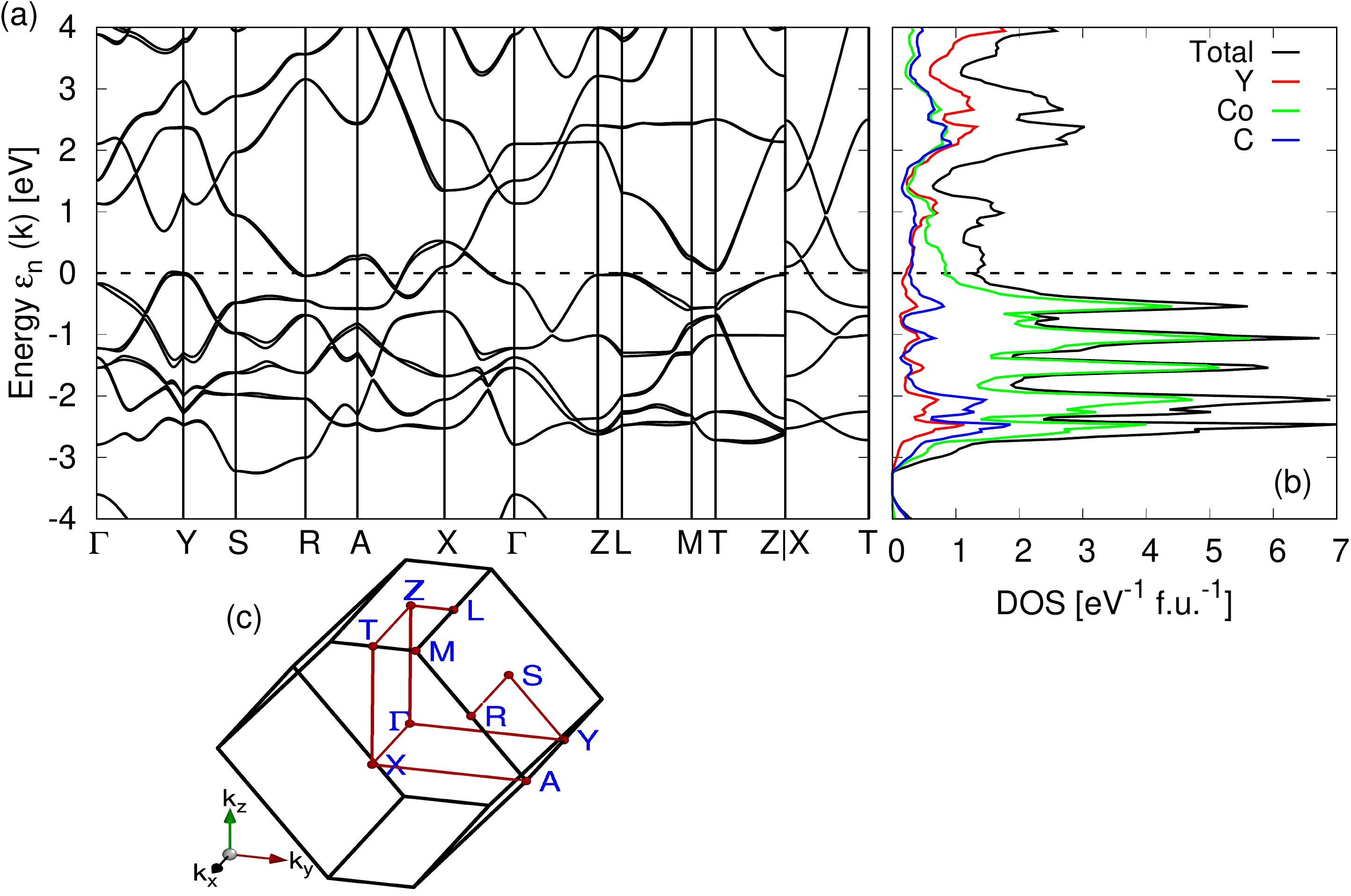}
        \caption{Electronic properties of {\yco}: (a) bandstructure, and
        (b) total and element-resolved density of states (DOS) with spin
        orbit effects included. The
        high-symmetry points are shown in (c).}
    \label{fig:elprop_ycoc2}
\end{figure*}

\begin{figure*}[h!]
    \centering
        \includegraphics[angle=0,width=0.895\textwidth]{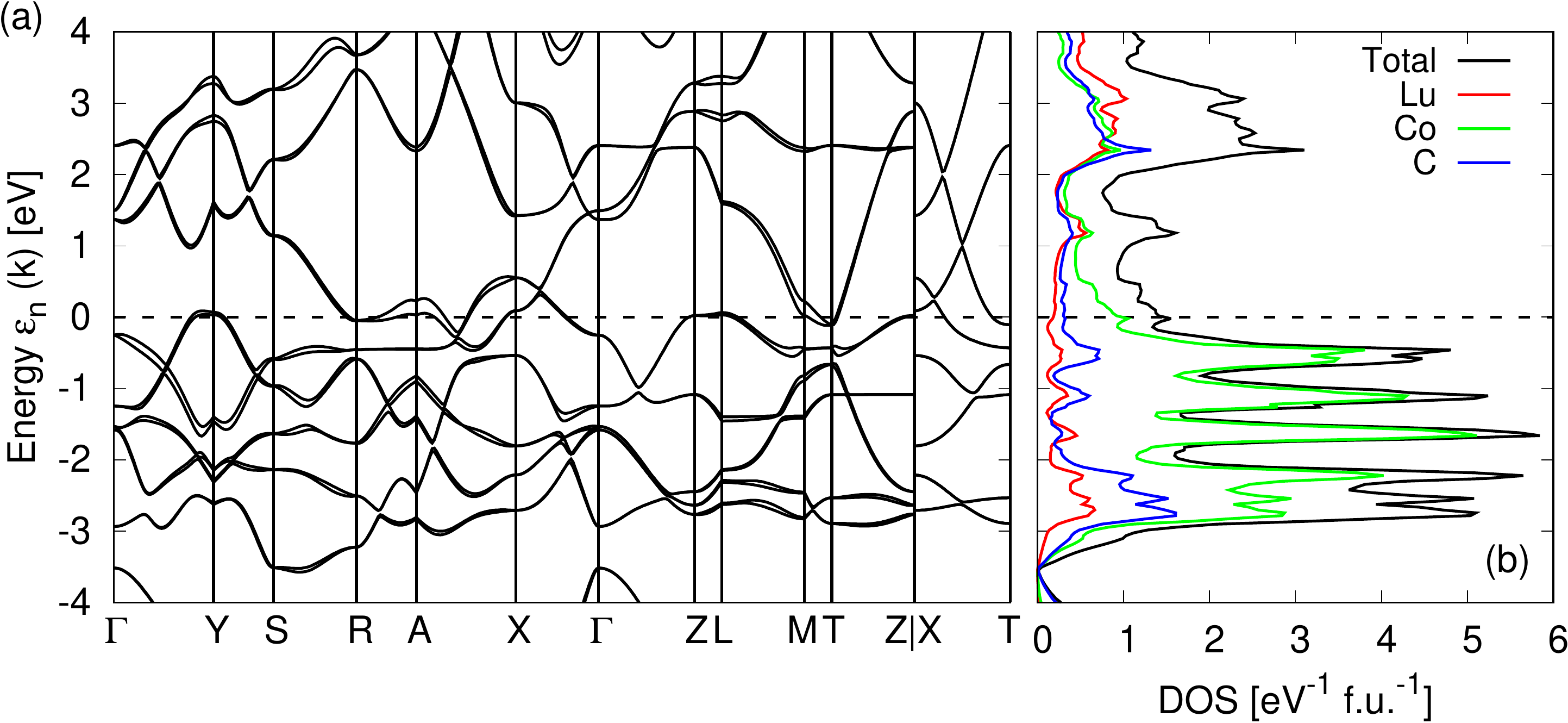}
        \caption{Electronic properties of {\luco}: (a)
            bandstructure, and (b) total and element-resolved contributions to DOS with spin-orbit effects included. The BZ and the high symmetry
    points are similar to that of {\yco}, shown in Fig. \ref{fig:elprop_ycoc2}(c).
    }
    \label{fig:elprop_lucoc2}
\end{figure*}

\begin{figure*}[h!]
    \centering
        \includegraphics[angle=0,width=0.795\textwidth]{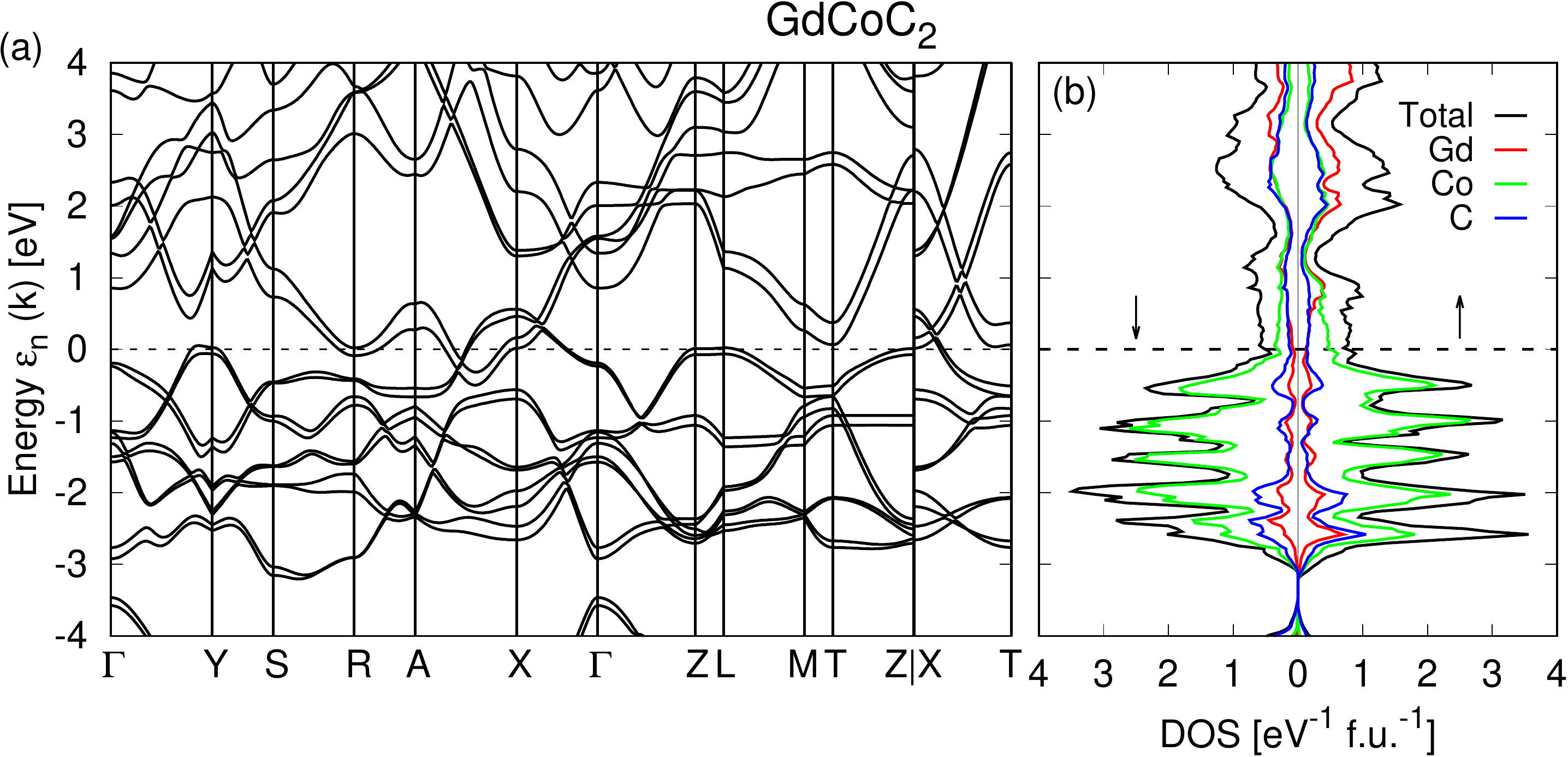}
        \includegraphics[angle=0,width=0.795\textwidth]{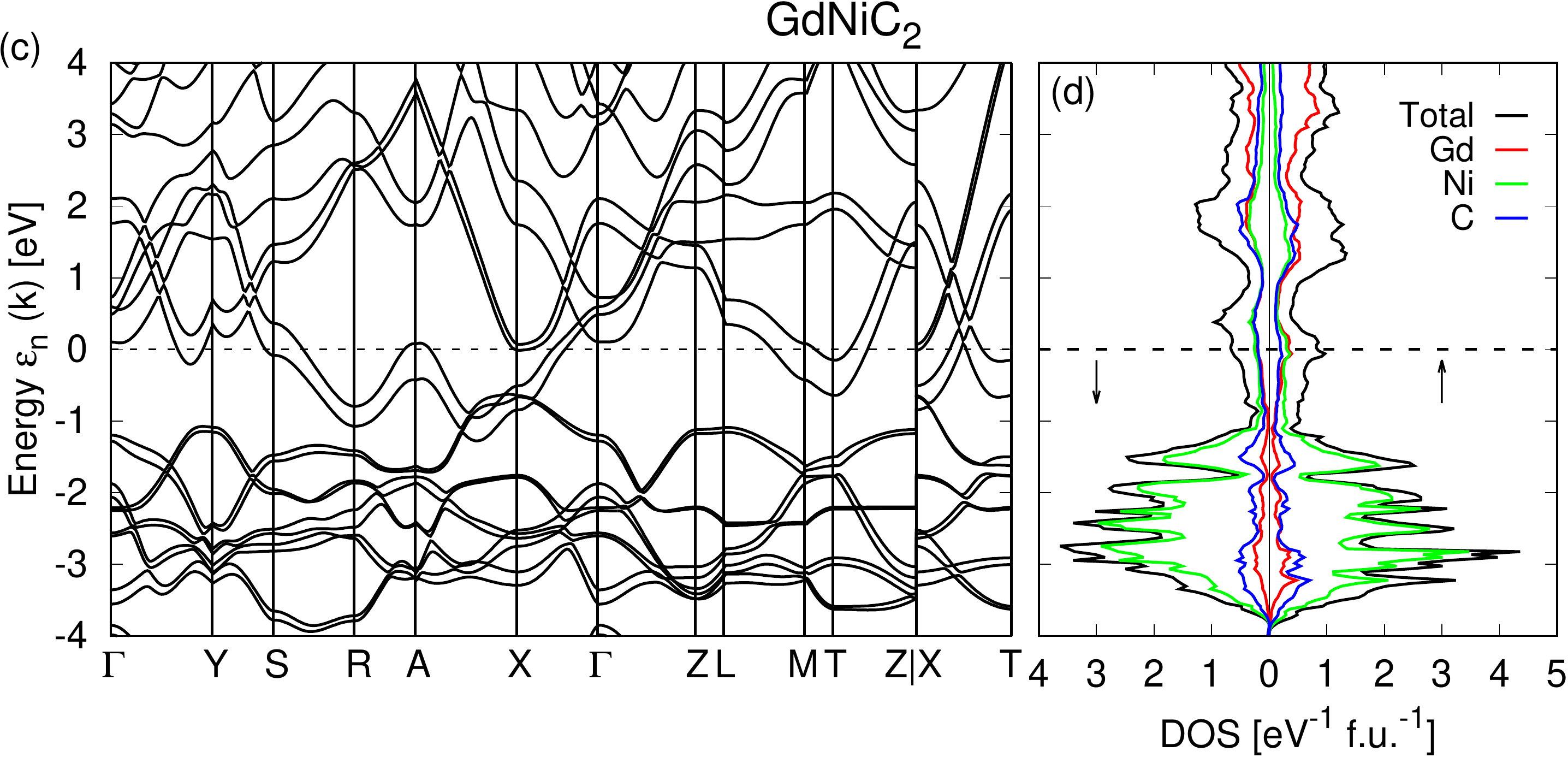}
        \includegraphics[angle=0,width=0.495\textwidth]{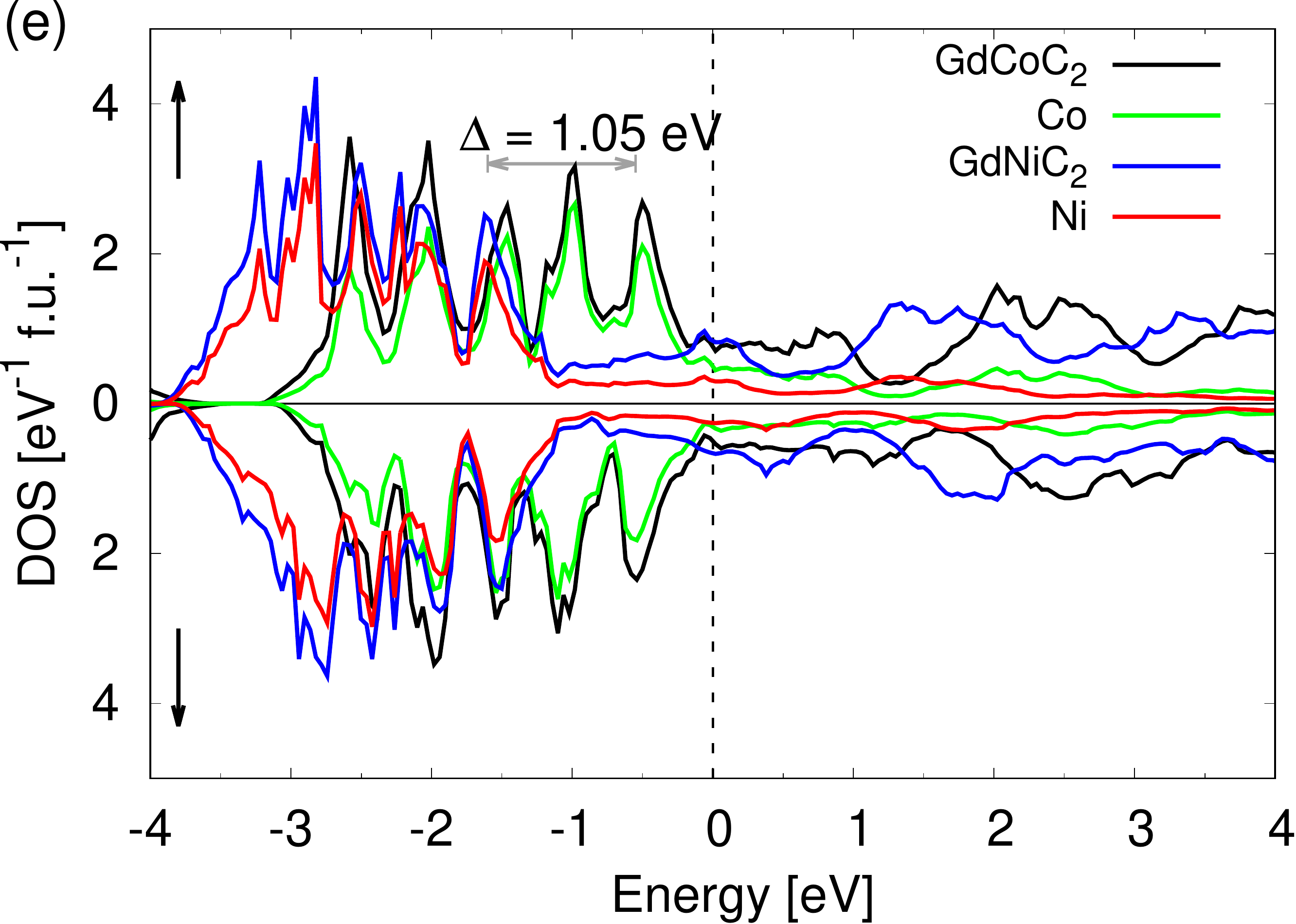}
        \caption{Electronic properties of GdMC$_2$: 
            bandstructures, and
            DOS for
            (a)-(b) {\gdco}, (c)-(d) {\gdni} for the FM state with spin
            quantization axis $\mathbf{m} \parallel [0 0 1]$, as obtained
        within the OC approach and spin-orbit effects included. The BZ and the
            corresponding high-symmetry points are similar to that of
            {\yco} [see Fig. \ref{fig:elprop_ycoc2}(c)]. The spin- and element-resolved
            partial DOS is also shown, where the spin components are
            indicated by arrows. A comparison of the total
            DOS and M-$3d$ contribution to the DOS for both the compounds is shown in (e), clearly
        showing that the relative shift is only nomimal $\sim 1\,$eV. }
    \label{fig:elprop_gdnico}
\end{figure*}

\begin{figure*}[h!]
    \centering
        \includegraphics[angle=0,width=0.995\textwidth]{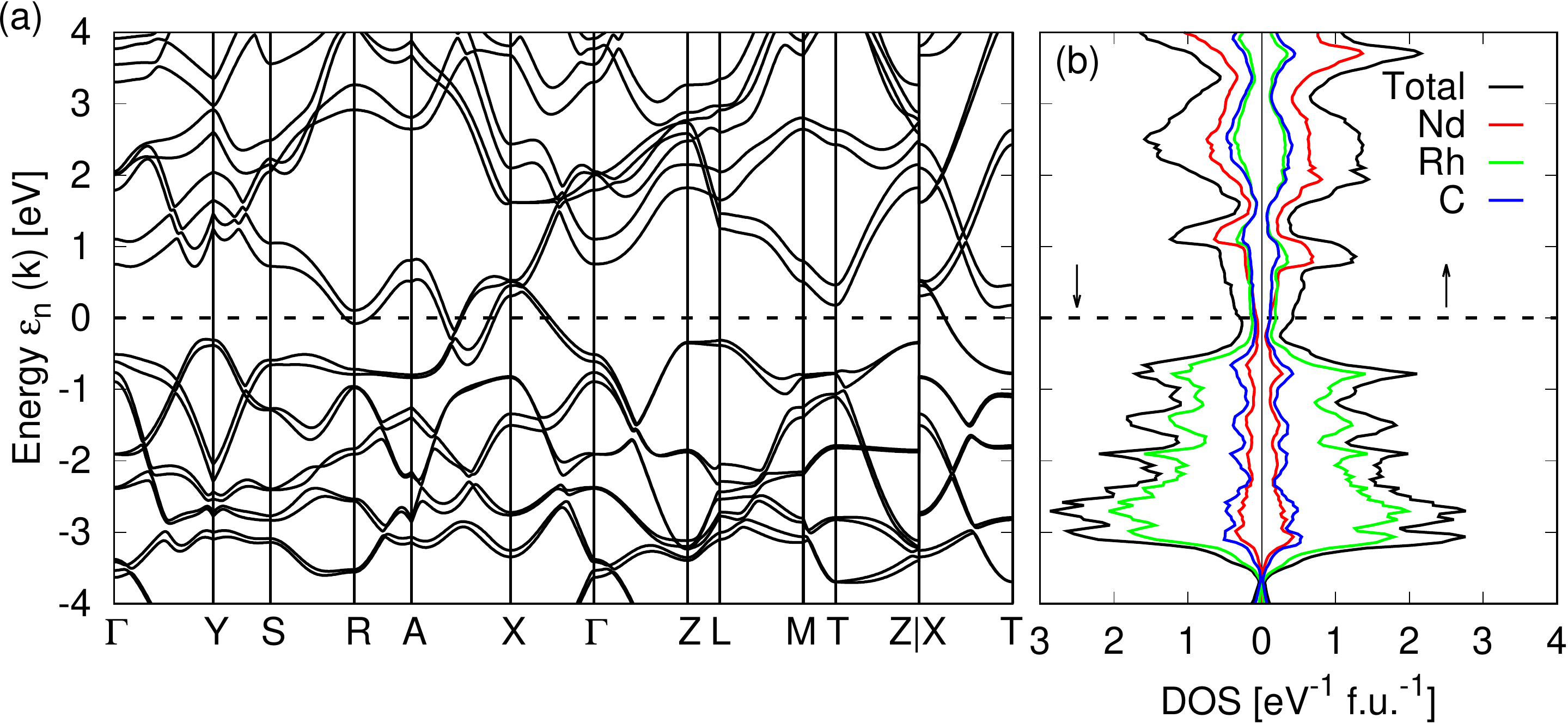}  
        \caption{Electronic properties of {\nd}: bandstructure, and
        total and element-resolved DOS for the FM
        configuration with $\mathbf{m} \parallel [0 0 1]$, as obtained within the OC
        approach and spin-orbit effects included. For the position of the
        high-symmetry points in the Brillouin zone, see Fig.
    \ref{fig:elprop_ycoc2}(c).}
    \label{fig:elprop_ndrhc2}
\end{figure*}

\begin{figure*}[h!]
    \centering
        \includegraphics[angle=0,width=0.995\textwidth]{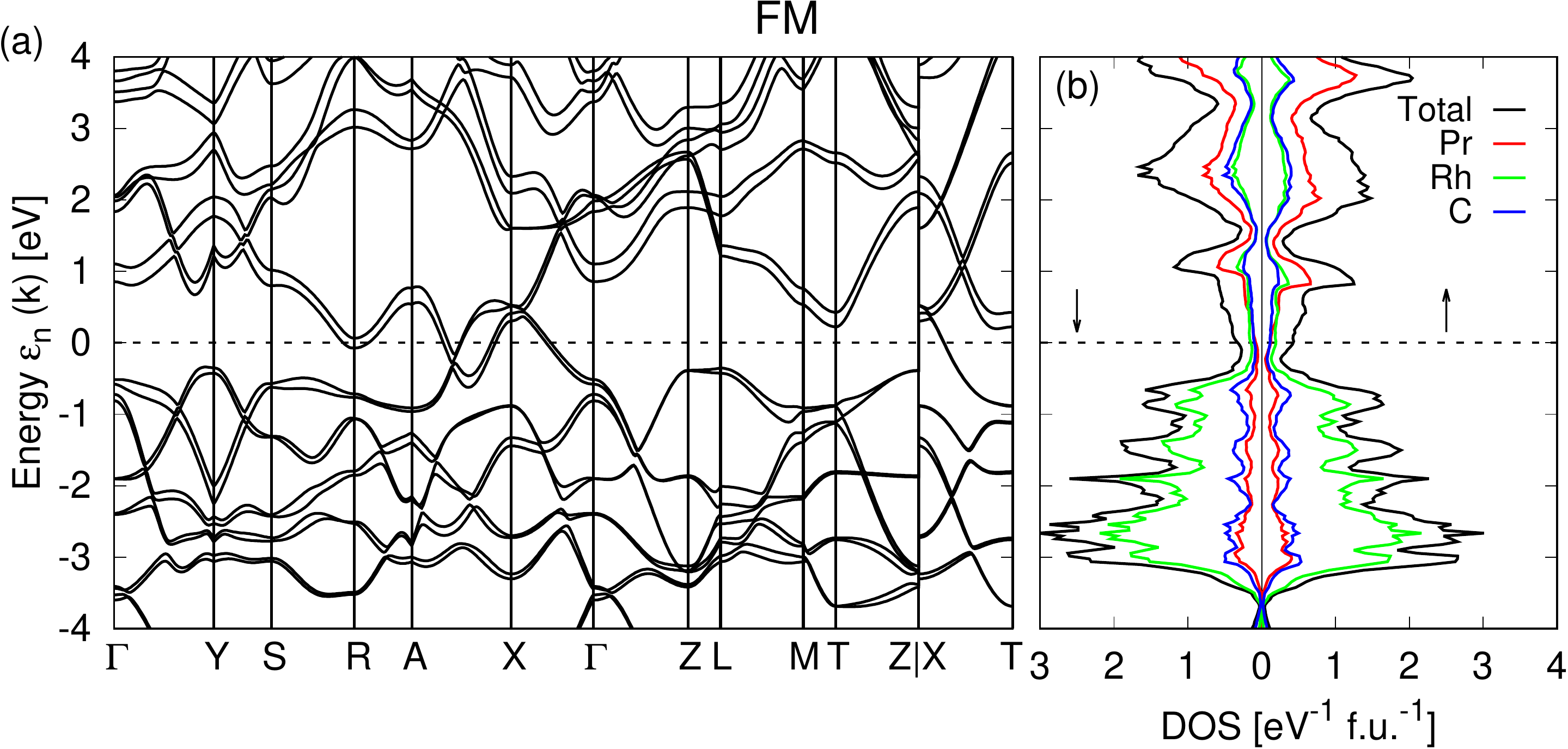}
        \includegraphics[angle=0,width=0.995\textwidth]{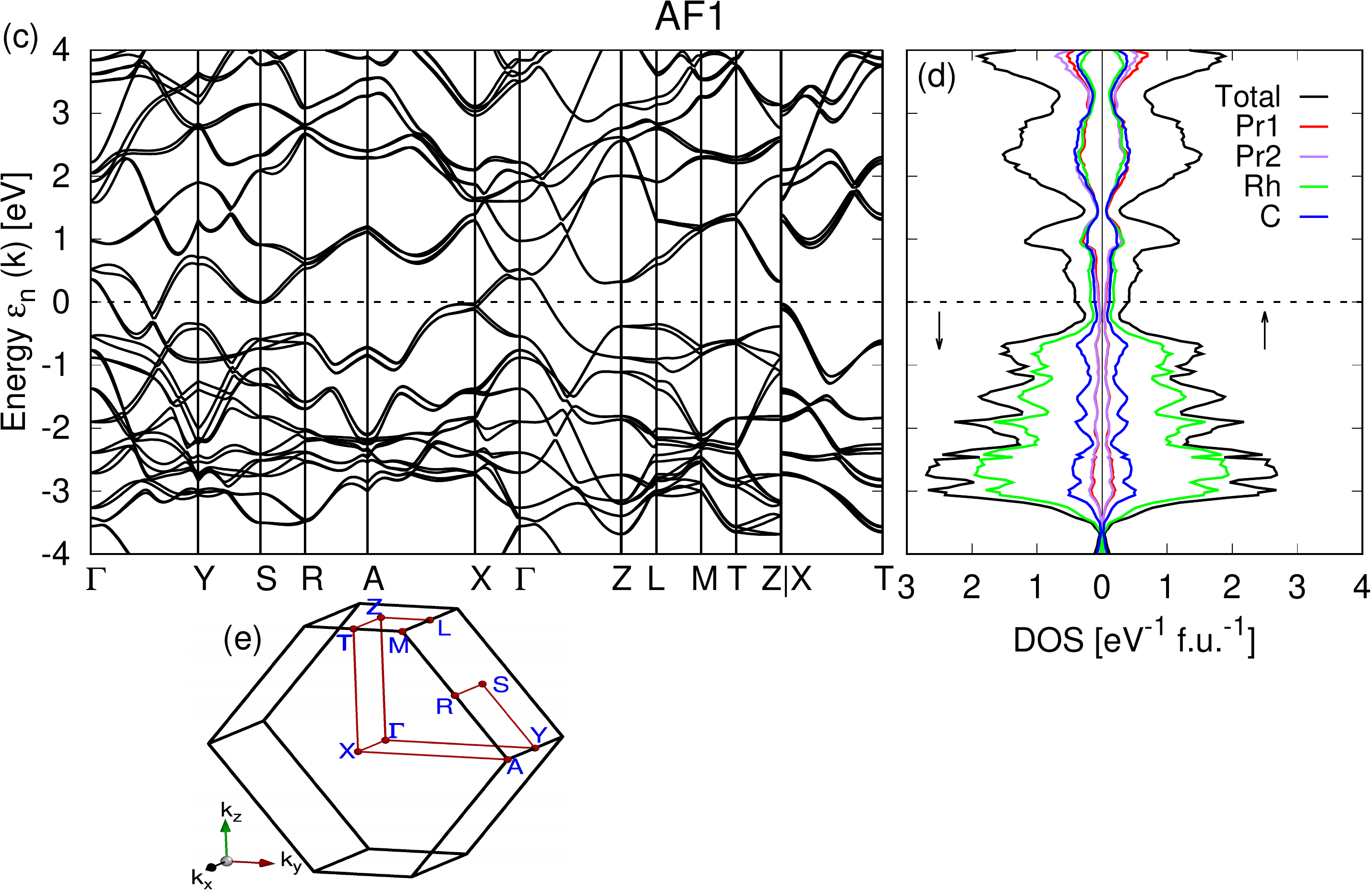}
        \caption{Electronic properties of {\pr}: bandstructure, and
        total and element-resolved DOS for (a)-(b) FM,
        and (c)-(d) AF1 magnetic configurations with $\mathbf{m} \parallel [0 0 1]$, as obtained within the OC
        approach and spin-orbit effects included. The BZ and the 
        high-symmetry points for both the AF1 magnetic configuration is
        similar to FM, since the AF1 state corresponds to doubling of the unit
cell along $a$ (see Fig. \ref{fig:str_af}), shown in (e).}
    \label{fig:elprop_prrhc2}
\end{figure*}

\begin{figure*}[h!]
    \centering
        \includegraphics[angle=0,width=0.995\textwidth]{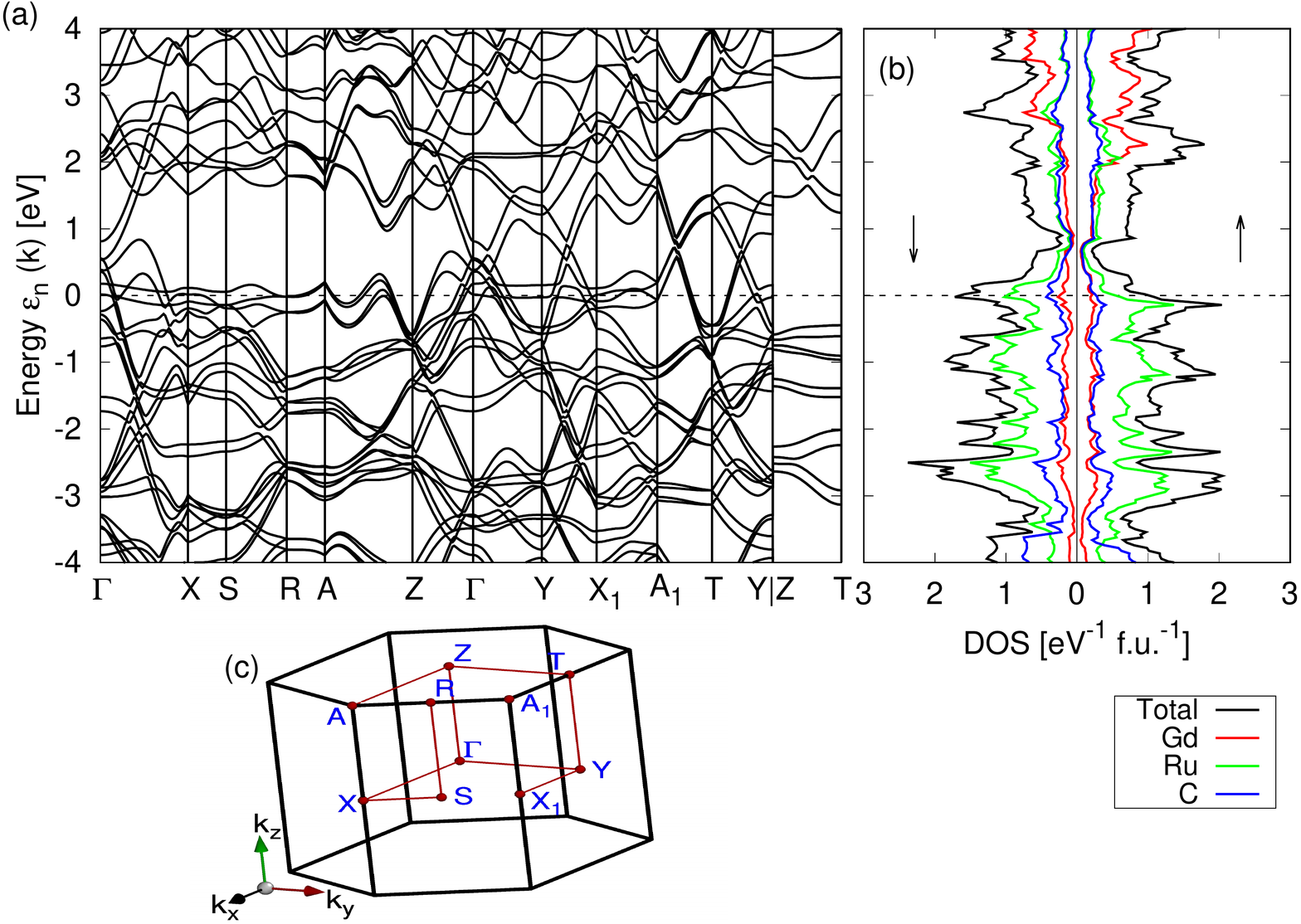}
        \caption{Electronic properties of {\gdru}: (a) bandstructure, and
        (b) total and element-resolved DOS, as obtained
        within the OC approach and spin-orbit effects included. The
        high-symmetry points are shown in (c). Note that due to the different number of valence
    electrons as
    compared to {\gdco}, the band structure here is noticeably more
complex at the Fermi energy and, therefore, {\gdru} should not be considered a semimetal.}
    \label{fig:elprop_gdruc2}
\end{figure*}

\begin{figure*}[h!]
    \centering
        \includegraphics[angle=0,width=0.795\textwidth]{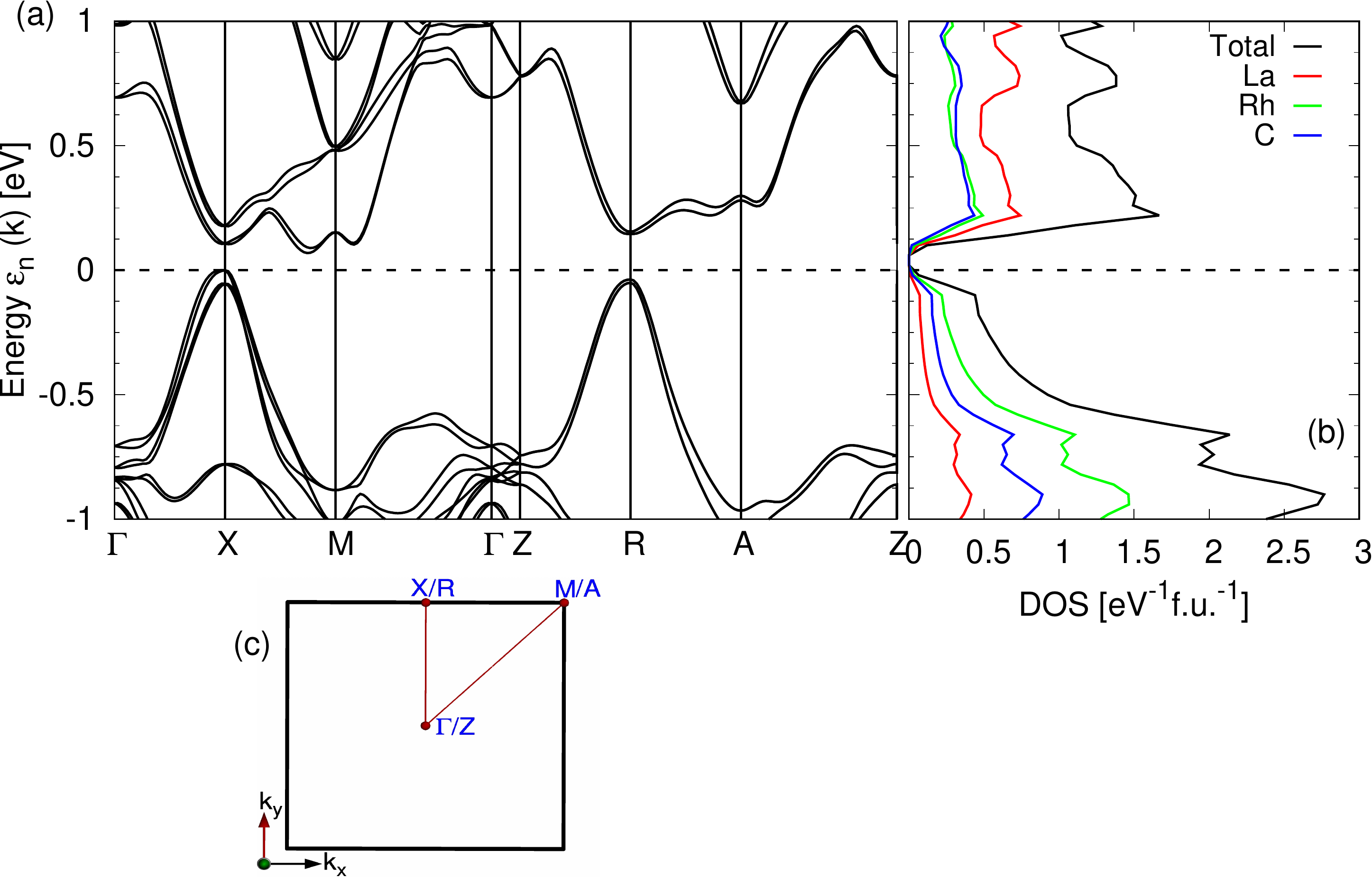}
        \caption{(a) Bandstructure and (b) DOS for
            {\la}, as obtained using the GGA functional with spin-orbit effects included. The band gap is approximately $\sim 84$ meV. The high
            symmetry points in the Brillouin zone are depicted in (c).}
    \label{fig:elprop_larhc2}
\end{figure*}

Fig. \ref{fig:wn_gdcoc2} shows the splitting in energy of the Weyl node closest to the Fermi energy in GdCoC$_2$ as the magnetization is canted from $[001]$ to $[111]$. 
Notice that energy splittings of tens of meV are obtained, similar to the results for NdRhC$_2$ in the main text, regardless of the smaller magnetic anisotropy energy expected in GdCoC$_2$ due to the half-filled 4f shell.

It is important to realize, as these results indicate, that the energy scale associated with the splitting in energy of the Weyl nodes need not scale with the magnetic anisotropy energy. Thus, sizeable splittings can occur in materials which require significantly different external magnetic field to cant the magnetization.

\section{Electronic properties}
\label{sec_EP}

For completeness, below  we provide the bandstructures and element-resolved density of states (DOS)
for the different compounds considered in this work and list all the respective Weyl nodes found in
the energy window between $\pm 120\,$meV.

\begin{table}[h!]
    \small
    \caption{Weyl points (WPs) in {\yco} arising from crossing of bands `Band' and `Band+1'. $N$ refers to the no.
        of electrons in the
    system and, therefore, also to the valence band lying closest to the
    Fermi energy. Other equivalent WPs can be generated by considering the
crystal symmetries $\{E, m(x), m(y), C_2(z)\}$ and the time-reversal symmetry.}
    \begin{tabular*}{0.48\textwidth}{ p{1.5cm} p{4.0cm} p{7.5cm} p{1.0cm}  p{1.5cm} }
        \hline\hline
        {\bf Band } & \multicolumn{1}{c}{\bf Energy} & \multicolumn{1}{c}{\bf Position} & \multicolumn{1}{c}{$\chi$}  & {\bf Deg}\\
                    &  \multicolumn{1}{c}{(meV)}             & \multicolumn{1}{c}{($k_x/a$, $k_y/b$, $k_z/c$)/$2\pi$} & & \\
        \hline
        $N-1$    & \multicolumn{1}{c}{ 60.7}   & \multicolumn{1}{c}{($0.366, 0.173, 0$)} &
\multicolumn{1}{c}{$\enspace 1$} & 4 \\
            & \multicolumn{1}{c}{103.7}   & \multicolumn{1}{c}{($-0.382,  0.159,  0.158$)}   & 
\multicolumn{1}{c}{$\enspace 1$} & 8  \\

        \hline
    \hline
    \end{tabular*}
    \label{table:WPs_ycoc2}
\end{table}
\begin{table}[h!]
    \small
    \caption{WPs in {\luco}. Other details and conventions are same as in Table
    \ref{table:WPs_ycoc2}.}
    \begin{tabular*}{0.48\textwidth}{ p{1.5cm} p{4.0cm} p{7.5cm} p{1.0cm}  p{1.5cm} }
    
        \hline\hline
        {\bf Band } & \multicolumn{1}{c}{\bf Energy} & \multicolumn{1}{c}{\bf Position} & \multicolumn{1}{c}{$\chi$}  & {\bf Deg}\\
                    &  \multicolumn{1}{c}{(meV)}  & \multicolumn{1}{c}{($k_x/a$, $k_y/b$, $k_z/c$)/$2\pi$} & & \\
        \hline
        $N-1$ & \multicolumn{1}{c}{ 89.5}   & \multicolumn{1}{c}{($-0.362, -0.163, 0$)} &
\multicolumn{1}{c}{$-1$} & 4 \\
        \hline
    \hline
    \end{tabular*}
    \label{table:WPs_lucoc2}
\end{table}
\begin{table}[h!]
    \small
    \caption{WPs in the FM-$[0 0 1]$ phase of {\gdco}, arising out of the crossing of the bands
    `Band' and `Band +1'. The OC solution was considered. $N$ refers to the number of valence electrons
in the system and, therefore, also to the valence band closest to the
Fermi energy. Other equivalent WPs can be generated by considering the
symmetries: $\{ E, C_2(z), m(x)\Theta, m(y) \Theta \}$.}

    \begin{tabular*}{0.48\textwidth}{ p{1.5cm} p{3.5cm} p{4.0cm} p{1.0cm}  p{1.5cm} }
        
            \hline\hline
        {\bf Band } & \multicolumn{1}{c}{\bf Energy} & \multicolumn{1}{c}{\bf Position}
        & \multicolumn{1}{c}{$\chi$}  & {\bf Deg}\\
       &  \multicolumn{1}{c}{(meV)}             & \multicolumn{1}{c}{($k_x/a$, $k_y/b$, $k_z/c$)/$2\pi$} & & \\
            \hline
        $N-1$ & \multicolumn{1}{c}{34.5}   & \multicolumn{1}{c}{($0.347, 0.206, 0.171$)} & 
\multicolumn{1}{c}{$\enspace 1$} & 4  \\
\\
        $N$  & \multicolumn{1}{c}{66.1}   & \multicolumn{1}{c}{($0.483, -0.114, -0.282$)} &
\multicolumn{1}{c}{$\enspace 1$} & 4 \\
             & \multicolumn{1}{c}{115.0}   & \multicolumn{1}{c}{($0.373, 0.129, 0.187$)} &
\multicolumn{1}{c}{$\enspace 1$} & 4 \\

        \hline
    \hline
    \end{tabular*}
    \label{table:WPs_gdcoc2}
\end{table}
\begin{table}
    \small
    \caption{WPs in {\gdni}. Other details and conventions are same as
    in Table \ref{table:WPs_gdcoc2}.}

    \begin{tabular*}{0.48\textwidth}{ p{1.5cm} p{3.5cm} p{4.0cm} p{1.0cm}  p{1.5cm} }
    
        \hline\hline
        {\bf Band } & \multicolumn{1}{c}{\bf Energy} & \multicolumn{1}{c}{\bf Position}
        & \multicolumn{1}{c}{$\chi$}  & {\bf Deg}\\
        &  \multicolumn{1}{c}{(meV)}   & \multicolumn{1}{c}{($k_x/a$, $k_y/b$, $k_z/c$)/$2\pi$} & & \\
        \hline
        $N+1$ & \multicolumn{1}{c}{108.3}   & \multicolumn{1}{c}{($-0.5, 0, 0.487$)}  &
\multicolumn{1}{c}{$-1$} & 2  \\

        \hline
    \hline
    \end{tabular*}
    \label{table:WPs_gdnic2}
\end{table}
\begin{table}
    \small
    \caption{WPs in {\nd}. Other details and conventions are same as
    in Table \ref{table:WPs_gdcoc2}.}

    \begin{tabular*}{0.48\textwidth}{ p{1.5cm} p{3.5cm} p{5.0cm} p{1.0cm}  p{1.5cm} }
    
        \hline\hline
        {\bf Band } & \multicolumn{1}{c}{\bf Energy} & \multicolumn{1}{c}{\bf Position}
        & \multicolumn{1}{c}{$\chi$}  & {\bf Deg}\\
        &  \multicolumn{1}{c}{(meV)}   & \multicolumn{1}{c}{($k_x/a$, $k_y/b$, $k_z/c$)/$2\pi$} & & \\
        \hline

        $N$   & \multicolumn{1}{c}{-114.1}  & \multicolumn{1}{c}{($-0.414, -0.328, 0.220$)} &
\multicolumn{1}{c}{$\enspace 1$} & 4 \\
            & \multicolumn{1}{c}{43.4}   & \multicolumn{1}{c}{($-0.393, -0.233, -0.199$)} &
\multicolumn{1}{c}{$\enspace 1$} & 4 \\

        \hline
    \hline
    \end{tabular*}
    \label{table:WPs_ndrhc2}
\end{table}
\begin{table}
    \small
    \caption{WPs in the AF1-[001] state of {\pr}. Other details and conventions are same as
        in Table \ref{table:WPs_gdcoc2}. The near-degeneracy of WPs between bands ($N$,
    $N+1$) and ($N+1$, $N+2$) reflect that $\Theta \mathcal{T}$, where $\mathcal{T}$ is the 
    translation operator for shift by half a unit cell along $a$, is only marginally broken
around these WPs. }
    
    \begin{tabular*}{0.48\textwidth}{ p{1.1cm} p{2.0cm} p{5.0cm} p{1.0cm}  p{1.0cm} }
    
        \hline\hline
        {\bf Band } & \multicolumn{1}{c}{\bf Energy} & \multicolumn{1}{c}{\bf Position} & \multicolumn{1}{c}{$\chi$}  & {\bf Deg}\\
        &  \multicolumn{1}{c}{(meV)}    & \multicolumn{1}{c}{($k_x/a$, $k_y/b$, $k_z/c$)/$2\pi$} & & \\
        \hline
        $N-1$ & \multicolumn{1}{c}{ 9.5}   & \multicolumn{1}{c}{($0, -0.158, -0.222$)} & 
\multicolumn{1}{c}{$\enspace 1$} & 4 \\
            & \multicolumn{1}{c}{75.1}   & \multicolumn{1}{c}{($-0.231, -0.203, 0$)}
&\multicolumn{1}{c}{$\enspace 1$} & 4 \\

        \\
        $N$  & \multicolumn{1}{c}{-100.03} & \multicolumn{1}{c}{($0.1952, 0.3276, 0.2216$)}
&\multicolumn{1}{c}{$\enspace 1$} & 4 \\
             & \multicolumn{1}{c}{-99.55} & \multicolumn{1}{c}{($-0.1954, -0.3274,-0.2214$)}
&\multicolumn{1}{c}{$\enspace 1$} & 4 \\
        \\
        $N+1$ & \multicolumn{1}{c}{-36.19}  & \multicolumn{1}{c}{($ 0.1215, 0.3542, 0.1725$)} &
\multicolumn{1}{c}{$-1$} & 4 \\
            & \multicolumn{1}{c}{-36.18}  & \multicolumn{1}{c}{($-0.1230, -0.3540, -0.1726$)} &
\multicolumn{1}{c}{$-1$} & 4 \\
            & \multicolumn{1}{c}{-27.3}  & \multicolumn{1}{c}{($-0.004, -0.363, 0.163$)} &
\multicolumn{1}{c}{$\enspace 1$} & 4 \\

        \hline
    \hline
    \end{tabular*}
    \label{table:WPs_prrhc2}
\end{table}

\clearpage
\newpage

\end{document}